\documentclass[preprint2]{aastex}
\usepackage{spr-astr-addons}
\usepackage{url}
\usepackage{lscape}
\bibpunct{(}{)}{,}{n}{,}{,}

\RequirePackage{color}

\begin{document}

 \title{Homing in for New Year: impact parameters and pre-impact 
        orbital evolution of meteoroid 2014~AA}

 \shorttitle{Pre-impact orbital evolution of 2014~AA}
 \shortauthors{de la Fuente Marcos et al.}

 \author{C.~de~la~Fuente Marcos}
  \and
 \author{R.~de~la~Fuente Marcos}
 \affil{Apartado de Correos 3413, E-28080 Madrid, Spain}
  \and
 \author{P. Mialle}
 \affil{Provisional Technical Secretariat,
        Comprehensive Nuclear-Test-Ban Treaty Organisation,
        PO Box 1200, Vienna 1400, Austria}
 \email{carlosdlfmarcos@gmail.com}

  \begin{abstract}
     On 2008 October 7, small asteroid 2008~TC$_{3}$ turned itself into the 
     parent body of the first meteor ever to be predicted before entering 
     the Earth's atmosphere. Over five years later, the 2014~AA event became 
     the second instance of such an occurrence. The uncertainties associated 
     with the pre-impact orbit of 2008~TC$_{3}$ are relatively small because 
     thousands of observations were made during the hours preceding the 
     actual meteor airburst. In sharp contrast, 2014~AA was only observed 
     seven times before impact and consequently its trajectory is somewhat 
     uncertain. Here, we present a recalculation of the impact parameters 
     ---location and timing--- of this meteor based on infrasound recordings. 
     The new values ---($\lambda_{\rm impact}$, $\phi_{\rm impact}$, $t_{\rm 
     impact}$) = (-44\degr, +11\degr, 2456659.618 JD UTC)--- and their
     uncertainties together with Monte Carlo and $N$-body techniques, are 
     applied to obtain an independent determination of the pre-impact orbit 
     of 2014~AA: $a$~=~1.1623 AU, $e$~=~0.2116, $i$~=~1$\fdg$4156, 
     $\Omega$~=~101$\fdg$6086, and $\omega$~=~52$\fdg$3393. Our orbital 
     solution is used to investigate the possible presence of known 
     near-Earth objects (NEOs) moving in similar orbits. Among the objects 
     singled out by this search, the largest is 2013~HO$_{11}$ with an 
     absolute magnitude of 23.0 (diameter 75--169~m) and a MOID of 0.006~AU. 
     Prior to impact, 2014~AA was subjected to a web of overlapping secular 
     resonances and it followed a path similar to those of 2011~GJ$_{3}$, 
     2011~JV$_{10}$, 2012~DJ$_{54}$, and 2013~NJ$_{4}$. NEOs in this 
     transient group have their orbits controlled by close encounters with 
     the Earth--Moon system at perihelion and Mars at aphelion, perhaps 
     constituting a dynamical family. Extensive comparison with other studies 
     is also presented.
  \end{abstract}


  \section{Introduction}
     On 2014 January 2, asteroid 2014~AA became the second example of an object discovered just prior to hitting the Earth (Brown 2014; 
     Jenniskens 2014). Over five years before this event occurred, similarly small 2008~TC$_{3}$ had stricken our planet hours after being 
     first spotted (Chodas et al. 2009; Jenniskens et al. 2009; Oszkiewicz et al. 2012). The fireball caused by the entry of 2008~TC$_{3}$ 
     was observed by the Meteosat 8 weather satellite (Borovi\v{c}ka \& Charv\'at 2009); no images of the atmospheric entry of 2014~AA have 
     emerged yet, although there is robust evidence for an impact over the Atlantic Ocean (Chesley et al. 2015; Farnocchia et al. 2016) less 
     than a day after this small asteroid was discovered. 

     The uncertainty associated with the pre-impact orbit of 2008~TC$_{3}$ is relatively small because thousands of observations were made 
     during the hours preceding the actual strike (Chodas et al. 2009; Jenniskens et al. 2009; Kozubal et al. 2011; Oszkiewicz et al. 2012). 
     In sharp contrast, 2014~AA was only observed seven times before impact (Kowalski et al. 2014) and consequently its pre-impact orbit is 
     somewhat uncertain (Chesley et al. 2014, 2015; Farnocchia et al. 2016). Both objects hit the Earth about 20 hours after they were first 
     detected (Farnocchia et al. 2015). Asteroid 2008~TC$_{3}$ completely broke up over northern Sudan on 2008 October 7 (Jenniskens et al. 
     2009); asteroid 2014~AA probably met a similar end over the Atlantic Ocean. Meteorites were collected from 2008~TC$_{3}$ (Jenniskens et 
     al. 2009); any surviving meteorites from 2014~AA were likely lost to the sea. 

     Both 2008~TC$_{3}$ and 2014~AA had similar sizes of a few metres. Such small asteroids or meteoroids (diameter $<$ 10 m) are probably 
     fragments of larger objects, which may also be fragments themselves. The study of the orbital dynamics of such fragments is a subject 
     of considerable practical interest because small bodies dominate the risk of unanticipated Earth impacts with just local effects (Brown 
     et al. 2013). Asteroid fragmentation could be induced by collisional processes (e.g. Dorschner 1974; Ryan 2000) but also be the 
     combined result of thermal fatigue (e.g. \v{C}apek \& Vokrouhlick\'y 2010) and rotational (e.g. Walsh et al. 2008) or tidal stresses 
     (e.g. Richardson et al. 1998; T\'oth et al. 2011). The present-day rate of catastrophic disruption events of asteroids in the main belt 
     has been recently studied by Denneau et al. (2015). These authors have found that the frequency of this phenomenon is much higher than 
     previously thought, with rotational disruptions being the dominant source of fragments. Production of fragments can be understood 
     within the context of active asteroids (see e.g. Jewitt 2012; Jewitt et al. 2015; Drahus et al. 2015; Agarwal et al. 2016). 

     Here, we revisit the topic of the impact parameters of 2014~AA, then apply Monte Carlo and $N$-body techniques to obtain an independent 
     determination of the pre-impact orbit of 2014~AA. The computed orbital solution is used to investigate the existence of near-Earth 
     objects (NEOs) moving in similar orbits. This paper is organized as follows. In Sect. 2, we review what is currently known about 
     2014~AA. The atmospheric entry of 2014~AA is revisited in Sect. 3; an improved impact solution (location and timing), that is based 
     on a refined analysis of infrasound recordings, is presented. A Monte Carlo technique is used in Sect. 4 to estimate the most probable, 
     in geometric terms, pre-impact orbit. An $N$-body approach is described and applied in Sect. 5 to obtain a more realistic orbital 
     solution. In Sect. 6, we provide an extensive and detailed comparison with results obtained by other authors and show that all the 
     solutions published so far are reasonably consistent. Based on the new solution, the recent past orbital evolution of 2014~AA is 
     reconstructed in Sect. 7. A number of perhaps dynamically-related small bodies are discussed in Sect. 8. Section 9 summarizes our 
     conclusions.
%
%
     \begin{table*}
      \centering
      \fontsize{8}{11pt}\selectfont
      \tabcolsep 0.15truecm
      \caption{Heliocentric Keplerian orbital elements of 2014~AA from the JPL Small-Body Database and \textsc{Horizons} On-Line Ephemeris 
               System. Values include the 1$\sigma$ uncertainty. The orbits are computed at epoch JD 2456658.5 that corresponds to 0:00 TDB 
               on 2014 January 1. The orbit in the left-hand column was computed on 2014 June 13 18:59:39 \textsc{ut} and it is based on 
               seven astrometric observations; the orbit in the column next to it was computed on 2015 April 13 00:35:03 \textsc{ut} and it 
               is based on eight observations (seven astrometric and one infrasounds-based). The third orbit is the one currently available; 
               it was computed on 2016 January 13 11:17:04 \textsc{ut} and, as the previous one, is based on seven astrometric observations 
               and one infrasounds-based (Farnocchia et al. 2016). Values in parentheses are referred to epoch JD 2456658.628472222 that 
               corresponds to 03:05:00.0000 TDB on 2014 January 1 (or nearly 24 h before impact time) and are based on the orbital solution 
               displayed in the column next to it (J2000.0 ecliptic and equinox).
              }
      \begin{tabular}{llllll}
       \hline
        Semi-major axis, $a$ (AU)                         & = &   1.16$\pm$0.03   &   1.163$\pm$0.011   &   1.162$\pm$0.004   & (1.162312786616874)    \\
        Eccentricity, $e$                                 & = &   0.21$\pm$0.03   &   0.213$\pm$0.011   &   0.211$\pm$0.004   & (0.2116141752291786)   \\
        Inclination, $i$ (\degr)                          & = &   1.4$\pm$0.2     &   1.42$\pm$0.07     &   1.41$\pm$0.03     & (1.415646256117421)    \\
        Longitude of the ascending node, $\Omega$ (\degr) & = & 101.58$\pm$0.12   & 101.61$\pm$0.02     & 101.613$\pm$0.010   & (101.6086439360293)    \\
        Argument of perihelion, $\omega$ (\degr)          & = &  52.3$\pm$1.2     &  52.4$\pm$0.5       &  52.3$\pm$0.2       & (52.33920188906649)    \\
        Mean anomaly, $M$ (\degr)                         & = & 324.2$\pm$1.3     & 324.1$\pm$0.4       & 324.0$\pm$0.2       & (324.1460200866331)    \\
        Time of perihelion passage, $\tau$ (JD TDB)       & = & 2456245$\pm$16    & 2456704.24$\pm$0.06 & 2456704.22$\pm$0.02 & (2456704.213037788402) \\
        Perihelion, $q$ (AU)                              & = &   0.916$\pm$0.009 &   0.916$\pm$0.004   &   0.917$\pm$0.002   & (0.9163509249186160)   \\
        Aphelion, $Q$ (AU)                                & = &   1.41$\pm$0.03   &   1.411$\pm$0.013   &   1.407$\pm$0.005   & (1.408274648315132)    \\
        Absolute magnitude, $H$ (mag)                     & = &  30.9             &                     &                     &                        \\
       \hline
      \end{tabular}
      \label{elements}
     \end{table*}
%
%

  \section{Asteroid 2014 AA}
     Asteroid 2014~AA was discovered on 2014 January 1 by R.~A. Kowalski using the 1.5-m telescope of the Mount Lemmon Survey in Arizona 
     (Kowalski et al. 2014), becoming the first asteroid identified in 2014. It was initially observed at a $V$-magnitude of 19.1 and found 
     to be a very small body with $H$ = 30.9 mag which translates into a diameter in the range 1--4 m for an assumed albedo of 0.20--0.04. 
     The available orbits of this Apollo meteoroid are based on just seven astrometric observations for a data-arc span of 1 hour and 9.5 
     minutes; therefore, its actual path is poorly constrained (see Table \ref{elements} for the orbits computed by the Solar System 
     Dynamics Group or SSDG).\footnote{The orbit available from the Minor Planet Center is: $a$ = 1.1605495 AU, $e$ = 0.2092087, $i$ = 
     1\fdg39894, $\Omega$ = 101\fdg70409, and $\omega$ = 52\fdg02425, referred to the epoch 2456600.5 JD TDB.}$^{,}$\footnote{The orbit 
     available from NEODyS-2 is: $a$ = 1.17$\pm$0.03 AU, $e$ = 0.22$\pm$0.03, $i$ = 1\fdg4$\pm$0\fdg2, $\Omega$ = 101\fdg57$\pm$0\fdg12, and 
     $\omega$ = 52\degr$\pm$1\degr, referred to the epoch 2456658.3 JD TDB.} With a value of the semi-major axis of 1.16 AU, its 
     eccentricity was moderate, $e$~=~0.21, and its inclination very low, $i$~=~1\fdg4. Its Minimum Orbit Intersection Distance (MOID) with 
     our planet was 4.5$\times$10$^{-7}$ AU and it orbited the Sun with a period of 1.25 yr. This type of orbit is only directly perturbed 
     by the Earth--Moon system (at perihelion) and Mars (at aphelion). 

     In spite of the uncertain orbit, independent calculations carried out by Bill Gray, the Minor Planet Center (MPC), and Steven R. 
     Chesley at the Jet Propulsion Laboratory (JPL) all claimed a virtually certain collision between 2014~AA and our planet to occur on 
     2014 January 2.2$\pm$0.4 (Kowalski et al. 2014); in particular, Steven R. Chesley predicted impact locations along an arc extending 
     from Central America to East Africa.\footnote{\url{http://neo.jpl.nasa.gov/news/news182.html}} Using data from the Comprehensive 
     Nuclear-Test-Ban Treaty Organization (CTBTO) infrasound sensors, Steven R. Chesley, Peter Brown, and Peter Jenniskens computed a 
     probable impact location; Steven R. Chesley pointed out that the impact time could have been 2014 January 2 at 4:02 UTC, with a 
     temporal uncertainty of tens of minutes, and the impact location coordinates could have been 11\fdg7 N, 318\fdg7 E (or 41\fdg3 W) with 
     a spatial error of a few hundred kilometres.\footnote{\url{http://neo.jpl.nasa.gov/news/news182a.html}}

     Chesley et al. (2015) have released the hypocentre location solution for the 2014~AA impact (see their table 1) as included in the 
     Reviewed Event Bulletin (REB) of the International Data Centre (IDC) of the CTBTO for 2014 January 2. This impact solution has been 
     utilized in Farnocchia et al. (2016) to further improve the trajectory of 2014~AA. The impact time was 2014 January 2 at 3:05:25 UTC 
     with an uncertainty of 632 s (epoch JD 2456659.628762$\pm$0.007315). The impact location coordinates were latitude ({\degr}N) equal to 
     +14\fdg6326 and longitude ({\degr}E) of $-$43\fdg4194 with an uncertainty of about 3\fdg5$\times$1\fdg4 and a major axis azimuth of 
     76\degr (clockwise from N, see fig. 5 in Chesley et al. 2015). These values and those from the improved impact solution presented in 
     the following section are used here as constraints to compute the pre-impact orbit of 2014~AA. Our approaches do not initially rely on 
     actual astrometric observations of this meteoroid obtained prior to its impact, but on Keplerian orbits (geometry) and Newtonian 
     gravitation ($N$-body calculations). None the less, the available astrometry is used later to further refine our orbital solution. 
     Therefore, our favoured orbital solution combines the original, ground-based optical astrometry and infrasound data.

  \section{An improved determination of the hypocentre location}
     The IDC of the CTBTO in Vienna processes automatically and in near real time continuous recordings from the globally deployed 
     International Monitoring System (IMS) infrasound stations. The IDC automatic system is designed to detect close-to-the-ground, 
     explosion-like signals. Station detections are associated to form events. The system can automatically associate signal detections at 
     distances up to 6,700 km (or 60\degr) from the source location; for larger propagation distances the signals are manually associated
     with the event. The signal from the impact of 2014 AA was automatically detected by the IDC automatic system. The reviewed analysis 
     carried out in the hours following the event (published in the REB) provided a refined list of infrasound signals associated with the 
     meteor as well as an improved source location based on infrasound recordings. 

\begin{figure}
        \centering
        \includegraphics[width=\linewidth]{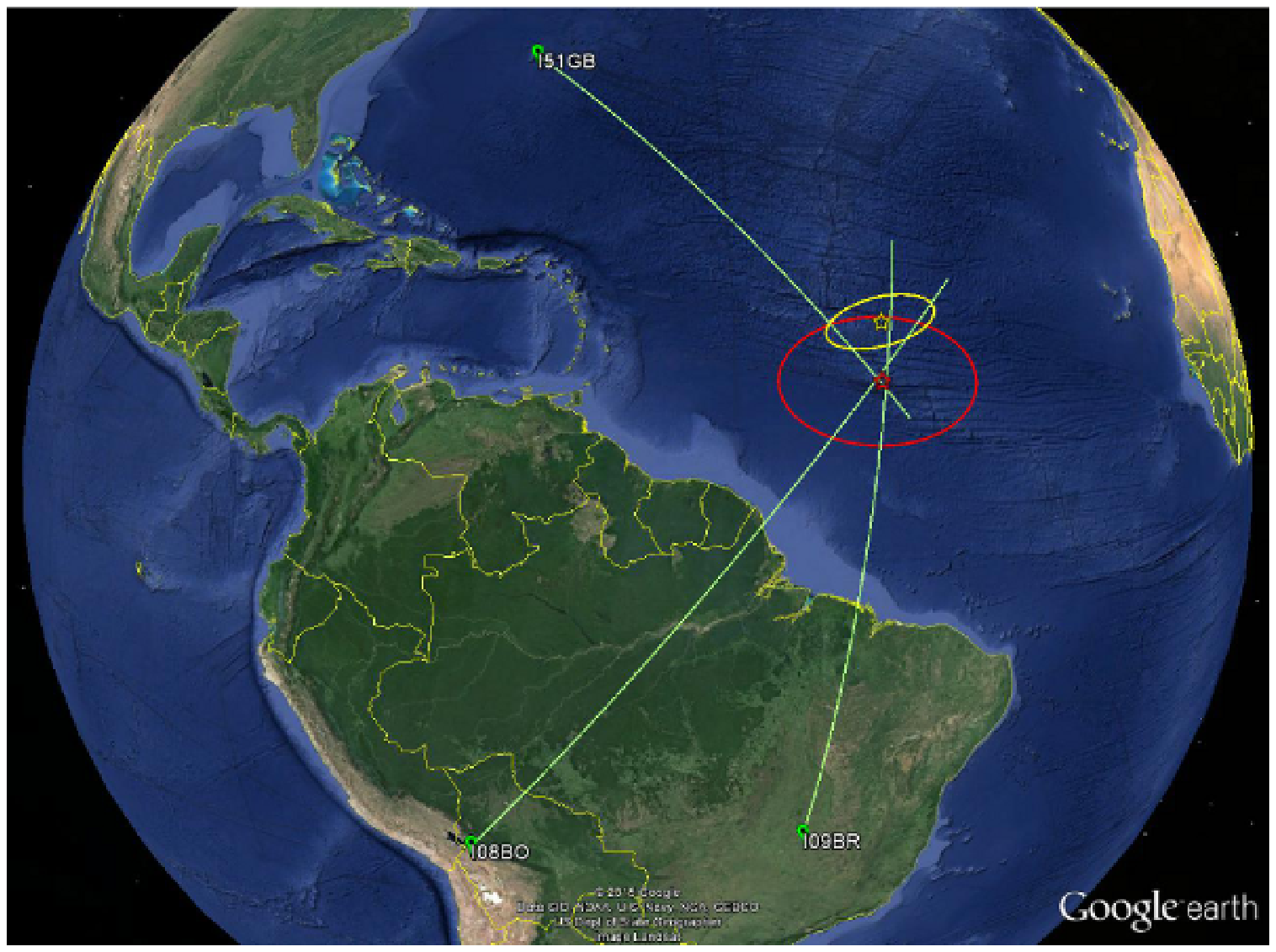}
        \caption{Hypocentre location of the 2014~AA event on the surface of the Earth (red star and error ellipse) as given in Table 
                 \ref{pierrick} and assuming specific source properties (see the text for details). The previous REB determination (yellow 
                 star and error ellipse), and the locations and codes of the three detecting stations are also displayed. The error ellipses
                 show the 90th percentile; the estimated directions of arrival (back-azimuth) are displayed for each infrasound station 
                 (green lines).
                }
        \label{pmialle}
     \end{figure}

     In the REB of the IDC, signals recorded at three IMS infrasound stations were associated to build an event in the Atlantic, 1,450 km 
     to the north-east of French Guyana, at coordinates (14\fdg63 N, 43\fdg42 W) with an error ellipse of 390.4 km$\times$154.8 km 
     (semi-major axis$\times$semi-minor axis) and a major axis azimuth of 76\fdg4. The source origin time of the main blast was estimated at 
     03:05:25 UTC with an origin time error of about 630 s. The three IMS infrasound stations that recorded the airburst are located at 
     large distances, ranging from 2,900 to 4,400 km from the REB location, one located to the north-west in the Northern Hemisphere and 
     two to the south-west across the Equator (see Fig. \ref{pmialle}). For further technical details, see sect. 2.2 in Farnocchia et al. 
     (2016).

     The IDC system makes use of back-azimuths (or direction of arrivals) and times from the associated detections to localize and estimate 
     the origin time of infrasound events. In order to associate signals and localize acoustic events, travel-time tables are used and these 
     are based on an empirical celerity model (Brachet et al. 2010) that allows for fast computations as required by the IDC operational 
     system. Celerity, or propagation speed of the wave, is the horizontal propagation distance between its origin and the detecting station 
     divided by its associated travel-time in the atmosphere. The back-azimuths and travel-times for each individual station are not 
     corrected to account for atmospheric effects during the propagation of the waves. The combination of both parameters (back-azimuths and 
     times) for the localization explains the separation between the actual location in the REB and the area of cross bearing of the three 
     detections. The acoustic source altitude and its extension are not considered for the REB solution in space as the CTBTO infrasound 
     system has been built to monitor a close-to-the-ground, explosion-like source, i.e. a point source rather than a line source. However, 
     given the distance from the airburst to the detecting stations and the specificity of the acoustic source generated by the airburst of 
     2014 AA, a realistic approximation for the location and origin time estimations is to consider only the three directions of arrival and 
     the detection time of the closest station to the north-west as the travel-time model does not account for paths crossing the Equator. 
     This corrected solution leads to an event location at the intersection of the back-azimuth (11\fdg22 N, 43\fdg71 W) and an origin time 
     of 2014 January 2 02:49:36 UTC (epoch JD 2456659.617778$\pm$0.011087) with an error ellipse of dimensions 678 km$\times$404 km, major 
     axis azimuth 180\degr, and origin time error in excess of 1500 s. The new solution for the hypocentre location of the 2014 AA impact is 
     given in Table \ref{pierrick} and its location shown in Fig. \ref{pmialle}.
%
%
     \begin{table}
      \centering
      \fontsize{8}{10pt}\selectfont
      \tabcolsep 0.12truecm
      \caption{Hypocentre location solution for the 2014 AA impact on 2014 January 2. Impact coordinates include the 1$\sigma$ uncertainty. 
              }
      \begin{tabular}{lll}
       \hline
        Time (UTC)                            & = & 02:49:36.45              \\
        Time uncertainty (s)                  & = & 957.898                  \\
        Latitude ({\degr}N)                   & = & +11.2$\pm$2.8            \\
        Longitude ({\degr}E)                  & = & $-$43.7$\pm$1.7          \\
       \hline
        Confidence region at 0.90 level:      &   &                          \\
       \hline
        \ \ \ \ \ Semi-major axis (km, \degr) & = & 677.6, 6.09              \\
        \ \ \ \ \ Semi-minor axis (km, \degr) & = & 404.4, 3.64              \\
        \ \ \ \ \ Major axis azimuth (\degr)  & = & 179.7 (clockwise from N) \\
        \ \ \ \ \ Time uncertainty (s)        & = & 1576.9                   \\
       \hline
      \end{tabular}
      \label{pierrick}
     \end{table}
%
%
%
%
%
%

     The value of the time uncertainty illustrates the large variability of the infrasound event origin in time due to the heterogeneity of 
     the atmosphere in space and time, the source altitude, and the source displacement, which are currently not fully captured by the IDC 
     system. In principle, the location of the airburst could be refined using atmospheric propagation modelling with real-time accurate 
     atmospheric datasets. However, it would also be difficult to constrain the solution better given the limited number of observations 
     (three) and the numerous hypotheses made on the propagation ranges, the uncertainty of the meteorological models in the stratosphere, 
     and the source altitude and dimensions in space and time. 

  \section{Pre-impact orbit: geometric approach}
     Following the approach implemented in de la Fuente Marcos \& de la Fuente Marcos (2013, 2014), we have used the published impact time 
     and coordinates ---see sect. 3 in Chesley et al. (2015) and our own Sect. 3 above--- to investigate the pre-impact orbit of 
     2014~AA. The methodology applied in this section is simple and entirely geometric. 

     \subsection{A purely geometric Monte Carlo approach}
        Let us consider a planet and an incoming natural object, the parent body of a future meteor. Eventually a collision takes place, and 
        the impact time and coordinates of the impact site on the atmosphere of the planet are reasonably well determined. The objective is 
        computing the path of the impactor prior to hitting the planet. Let us assume that, instantaneously, both the orbit of the planet 
        and that of the putative impactor are Keplerian ellipses (their osculating trajectories). Under the two-body approximation, the 
        equations of the orbit around the Sun of any body (planetary or minor) in space are given by the expressions (see e.g. Murray \& 
        Dermott 1999):
        \begin{eqnarray}
           X & = & r \ (\cos \Omega \cos(\omega + f) - \sin \Omega \sin(\omega + f) \cos i)
                       \nonumber \\
           Y & = & r \ (\sin \Omega \cos(\omega + f) + \cos \Omega \sin(\omega + f) \cos i)
                       \label{orbit} \\
           Z & = & r \ \sin(\omega + f) \sin i \nonumber
        \end{eqnarray}
        where $r = a (1 - e^{2})/(1 + e \cos f)$, $a$ is the semi-major axis, $e$ is the eccentricity, $i$ is the inclination, $\Omega$ is 
        the longitude of the ascending node, $\omega$ is the argument of perihelion, and $f$ is the true anomaly. At the time of impact, the 
        osculating elements of the planet ($a_{\rm p}$, $e_{\rm p}$, $i_{\rm p}$, $\Omega_{\rm p}$, $\omega_{\rm p}$, and $f_{\rm p}$) are 
        well established. In a general case, the impact time, $t_{\rm impact}$, is known within an uncertainty interval ${\Delta}t_{\rm 
        impact}$. This implies that the osculating elements of the planet may also be affected by their respective uncertainties 
        (${\Delta}a_{\rm p}$, ${\Delta}e_{\rm p}$, ${\Delta}i_{\rm p}$, ${\Delta}\Omega_{\rm p}$, ${\Delta}\omega_{\rm p}$, and 
        ${\Delta}f_{\rm p}$). Let us assume that immediately before striking, the impactor was moving around the Sun in an orbit with 
        certain values of the osculating elements ($a$, $e$, $i$, $\Omega$, and $\omega$); in this case, the minimum distance between the 
        planet and the orbit of the virtual impactor at the time of impact can be easily estimated using Monte Carlo techniques (Metropolis 
        \& Ulam 1949; Press et al. 2007). 

        Given a set of osculating elements for a certain object, the above equations are randomly sampled in true anomaly for the object and 
        the position of the planet at the time of impact is used to compute the usual Euclidean distance between both points (one on the 
        orbit and the other one being the location of the planet) so the minimum distance is eventually found. This value coincides with the 
        MOID used in Solar System studies. In principle, the best orbit is the one with the smallest MOID at the recorded impact time but it
        depends on the actual values of the impact parameters (the score, see below). The position of the planet can be used with or without 
        taking into consideration the uncertainties pointed out above as this has only minor effects on the precision of the final solution 
        as long as ${\Delta}t_{\rm impact}$ is less than a few minutes. In de la Fuente Marcos \& de la Fuente Marcos (2013, 2014) only 
        ${\Delta}f_{\rm p}$ was sampled; here, we adopt a similar strategy allowing an uncertainty in $f_{\rm p}$ equivalent to about 180 s 
        (see Table \ref{Earth}, Appendix A). Using a resolution of about 2\farcs6 for this first phase of the Monte Carlo sampling is 
        generally sufficient to obtain robust candidate orbits after a few million trials. 

        Regarding the issue of time standards, $t_{\rm impact}$ is expressed as Coordinated Universal Time (UTC) which differs by no more 
        than 0.9 s from Universal Time (UT). Therefore and within the accuracy limits of this research, UTC and UT are equivalent. However,
        orbital elements and Cartesian state vectors (in general, Solar System ephemerides) are computed in a different time standard, the
        Barycentric Dynamical Time (TDB). UTC is discontinuous as it drifts with the addition of each leap second, which occur roughly once
        a year; in sharp contrast, TDB is continuous (for an extensive review on this important issue see Eastman et al. 2010). The JPL
        \textsc{Horizons} On-Line Ephemeris System shows that the difference between the uniform TDB and the discontinuous UTC was +67.18 s 
        on 2014 January 1. Taking into account this correction, the impact time in the REB is 2014 January 2 at 03:06:32 TDB and the one 
        computed in Sect. 3 is 2014 January 2 at 02:50:43.63 TDB.

        In general and given two orbits, our Monte Carlo algorithm discretizes both orbits (sampling in true anomaly) and computes the 
        distance between each and every pair of points (one on each orbit) finding the minimum value. If the discretization is fine enough, 
        that minimum distance matches the value of the MOID obtained by other techniques. The approach to compute the MOID followed here is 
        perhaps far more time consuming than other available algorithms but makes no a priori assumptions and can be applied to arbitrary 
        pairs of heliocentric orbits. It produces results that are consistent with those from other methods. Numerical routines to compute 
        the MOID have been developed by Baluev \& Kholshevnikov (2005), Gronchi (2005), \v{S}egan et al. (2011) and Wi\'sniowski \& Rickman 
        (2013), among others. Gronchi's approach is widely regarded as the de facto standard for MOID computations (Wi\'sniowski \& Rickman  
        2013).

        To further constrain the orbit, the coordinates of the impact point on the planet for our trial orbit are computed as described in 
        e.g. Montenbruck \& Pfleger (2000). In computing the longitude of impact, $\lambda_{\rm impact}$, it is assumed that the MOID takes 
        place when the object is directly overhead (is crossing the local meridian at the hypocentre). Under this approximation, the local 
        sidereal time corresponds to the right ascension of the object and its declination is the latitude of impact, $\phi_{\rm impact}$. 
        In other words and for the Earth, as the local hour angle is zero when the meteor is on the meridian, the longitude is the right 
        ascension of the meteor at the MOID minus the Greenwich meridian sidereal time (positive east from the prime meridian). Including 
        the coordinates of the impact point as constraints has only relatively minor influence on orbit determination, the main effect is in 
        the inclination that may change by up to $\sim$4\% (de la Fuente Marcos \& de la Fuente Marcos 2014). This second stage of the Monte 
        Carlo sampling applies a resolution of about 1{\arcsec} and produces relatively precise results after tens to hundreds of billion 
        trials. In this phase, only a short arc of a few degrees is used for the test impact orbit.

        Applying the above procedure and for a given orbit, both the minimum separation (with respect to the planet at impact time) and the 
        geocentric coordinates of the point of minimum separation can be estimated. Here, the MOID is synonymous of true minimal approach 
        distance because we are studying actual impact orbits. In general, a very small value of the MOID does not imply that the orbit will
        result in an impact because protective dynamical mechanisms, namely resonances, may be at work. The next step is using Monte Carlo 
        to find the optimal orbit: the one that places the object closest to the planet at impact time and also the one that reproduces the 
        coordinates of the impact point (hypocentre) on the planet. In order to do that, the set of orbital elements ($a$, $e$, $i$, 
        $\Omega$, and $\omega$) of the incoming body is randomly sampled within fixed (assumed) ranges following a uniform (or normal) 
        distribution; for each set, the procedure outlined above is repeated so the optimal orbit is eventually found. The use of uniformly 
        distributed random numbers or Gaussian ones does not affect the quantitative outcome of the algorithm, but in this section we 
        utilize a uniform distribution instead of a standard normal distribution because this speeds up the task of finding the optimal 
        orbit. Neglecting gravitational focusing and in order to have a physical collision, the MOID must be $<$ 0.00004336 AU (one Earth's 
        radius, $R_{\rm E}$, in AU plus the characteristic thickness of the atmosphere, 115 km). In this context, MOIDs $<$ 1 $R_{\rm E}$ 
        are regarded as unphysical. This approach is, in principle, computationally expensive but makes very few a priori assumptions about 
        the orbit under study and can be applied to cases where little or no astrometric information is available for the meteor. Within the 
        context of massively parallel processing, our approach is inexpensive though. Our algorithm usually converges after exploring 
        several billion (up to a few trillion) orbits. Seeking the optimal orbit can be automated using a feedback loop to accelerate 
        convergence in real time based on the criterion used in de la Fuente Marcos \& de la Fuente Marcos (2013). 

        A key ingredient in our algorithm is the procedure to decide when a candidate solution is better than other. Candidate solutions 
        must be ranked based on how well they reproduce the observed impact parameters, i.e. they must be assigned a score. A robust choice 
        to rank the computed candidate solutions is a combination of two bivariate Gaussian distributions: one for the actual impact and a 
        second one for its location. In order to rank the impact time we use:
        \begin{equation}
           \beta = {\Large e}^{-\frac{1}{2}\left[
                         \left(\frac{d - d_{\rm impact}}{\sigma_{d_{\rm impact}}}\right)^{2} +
                         \left(\frac{f_{\rm E} - f_{\rm E_{\rm impact}}}{\sigma_{f_{\rm E_{\rm impact}}}}\right)^{2}
                      \right]} \,,
                      \label{estimate1}
        \end{equation}
        where $d$ is the MOID of the test orbit in AU, $d_{\rm impact}$ = 0 AU is the minimum possible MOID, $\sigma_{d_{\rm impact}}$ is 
        assumed to be the radius of the Earth in AU, $f_{\rm E}$ is the value of Earth's true anomaly used in the computation, $f_{\rm 
        E_{\rm impact}}$ is the value of Earth's true anomaly at the impact time (see Table \ref{Earth}, Appendix A), and $\sigma_{f_{\rm 
        E_{\rm impact}}}$ is half the angle subtended by the Earth from the Sun (0$\fdg$00488). We assume that there is no correlation 
        between $d$ and $f_{\rm E}$. However, the values of the impact time and Earth's true anomaly at that time are linearly correlated 
        in the neighbourhood of the impact point; i.e. $f_{\rm E_{\rm impact}}$ is a proxy for the impact time. If $\beta > 0.368$, a 
        collision is possible. Our best solutions have $\beta > 0.9999$. The MOID can be used to compute the altitude above the ground if
        $R_{\rm E}$ is considered. The altitude above the surface of the Earth is in the range 0--115 km (upper atmosphere limit). As for 
        the impact location we use:
        \begin{equation}
           \Psi = {\Large e}^{-\frac{1}{2}\left[
                        \left(\frac{\lambda - \lambda_{\rm impact}}{\sigma_{\lambda_{\rm impact}}}\right)^{2} +
                        \left(\frac{\phi - \phi_{\rm impact}}{\sigma_{\phi_{\rm impact}}}\right)^{2}
                     \right]} \,,
                     \label{estimate2}
        \end{equation}
        where $\lambda$ and $\phi$ are the impact coordinates for a given test orbit (if $\beta > 0.368$), and $\sigma_{\lambda_{\rm 
        impact}}$ and $\sigma_{\phi_{\rm impact}}$ are the standard deviations associated with $\lambda_{\rm impact}$ and $\phi_{\rm impact}$ 
        supplied with the actual (observational) impact values. Again, we assume that there is no correlation between $\lambda$ and $\phi$, 
        and our best solutions have $\Psi > 0.9999$. The use of $\sigma_{\lambda_{\rm impact}}$ and $\sigma_{\phi_{\rm impact}}$ implicitly 
        inserts the direction of the local tangent (or its projection) into the calculations (see below).

        Impact events are defined by a number of parameters. Observational parameters are specified by a mean value and a standard deviation 
        or uncertainty; they are assumed to be independent. Numerical experiments generate virtual impacts, if successful. The parameters 
        associated with a virtual impact must be checked against the observational values in order to decide if a given pre-impact orbit can 
        reproduce them. An uncertainty model must be applied to rank the tested pre-impact orbits. For this purpose, we use a Gaussian 
        uncertainty model with multidimensional relevance ranking metrics. Equations (\ref{estimate1}) and (\ref{estimate2}) let us assign a 
        score to any given candidate solution. Assuming independence, the score can be computed as $\beta\times\Psi$. The higher the score, 
        the better the orbit. Equations (\ref{estimate1}) and (\ref{estimate2}) provide a simple but useful estimate of the probability that 
        a given candidate solution could reproduce the impact parameters. Reproducing the observed values of the impact parameters (and 
        those of any other available observational data) is the primary goal of our approach. Similar techniques are used in other 
        astronomical contexts; see e.g. sect. 4.2 in Scholz et al. (1999) or sect. 4 in Sariya \& Yadav (2015). The probability of being 
        able to reproduce the impact parameters is different from the probability of impact; a certain pre-impact orbit may have an 
        associated impact probability virtually equal to 1 and still be unable to reproduce the impact parameters if, for instance, the 
        timing deviates significantly from the recorded impact time ($\pm9\sigma$). Here, we assume that the probability of impact is 
        computed in the usual way or number of successes divided by number of trials.

        Our geometric approach is implemented iteratively as some initial guess for the pre-impact orbit is made based on some a priori 
        observational knowledge; the pre-impact orbit is improved by inspecting the reconstruction of the impact and its rank. The procedure 
        is iterated until an optimal solution is found. At this point, one may wonder how reliable our approach could be and what its 
        intrinsic limitations are. The orbital elements and therefore the position of the target planet at the time of impact are assumed to 
        be well known; if the input data are reliable enough then the computed solution must be equally robust. The time of impact and the 
        coordinates of the impact point are the observables used to constrain both the input data (the Earth's ephemerides in our case) and 
        the eventual solution. If the time of impact and/or the coordinates of the impact point are uncertain or wrong, then the solution 
        obtained will be equally unreliable or incorrect. The time of impact is by far the most critical parameter. In our present case, it 
        is a very reasonable assumption to consider that the available observational data ($t_{\rm impact}$, $\lambda_{\rm impact}$ and 
        $\phi_{\rm impact}$) are sufficiently robust to produce an equally sound orbital solution. 

        If information on the pre-impact velocity of the object is available, it can be used to further refine the candidate orbital 
        solution, again by iteration. The observational pre-impact speed is the velocity at atmospheric entry, $v_{\rm impact}$. As a 
        by-product of our geometric reconstruction, we obtain the relative velocity at atmospheric entry neglecting the acceleration caused 
        by the Earth's gravitational field; this is called the hyperbolic excess velocity, $v_{\infty}$, or the characteristic geocentric 
        velocity of the meteor's radiant, $v_{\rm g}$. The velocity at atmospheric entry and the hyperbolic excess velocity are linked by 
        the expression 
        \begin{equation}
           v_{\infty}^{2} = v_{\rm impact}^{2} - v_{\rm escape}^{2}\,, 
                             \label{vimp}
        \end{equation}
        where $v_{\rm escape}\sim11.2$~km~s$^{-1}$ is the Earth's escape velocity at atmospheric entry. Therefore, for any geometrically 
        reconstructed pre-impact orbit, $v_{\rm impact}$ can be easily estimated in order to compare with the observational data, even if 
        our geometric approach is entirely non-collisional. The amount $v_{\infty}$ can also be interpreted as the velocity of the object
        relative to an assumed massless Earth. Although not standard in meteor astronomy, here we have followed the terminology discussed by 
        Chodas and Blake.\footnote{\url{http://neo.jpl.nasa.gov/risks/}}$^{,}$\footnote{\url{http://neo.jpl.nasa.gov/risks/a29075.html}} In
        Ceplecha (1987), a classic work in meteor astronomy, $v_{\infty}$ is the velocity of the meteoroid corrected for the atmospheric 
        drag and referred to the entry point in the Earth's atmosphere, i.e. our $v_{\rm impact}$. 

        Our geometric approach works because, for an object orbiting around the Sun, it is always possible to find an instantaneous 
        Keplerian orbit that fits its instantaneous position and during a close encounter the largest orbital changes take place during the 
        time interval immediately after reaching the distance of closest approach. In principle, degenerate orbital solutions (two or more 
        very different impact orbits being compatible with a given set of impact data) are possible, but additional observational 
        information such as how the meteor was travelling across the sky (e.g. north to south) and velocity-related data should be 
        sufficient to break any degeneracy unless the orbits are part of the same family (very similar orbital parameters). However, if the 
        orbits are so similar they belong to the same meteoroid stream (see e.g. Jopek \& Williams 2013; Schunov\'a et al. 2014).

     \subsection{Validation: the case of the Almahata Sitta event}
        The algorithm described above, in its simplest form, was tested in the case of the Almahata Sitta event caused by the meteoroid 
        2008~TC$_{3}$ (Jenniskens et al. 2009; Oszkiewicz et al. 2012) and it was found to be able to generate an orbital solution 
        consistent with those from other authors (de la Fuente Marcos \& de la Fuente Marcos 2013). Applying the modified algorithm that 
        includes the location of the impact point we obtain: $a=1.3085\pm0.0003$ AU, $e=0.3124\pm0.0002$, $i=2\fdg525\pm0\fdg002$, 
        $\Omega=194\fdg10618\pm0\fdg00007$, $\omega=234\fdg466\pm0\fdg008$ and $M=330\fdg840\pm0\fdg013$, with $\lambda_{\rm 
        impact}=31\fdg37\pm0\fdg04$, $\phi_{\rm impact}=+20\fdg87\pm0\fdg06$ at an altitude of 27$\pm$20 km on 2008 October 07 02:45:40$\pm$5 
        UTC (at the time of the Almahata Sitta event the difference between TDB and UTC was +65.18 s). These are the average values of 11 
        best solutions ranked using Eqs. (\ref{estimate1}) and (\ref{estimate2}). In this and subsequent calculations the errors quoted 
        correspond to one standard deviation (1$\sigma$) computed applying the usual expressions (see e.g. Wall \& Jenkins 2012). The 
        relative differences between this geometric orbital solution and the one computed by Steven R. Chesley, and available from the JPL 
        Small-Body Database, are: 0.023\% in $a$, 0.11\% in $e$, 0.68\% in $i$, 0.0026\% in $\Omega$, 0.0073\% in $\omega$, and 0.0026\% in 
        $M$. Therefore, the orbital solution has relatively small uncertainties when compared against a known robust determination. The 
        values of the impact parameters are consistent with those available from the 
        JPL.\footnote{\url{http://neo.jpl.nasa.gov/fireballs/}}$^{,}$\footnote{\url{http://neo.jpl.nasa.gov/news/2008tc3.html}}

     \subsection{The most probable pre-impact orbit of 2014~AA in a strict geometric sense}
\begin{figure}
        \centering
        \includegraphics[width=0.49\linewidth]{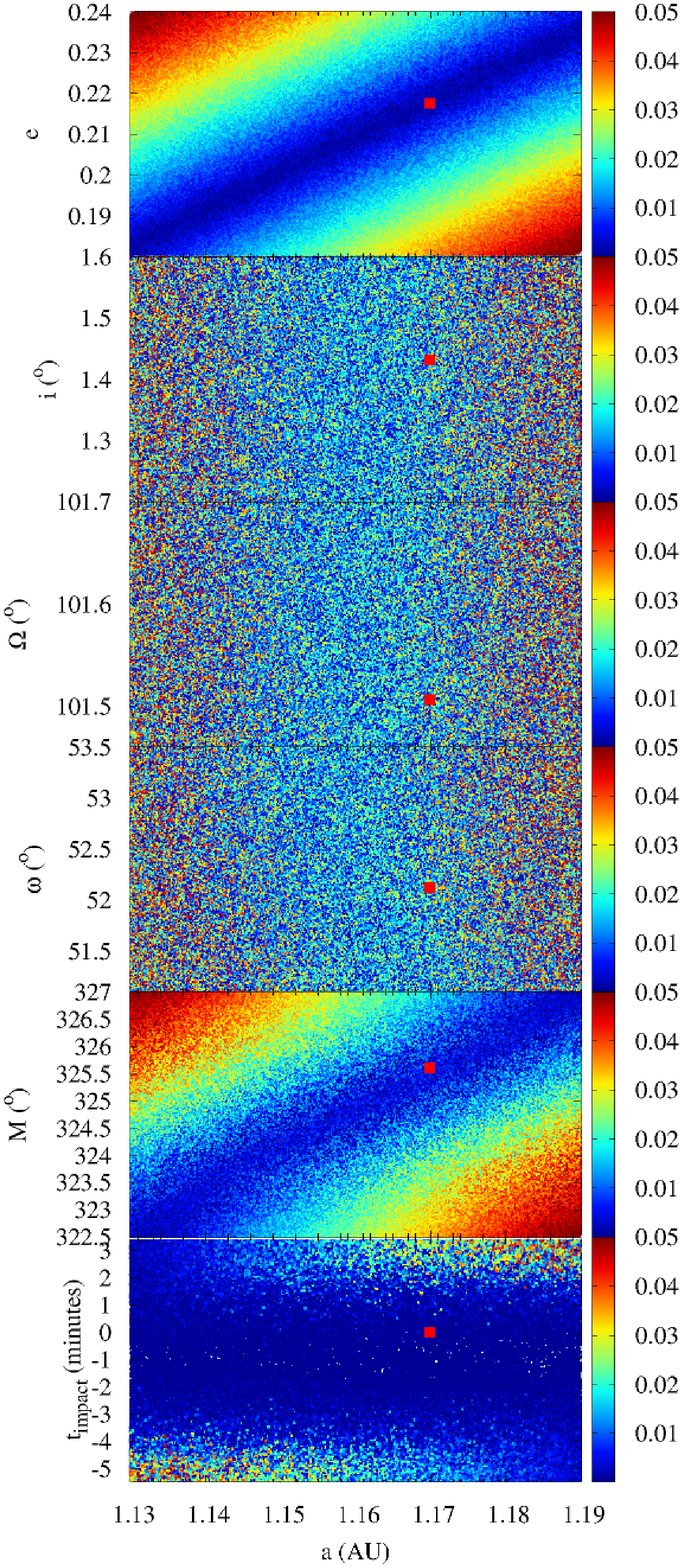}
        \includegraphics[width=0.49\linewidth]{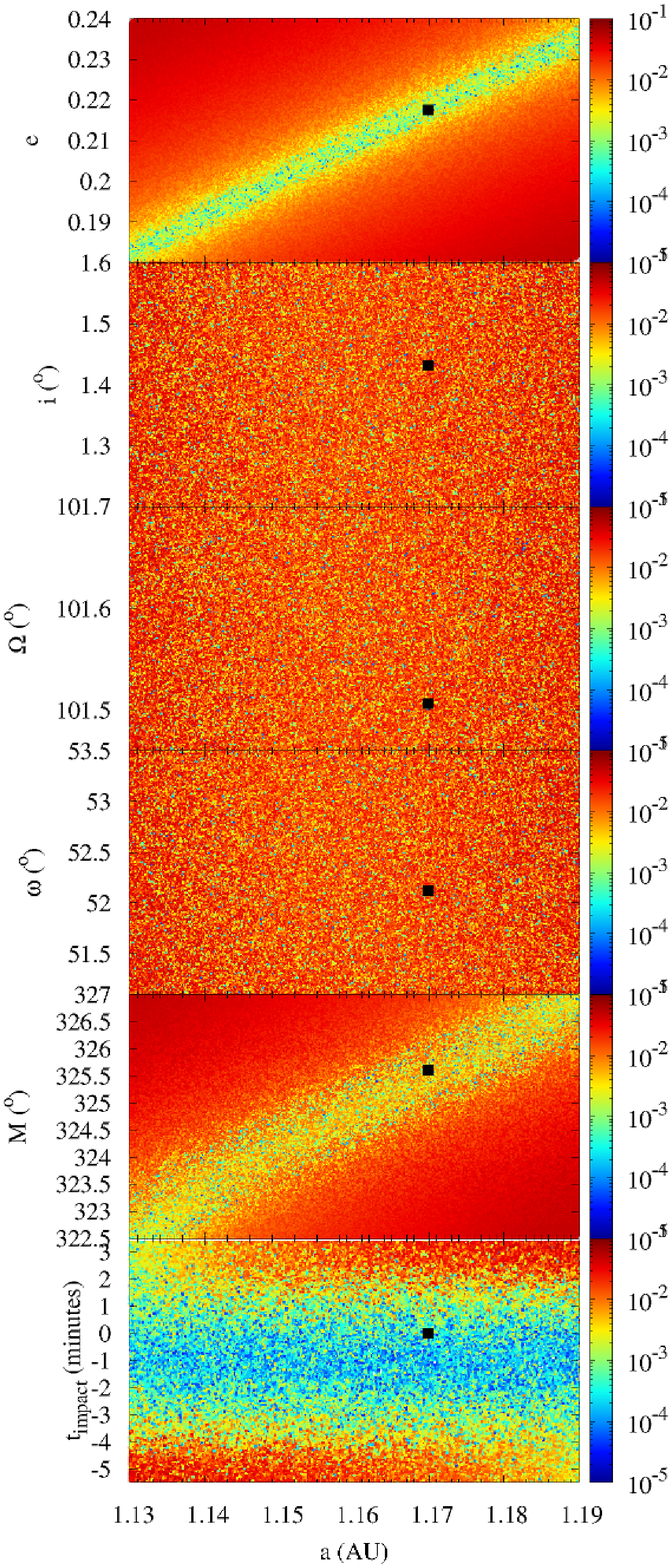}\\
        \caption{Results from the geometric approach described in the text for the 2014~AA impactor. The colours in the colour maps are 
                 proportional to the value of the minimal approach distance in AU for a given test orbit following the associated colour 
                 box (linear scale, left-hand panels; logarithmic scale, right-hand panels). Only test orbits with MOIDs under 0.05 AU are 
                 displayed; the solution in Table \ref{ours1} (left-hand column) is also indicated (red/black squares). A representative 
                 sample of 10$^7$ test orbits is plotted. In this figure, $t_{\rm impact}$ is referred to the epoch JD TDB 2456659.629537, 
                 the one used in Chesley et al. (2015) and Farnocchia et al. (2016).}
        \label{moid0}
     \end{figure}

        We have applied the procedure described above using, as input, data from Table \ref{Earth} (epoch 2456659.629537) and the 
        coordinates of the impact point as given in Chesley et al. (2015). Figure \ref{moid0} shows a representative sample (10$^7$ points) 
        of our results as well as the best solution (red/black squares). Our best solution, found after about 10$^{10}$ trials, appears in 
        Table \ref{ours1} (left-hand column) and produces a virtual impact at ($\lambda_{\rm impact}$, $\phi_{\rm impact}$, $t_{\rm impact}$) 
        = ($-$43{\fdg}417$\pm$0{\fdg}007, +14{\fdg}632$\pm$0{\fdg}008, 2456659.62954$\pm$0.00002 JD TDB or 2014 January 2 at 3:05:25 UTC). 
        From the MOID, the altitude is 47$\pm$31 km. The orbital solution displayed in Table \ref{ours1}, left-hand column, is the average 
        of 15 representative good solutions ranked using Eqs. (\ref{estimate1}) and (\ref{estimate2}). The geometric impact probability for 
        this orbital solution is virtually 1, given the very large number of trials. Two values of $\tau$ are provided because it is 
        customary to quote the $\tau$ closest to the epoch under study but it is also true that, as a result of the impact, 2014~AA never 
        reached the 2456703.79 JD TDB perihelion so the previous one is also given for consistency.  

        Using the new determination presented in Sect. 3 and consistent data analogous to those in Table \ref{Earth}, we obtain (see Table 
        \ref{ours1}, right-hand column): $a$~=~1.168706 AU, $e$~=~0.216553, $i$~=~1\fdg36251, $\Omega$~=~101\fdg50175, $\omega$~=~52\fdg1144, 
        and $M$~=~325\fdg5325. This is the average of 27 representative good solutions. The virtual impact is now at ($\lambda_{\rm impact}$, 
        $\phi_{\rm impact}$, $t_{\rm impact}$) = ($-$43{\fdg}714$\pm$0{\fdg}008, +11{\fdg}218$\pm$0{\fdg}007, 2456659.61856$\pm$0.00002 JD 
        TDB or 2014 January 2 at 2:49:36.5 UTC); the altitude is 39$\pm$26~km. As for the entry velocity, the value obtained in Chesley et 
        al. (2015) is 12.23~km~s$^{-1}$; the $v_{\rm impact}$ from our geometric approach ---see Eq. (\ref{vimp})--- is 12.33~km~s$^{-1}$. 
%
%
     \begin{table*}
      \centering
      \fontsize{8}{10pt}\selectfont
      \tabcolsep 0.10truecm
      \caption{Heliocentric Keplerian orbital elements of 2014~AA from our geometric approach. Values include the 1$\sigma$ uncertainty. The 
               orbit is computed at an epoch arbitrarily close to the impact time. The two values quoted for $\tau$ are separated by one 
               full orbital period. The orbit on the left-hand column produces a virtual impact with parameters consistent with those used 
               in Chesley et al. (2015) or Farnocchia et al. (2016), the orbit in the right-hand column is consistent with the new 
               determination presented in Sect. 3 (see the text for details).
              }
      \begin{tabular}{llll}
       \hline
        Semi-major axis, $a$ (AU)                         & = &   1.16995$\pm$0.00009      &   1.168706$\pm$0.000010    \\
        Eccentricity, $e$                                 & = &   0.21763$\pm$0.00008      &   0.216553$\pm$0.000008    \\
        Inclination, $i$ (\degr)                          & = &   1.4319$\pm$0.0002        &   1.36251$\pm$0.00006      \\
        Longitude of the ascending node, $\Omega$ (\degr) & = & 101.50618$\pm$0.00003      & 101.50175$\pm$0.00003      \\
        Argument of perihelion, $\omega$ (\degr)          & = &  52.1249$\pm$0.0005        &  52.1144$\pm$0.0002        \\
        Mean anomaly, $M$ (\degr)                         & = & 325.609$\pm$0.006          & 325.5325$\pm$0.0006        \\
        Time of perihelion passage, $\tau$ (JD TDB)       & = & 2456241.57$\pm$0.06        & 2456242.327$\pm$0.006      \\
                                                          &   & 2012-Nov-10 01:39:21.6 UTC & 2012-Nov-10 19:49:26.4 UTC \\
                                                          & = &    or 2456703.79           &    or 2456703.802          \\
                                                          &   & 2014-Feb-15 06:56:09.6 UTC & 2014-Feb-15 07:13:26.4 UTC \\
        Perihelion, $q$ (AU)                              & = &   0.91534$\pm$0.00002      &   0.915619$\pm$0.000002    \\
        Aphelion, $Q$ (AU)                                & = &   1.4246$\pm$0.0002        &   1.42179$\pm$0.00002      \\
       \hline
      \end{tabular}
      \label{ours1}
     \end{table*}
%
%
%
%
%
%

        Our approach provides the most probable solution, in a strict geometric sense, for the pre-impact orbit of 2014~AA. This solution is 
        fully consistent with those in Table \ref{elements} even if no astrometry has been used to perform the computations. The high degree 
        of coherence between the orbital elements derived using actual observations and those produced following the methodology described 
        in this section clearly vindicates this geometric Monte Carlo approach as a valid method to compute low-precision ---but still 
        reliable--- pre-impact orbits of the parent bodies of meteors. The values of the errors quoted represent the standard deviations
        associated with the selected sample of best orbits, they do not include the larger, systematic component linked to the observational
        values of impact coordinates and time (see above) that in the case of 2014~AA is dominant. It is true that the orbital solutions in 
        Table \ref{elements} are rather uncertain, but in the case of 2008~TC$_{3}$ ---which is much better constrained, including the 
        values of the impact parameters--- our approach also provides very good agreement with astrometry-based solutions (see de la Fuente 
        Marcos \& de la Fuente Marcos 2013 and above). 

  \section{Pre-impact orbit: \textit{N}-body approach}
     It can be argued that the Monte Carlo technique used in the previous section may not be adequate to make a proper determination of the 
     pre-impact orbit of the parent body of an observed meteor. In particular, one may say that the use of two two-body orbits is absolutely 
     inappropriate for this problem as three-body effects are fundamental in shaping the relative dynamics during the close encounter that 
     leads to the impact. Let us assume that this concern is warranted. 

     \subsection{A full \textit{N}-body approach}
        In absence of astrometry, the obvious and most simple (but very expensive in terms of computing time) alternative to the method used 
        in the previous section is to select some reference epoch (preceding the impact time), assume a set of orbital elements ($a$, $e$, 
        $i$, $\Omega$, $\omega$, and $\tau$), generate a Cartesian state vector for the assumed orbit at the reference epoch, and use 
        $N$-body calculations to study the orbital evolution of the assumed orbit until an impact or a miss, within a specified time frame, 
        occurs. If enough orbits are studied, the best pre-impact orbit can be determined. This assumption is based on the widely accepted 
        notion that statistical results of an ensemble of collisional $N$-body simulations are accurate, even though individual simulations 
        are not (see e.g. Boekholt \& Portegies Zwart 2015). Given the fact that $N$-body simulations are far more CPU time consuming than 
        the calculations described in the previous section, having a rough estimate of the orbital solution is essential to make this 
        approach feasible in terms of computing time. The geometric Monte Carlo approach described above is an obvious candidate to supply 
        an initial estimate for the orbit under study.

        The method just described corresponds to that of an inverse problem where the model inputs (the pre-impact orbit) are unknown while 
        the model outputs (the impact parameters) are known (see e.g. Press et al. 2007). Our simulation-optimization approach searches for 
        the best inputs from among all the possible ones without explicitly evaluating all of them. We seek an optimal solution ---fitting 
        the data or model outputs within given constraints--- and also enforce that the resulting data fit within certain tolerances, given 
        by data uncertainties (those of the original, observational data). Our simulation model (the $N$-body calculations) is coupled with 
        optimization techniques ---based on Gaussian distributions analogous to Eqs. (\ref{estimate1}) and (\ref{estimate2}), see above--- 
        to determine the model inputs that best represent the observed data in an iterative process. The observational data are noisy (have 
        errors) and the uncertainty is incorporated into the optimization (via the Gaussian distributions), but we also deal with the 
        uncertainty by analysing multiple incarnations of the model inputs. Our hybrid optimization implementation maximizes the number of 
        successful trials resulting from a Monte Carlo simulation. Our best solutions have scores $> 0.9999$ (see the discussion in Sect. 
        4.1).

        The procedure described in this section can be seen as an inverse implementation of the techniques explored in Sitarski (1998, 1999, 
        2006). In these works, the author investigates the conditions for a hypothetical collision of a minor planet with the Earth creating 
        sets of artificial observations with the objective of finding out the time-scales necessary to realise that a collision is 
        unavoidable and to determine a precise impact area on the Earth's surface. In his work, the emphasis is made in how the 
        uncertainties in the observations affect our ability to be certain of an impending collision and, in the case of a confirmed one, to 
        be able to compute the correct impact location. In our case, we assume that (at least initially) no pre-impact observations are 
        available, only the impact parameters are known. Sitarski's study uses the pre-impact information as input to develop his 
        methodology, but here we use the post-impact data as a starting point. Sitarski's work is an example of solving a forward problem, 
        ours of solving an inverse problem. The use of impact data to improve orbit solutions is not a new concept, it was first used by 
        Chodas \& Yeomans (1996) in the orbit determination of 16 of the fragments of comet Shoemaker-Levy 9 that collided with Jupiter in 
        July 1994.

        It may be argued that the type of inverse problem studied here (going from impact to orbit) has a multiplicity of solutions as we 
        seek six unknowns (the orbital parameters) and the impact parameters are only three ($t_{\rm impact}$, $\lambda_{\rm impact}$ and 
        $\phi_{\rm impact}$, but also $h_{\rm impact}$). In general, the solution to an inverse problem may not exist, be non-unique, or 
        unstable. However, it is a well known fact used in probabilistic curve reconstruction (see e.g. Unnikrishnan et al. 2010) that if a 
        curve is smooth, the data scatter matrix (that contains the values of the variances) will be elongated and that its major axis, or 
        principal eigenvector, will approximate the direction of the local tangent. It is reasonable to assume that the pre-impact orbit of 
        the parent body of a meteor in the neighbourhood of the impact point ---high in the atmosphere--- is smooth and therefore that the 
        dispersions in $\lambda_{\rm impact}$, $\phi_{\rm impact}$, and $h_{\rm impact}$ (supplied with their values) provide a suitable 
        approximation to the local tangent as the principal eigenvector of the data scatter matrix is aligned with the true tangent to the 
        impact curve. In this context and by using the values of the dispersions (as part of the candidate solution ranking process, see 
        above), we have indirect access to the direction of the instantaneous velocity and perhaps its value. In any case, if $v_{\rm 
        impact}$ is available the solution of the inverse problem is unique (for additional details, see de la Fuente Marcos et al. 2015).  

        In this section we implement and apply an $N$-body approach to solve the problem of finding the pre-impact orbit of the parent body 
        of a meteor. This approach is applied within a certain physical model. Our model Solar System includes the perturbations by the 
        eight major planets and treats the Earth and the Moon as two separate objects; it also incorporates the barycentre of the dwarf 
        planet Pluto--Charon system and the three most massive asteroids of the main belt, namely, (1) Ceres, (2) Pallas, and (4) Vesta. 
        Input data are reference epoch and initial conditions for the physical model at that epoch. We use initial conditions (positions and 
        velocities referred to the barycentre of the Solar System) provided by the JPL \textsc{horizons}\footnote{\url{http://ssd.jpl.nasa.gov/?horizons}} 
        system (Giorgini et al. 1996; Standish 1998; Giorgini \& Yeomans 1999; Giorgini et al. 2001) and relative to the JD TDB (Julian 
        Date, Barycentric Dynamical Time) 2456658.628472222 epoch which is the $t$ = 0 instant (see Table \ref{Cartesian}, Appendix B); in 
        other words, the integrations are started $\sim$24 h before $t_{\rm impact}$. 

        The $N$-body simulations performed here were completed using a code that implements the Hermite integration scheme (Makino 1991; 
        Aarseth 2003). The standard version of this direct $N$-body code is publicly available from the IoA web 
        site.\footnote{\url{http://www.ast.cam.ac.uk/~sverre/web/pages/nbody.htm}} Relative errors in the total energy are as low as 
        10$^{-14}$ to 10$^{-13}$ or lower. The relative error of the total angular momentum is several orders of magnitude smaller. The 
        systematic difference at the end of the integration, between our ephemerides and those provided by the JPL for the Earth, is about 1 
        km. As the average orbital speed of our planet is 29.78 km s$^{-1}$, it implies that the temporal systematic error in our virtual 
        impact calculations could be as small as 0.04 s. Non-gravitational forces, relativistic and oblateness terms are not included in the 
        simulations, additional details can be found in de la Fuente Marcos \& de la Fuente Marcos (2012). The Yarkovsky and 
        Yarkovsky--O'Keefe--Radzievskii--Paddack (YORP) effects (see e.g. Bottke et al. 2006) may be unimportant when objects are tumbling 
        or in chaotic rotation ---but see the discussion in Vokrouhlick\'y et al. (2015) for 99942 Apophis (2004 MN$_{4}$)--- as it could be 
        the case of asteroidal fragments. Relativistic effects, resulting from the theory of general relativity are insignificant when 
        studying the long-term dynamical evolution of minor bodies that do not cross the innermost part of the Solar System (see e.g. 
        Benitez \& Gallardo 2008). For a case like the one studied here, the role of the Earth's oblateness is rather negligible ---see the 
        analysis in Dmitriev et al. (2015) for the Chelyabinsk superbolide. The effect of the atmosphere is not included in the calculations
        as we are interested in the dynamical evolution prior to the airburst event. For the particular case of the Chelyabinsk superbolide,
        table S1 in Popova et al. (2013) shows that the value of the apparent velocity of the superbolide remained fairly constant between 
        the altitudes of 97.1 and 27 km. This fact can be used to argue that neglecting the influence of the atmosphere should not have any
        adverse effects on the results of our analysis.

     \subsection{Zeroing in on the best orbital solution}
        The actual implementation of the ideas outlined above is simple:
        \begin{enumerate}
           \item A Monte Carlo approach is used to generate sets of orbital elements. In this case, Gaussian random numbers are utilized to 
                 better match the results of astrometry-based solutions; the Box-Muller method (Box \& Muller 1958; Press et al. 2007) is 
                 applied to generate random numbers with a normal distribution. 
           \item For each set of orbital parameters, the Cartesian state vector is computed at $t$ = 0. 
           \item An $N$-body simulation is launched as described above, running from the JD TDB 2456658.628472222 epoch until JD TDB 
                 2456660.82. 
           \item The output is processed to check for an impact or a miss. 
           \item If an impact takes place, the impact time and the location of the impact point are recorded; the coordinates of the impact 
                 point are computed as described in the previous section. The altitude above the surface of the Earth is in the range 
                 0--115 km (upper atmosphere limit). This is consistent with infrasound propagation; in general, airbursts observed by 
                 global infrasound sensors occur below or around the stratosphere (ground up to 60--65 km). 
           \item If a miss, the value of the minimal approach distance is recorded for statistical purposes.
           \item Candidate impact solutions are ranked using expressions similar to Eqs. (\ref{estimate1}) and (\ref{estimate2}) and 
                 assigned a score. 
           \item As the score improves, the new sets of orbital elements generated in step \#1 are based on the newest best solution in 
                 order to speed up convergence towards the optimal orbital determination, further improving the score of the subsequent
                 candidate solutions.
        \end{enumerate}
        A large number of test orbits is studied. The volume of orbital parameter space explored by the algorithm is progressively reduced 
        as the optimal solution is approached. In general, the data output interval is nearly 5 s; therefore, the impact is properly 
        resolved in terms of time and space. At the typical impact speeds induced here ($\sim$12 km s$^{-1}$) an object travels the Earth's 
        diameter in over 17 minutes and crosses the atmosphere in less than 10 s. 

        The orbital elements of our test orbits are varied randomly, within the ranges defined by their mean values and standard deviations. 
        They represent a number of different virtual impactors moving in similar orbits, they do not attempt to incarnate a set of 
        observations obtained for a single impactor. If actual observations are utilized, we have to consider how the elements affect each 
        other using the covariance matrix or following the procedure described in Sitarski (1998, 1999, 2006). Due to the large 
        uncertainties affecting the orbit of 2014~AA, we decided to neglect any corrections based on the covariance matrix to generate 
        our test or control orbits at this stage but see Sect. \ref{mccm}. This arbitrary choice should not have any major effects on the 
        assessment of the orbits made here. 

        We have performed an initial search for an optimal orbit using the $N$-body approach and the solution from the previous section 
        (Table \ref{ours1}, left-hand column) assuming a normal distribution in orbital parameters. In this case, the probability of impact 
        is $\leq$0.043. The virtual impacts take place 30 minutes to 3 h earlier than the time resulting from the analysis of infrasounds 
        in Chesley et al. (2015) or Farnocchia et al. (2016), the longitude of impact has a range of nearly 180\degr centred at about 
        $-$28\degr, and the latitude is in the range ($-$11, 30)\degr centred at about 9\degr. Using the solution in Table \ref{ours1}, 
        right-hand column, the probability of impact is 0.455, the recorded impact time is 102$\pm$42 minutes earlier than the one given in 
        Table \ref{pierrick}, the longitude of impact has a range close to 180\degr centred at about $-$51\degr, and the latitude is in the 
        range ($-$5, 18)\degr centred at 10\degr. These results confirm that the approach discussed in the previous section is able to 
        produce reasonably correct low-precision pre-impact orbits of meteors. In theory, the full \textit{N}-body approach is capable of 
        returning orbital determinations matching those from classical methods in terms of precision.

\begin{figure}
        \centering
        \includegraphics[width=\linewidth]{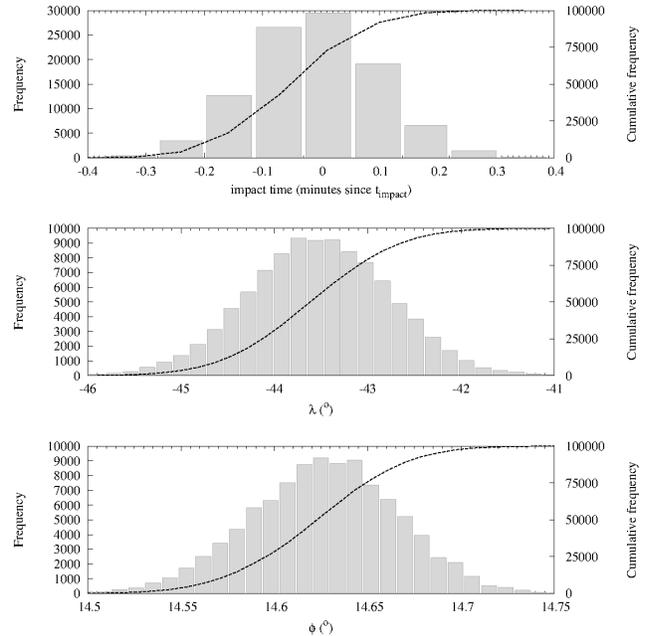}
        \caption{Resulting distributions in impact parameter space for an experiment using our initial solution from the $N$-body approach 
                 (see Table \ref{ours2}, left-hand column). The impact time is referred to the value based on infrasound data according to 
                 Chesley et al. (2015) or Farnocchia et al. (2016); the rather coarse distribution in impact time is induced by our data 
                 output interval of nearly 5~s. 
                 } 
        \label{ctbto}
     \end{figure}

        Using the impact parameters based on infrasound data as described in Chesley et al. (2015) or Farnocchia et al. (2016) and after a 
        few million trials, mostly automated, we find the solution displayed in Table \ref{ours2}, left-hand column, that is referred to 
        epoch JD TDB 2456658.628472222. It is the average of 23 good solutions ranked as explained above (score $> 0.9999$). The altitude 
        above the surface of the Earth at impact was 60$\pm$14 km. For this solution, the geocentric value of the entry velocity is 
        12.186$\pm$0.011~km~s$^{-1}$ which is reasonably consistent with that in Chesley et al. (2015), 12.23~km~s$^{-1}$, and also with the 
        one in Farnocchia et al. (2016), 12.17~km~s$^{-1}$. We consider that the orbit displayed in Table \ref{ours2} (left-hand column) is 
        the most probable pre-impact orbit of 2014~AA if the impact parameters in Chesley et al. (2015) or Farnocchia et al. (2016) are 
        assumed, and if the available astrometry is neglected. Figure \ref{ctbto} shows the results of an $N$-body experiment including 
        10$^{5}$ test orbits resulting from a Monte Carlo simulation with a normal distribution in orbital parameters according to Table 
        \ref{ours2}, left-hand column. In this experiment, the probability of impact is $>$0.99999. The rather coarse distribution in impact 
        time is the result of the unavoidable discretization of the output interval that also has an effect on the distribution in altitude 
        (not shown). When computers are used to produce a uniform random variable ---i.e. to seed the Box-Muller method to generate random 
        numbers from the standard normal distribution with mean 0 and standard deviation 1--- it will inevitably have some inaccuracies 
        because there is a lower boundary on how close numbers can be to 0. For a 64 bits computer the smallest non-zero number is $2^{-64}$ 
        which means that the Box-Muller method will not produce random variables more than 9.42 standard deviations from the mean.
%
%
     \begin{table*}
      \centering
      \fontsize{8}{11pt}\selectfont
      \tabcolsep 0.1truecm
      \caption{Heliocentric Keplerian orbital elements of 2014~AA from our $N$-body approach if the impact parameters in Chesley et al. 
               (2015) or Farnocchia et al. (2016) are assumed (left-hand column) and for the new impact solution presented in Sect. 3 
               (right-hand column). Values include the 1$\sigma$ uncertainty. The orbits are computed at epoch JD TDB 2456658.628472222 that 
               corresponds to 03:03:52.82 UTC on 2014 January 1, J2000.0 ecliptic and equinox.  
              }
      \begin{tabular}{llll}
       \hline
        Semi-major axis, $a$ (AU)                             & = &   1.1639272$\pm$0.0000003   &   1.1621932$\pm$0.0000004   \\
        Eccentricity, $e$                                     & = &   0.2136664$\pm$0.0000003   &   0.2115415$\pm$0.0000004   \\
        Inclination, $i$ (\degr)                              & = &   1.46070$\pm$0.00006       &   1.38009$\pm$0.00006       \\
        Longitude of the ascending node, $\Omega$ (\degr)     & = & 101.59934$\pm$0.00002       & 101.60617$\pm$0.00004       \\
        Argument of perihelion, $\omega$ (\degr)              & = &  52.58912$\pm$0.00003       &  52.34221$\pm$0.00006       \\
        Mean anomaly, $M$ (\degr)                             & = & 324.124173$\pm$0.000013     & 324.14040$\pm$0.00002       \\
        Time of perihelion passage, $\tau$ (JD TDB)           & = & 2456704.335877$\pm$0.000005 & 2456704.213146$\pm$0.000005 \\
        Perihelion, $q$ (AU)                                  & = &   0.91523502$\pm$0.00000010 &   0.9163412$\pm$0.0000002   \\
        Aphelion, $Q$ (AU)                                    & = &   1.4126194$\pm$0.0000007   &   1.4080453$\pm$0.0000009   \\
        Impact time, $t_{\rm impact}$ (JD UTC)                & = & 2456659.62878$\pm$0.00003   & 2456659.61773$\pm$0.00004   \\
        Longitude of impact, $\lambda_{\rm impact}$ (\degr)   & = & $-$43.425$\pm$0.010         & $-$43.704$\pm$0.007         \\ 
        Latitude of impact, $\phi_{\rm impact}$ (\degr)       & = & +14.630$\pm$0.003           & +11.228$\pm$0.014           \\ 
       \hline
      \end{tabular}
      \label{ours2}
     \end{table*}
%
%

        If the impact parameters presented in Sect. 3 are used instead of the values in Chesley et al. (2015) or Farnocchia et al. (2016), 
        a slightly different solution is obtained, see Table \ref{ours2}, right-hand column. This orbital determination is the average of 20 
        good solutions. For this solution, the value of the entry velocity is 12.172$\pm$0.008~km~s$^{-1}$ at an altitude above the surface 
        of the Earth at impact of 41$\pm$10 km.  
%
%
     \begin{table}
      \centering
      \fontsize{8}{11pt}\selectfont
      \tabcolsep 0.05truecm
      \caption{Same as Table \ref{ours2} but considering the available astrometry. The evolution of the best orbital determination presented 
               here matches (within 1$\sigma$) the impact parameters in the new impact solution presented in Sect. 3. There is no best 
               solution matching (within 1$\sigma$) the impact parameters in Chesley et al. (2015) or Farnocchia et al. (2016), such 
               solution cannot be found. 
              }
      \begin{tabular}{lll}
       \hline
        Semi-major axis, $a$ (AU)                           & = &       1.1623128$\pm$0.0000002   \\
        Eccentricity, $e$                                   & = &       0.2116144$\pm$0.0000002   \\
        Inclination, $i$ (\degr)                            & = &       1.41559$\pm$0.00002       \\
        Longitude of the ascending node, $\Omega$ (\degr)   & = &     101.608626$\pm$0.000014     \\
        Argument of perihelion, $\omega$ (\degr)            & = &      52.33925$\pm$0.00003       \\
        Mean anomaly, $M$ (\degr)                           & = &     324.146021$\pm$0.000008     \\
        Time of perihelion passage, $\tau$ (JD TDB)         & = & 2456704.213037$\pm$0.000004     \\
        Perihelion, $q$ (AU)                                & = &       0.91635067$\pm$0.00000013 \\
        Aphelion, $Q$ (AU)                                  & = &       1.4082749$\pm$0.0000005   \\
        Impact time, $t_{\rm impact}$ (JD UTC)              & = & 2456659.62830$\pm$0.00002       \\
        Longitude of impact, $\lambda_{\rm impact}$ (\degr) & = &   $-$44.663$\pm$0.013           \\ 
        Latitude of impact, $\phi_{\rm impact}$ (\degr)     & = &     +13.057$\pm$0.006           \\ 
       \hline
      \end{tabular}
      \label{ours3}
     \end{table}
%
%
%
%
%
%

     \subsection{Improving the solution using astrometry} \label{li}
        It may be argued that, for the particular case of 2014~AA, the astrometry is a piece of information far too important to be 
        neglected as we summarily did in the previous section. The MPC Database\footnote{\url{http://www.minorplanetcenter.net/db_search}} 
        includes seven astrometric observations; the JPL Small-Body Database computed its current solution using the same seven astrometric 
        positions and one derived from infrasound data. Until 2015 April 13 00:35:03 \textsc{ut}, the JPL computed its solution using only 
        the seven astrometric positions (see Table \ref{elements}). The seven observations from the MPC were published in Kowalski et al. 
        (2014) and they are topocentric values (see Table \ref{proof}). 

        Simulations provide geocentric equatorial coordinates directly. Observational topocentric values can be transformed into geocentric 
        values, but the conversion process is rather uncertain because the value of the geocentric distance associated with each pair of 
        topocentric equatorial coordinates is, in principle, unknown unless we adopt an orbital solution. Uncertainties in the plane of sky 
        as seen from the geocentre can be large, perhaps as large as 40{\arcsec} (=0\fdg011=0\fh00074). The other option, going from the 
        values of the geocentric equatorial coordinates obtained from simulations to topocentric values, is in theory less prone to error. 
        Unfortunately, high precision (deviations of a few arcseconds or smaller) conversion from geocentric equatorial coordinates to 
        topocentric is not exempt of problems itself when the values of the geocentric distance are as small as the ones considered here. 

        Formulae dealing with the conversion of the position of a body on the celestial sphere as viewed from the Earth's centre to the one 
        seen from a location on the surface of our planet of known longitude, latitude and altitude (parallax in right ascension and 
        declination) have been discussed in e.g. Maxwell (1932) or Smart (1977). These expressions depend on the value of the Greenwich Mean 
        Sidereal Time at 0 h UTC and also on the model used to describe the shape of the Earth. The conversion algorithm used to compute the 
        root-mean-square deviation in order to compare differences between the values of the topocentric (for observatory code G96, Mt. 
        Lemmon Survey) equatorial coordinates derived from simulations and those from observations has been validated/calibrated using MPC 
        data and, for the range of geocentric distances of interest here, it has been found to introduce systematic errors $<$1{\arcsec} in 
        both right ascension and declination; for geocentric distances $>$0.1 AU the systematic errors are completely negligible. The 
        original topocentric values given in Kowalski et al. (2014) are apparent values; they give us the position of 2014~AA when its light 
        left the asteroid. Geocentric equatorial coordinates derived from simulations give the true values of these coordinates. However, 
        our values can be adjusted for light-time, i.e. they can be made apparent values. The computed position will be observed at a later 
        time, $t_{\rm i} + \Delta_{\rm i}/c$, where $t_{\rm i}$ is the time when the light was reflected, $\Delta_{\rm i}$ is the 
        asteroid--Earth distance, and $c$ is the speed of light. For the range of distances associated with the available observations, the 
        epoch correction is $\sim1.35$~s, that is small enough to be neglected. 
%
%
     \begin{figure}
       \centering
        \includegraphics[width=\linewidth]{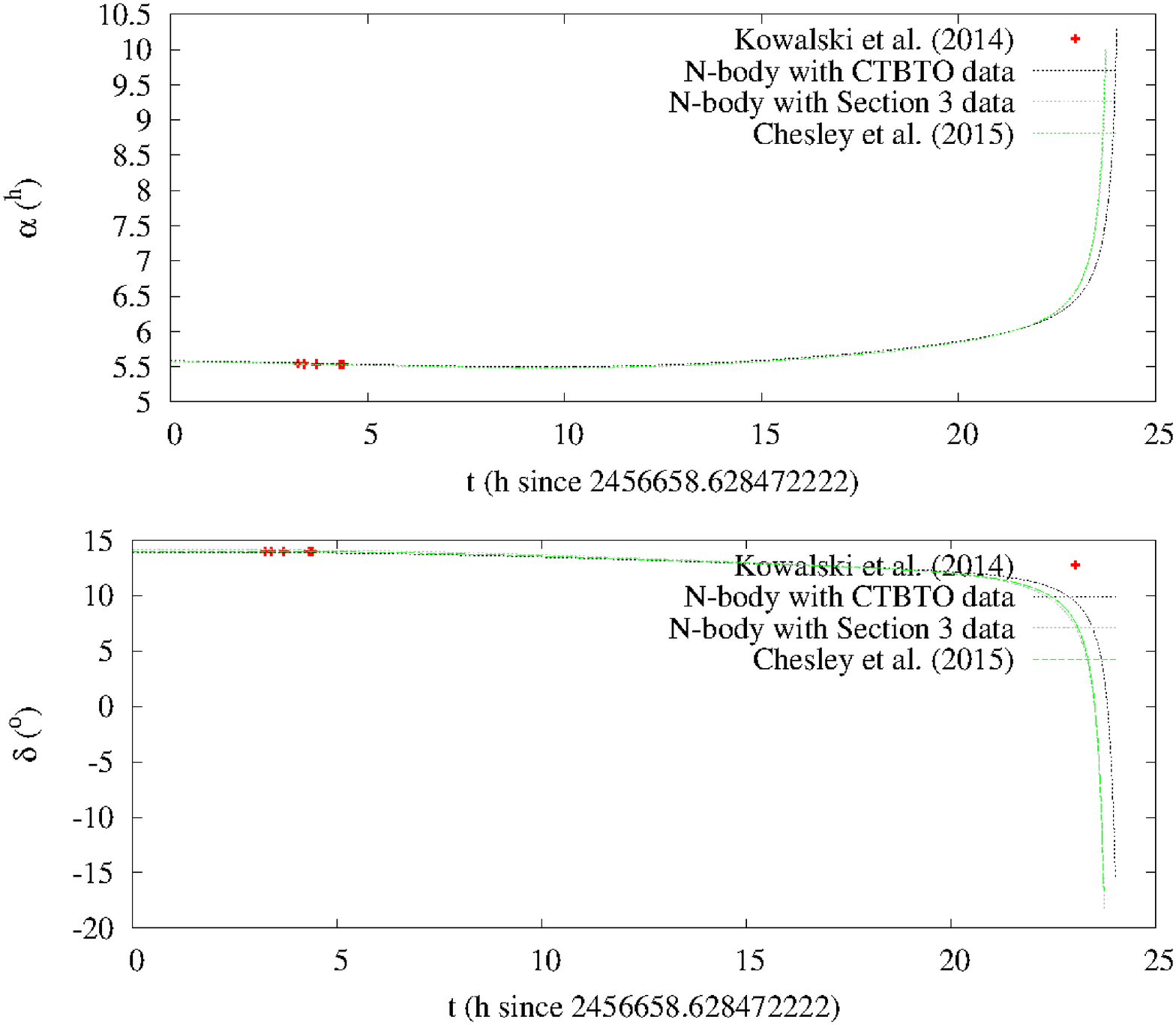}\\
        \includegraphics[width=\linewidth]{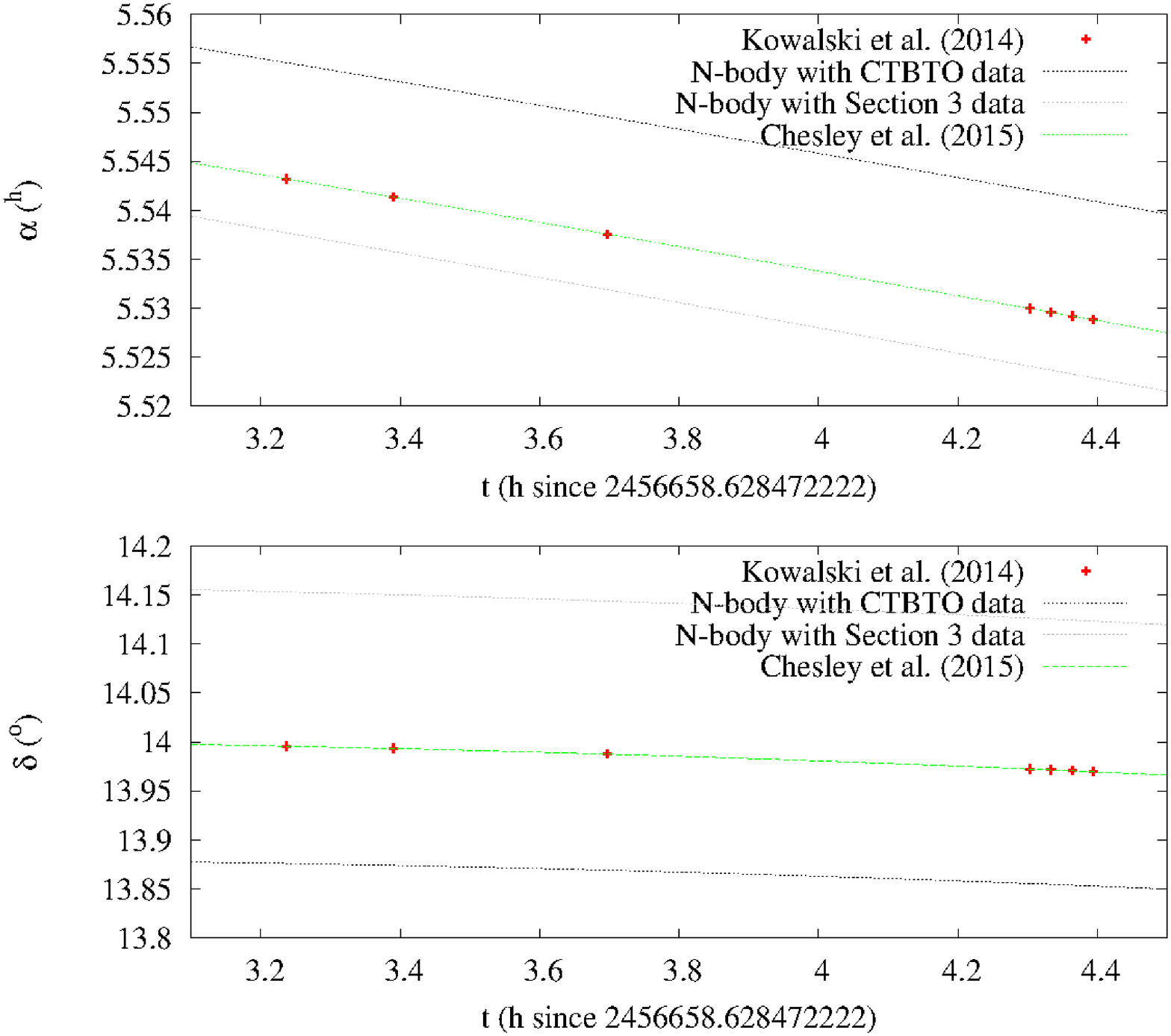}\\
        \caption{Time evolution of the topocentric equatorial coordinates for the various candidate orbital solutions of 2014~AA. The entire
                 integration is displayed on the first two panels and the time window defined by the observations in Kowalski et al. (2014)
                 is displayed on the other two (see the text for details).  
                }
        \label{astrometry}
     \end{figure}
%
%

        Figure \ref{astrometry} shows the evolution of the topocentric equatorial coordinates (right ascension, $\alpha$, and declination,
        $\delta$) during the integrations for three orbital solutions, including the two derived in the previous section ---see Table 
        \ref{ours2}, left-hand column, labelled as `$N$-body with CTBTO data' and right-hand column, labelled as `$N$-body with Section 3 
        data'--- and that discussed in Chesley et al. (2015). The orbital determinations labelled as `$N$-body with CTBTO data' and `Chesley 
        et al. (2015)' are based on the same values of the impact parameters, but the one in Chesley et al. (2015) was refined using the 
        available astrometry. The first two panels show the entire time evolution and the other two are restricted to the time window 
        defined by the observations in Kowalski et al. (2014); the actual observations are also displayed, their associated error bars are 
        too small to be seen. The limitations of our impact-parameters-only $N$-body determinations are clear from the figure, but the fact
        is that ---in the vast majority of cases--- meteoroid impacts do not have any associated pre-impact astrometry. For those cases, 
        the methodologies presented in this research could be very helpful for both finding pre-impact orbital solutions and assessing the
        quality of the ones obtained using other, more classical techniques.

\begin{figure}
        \centering
        \includegraphics[width=\linewidth]{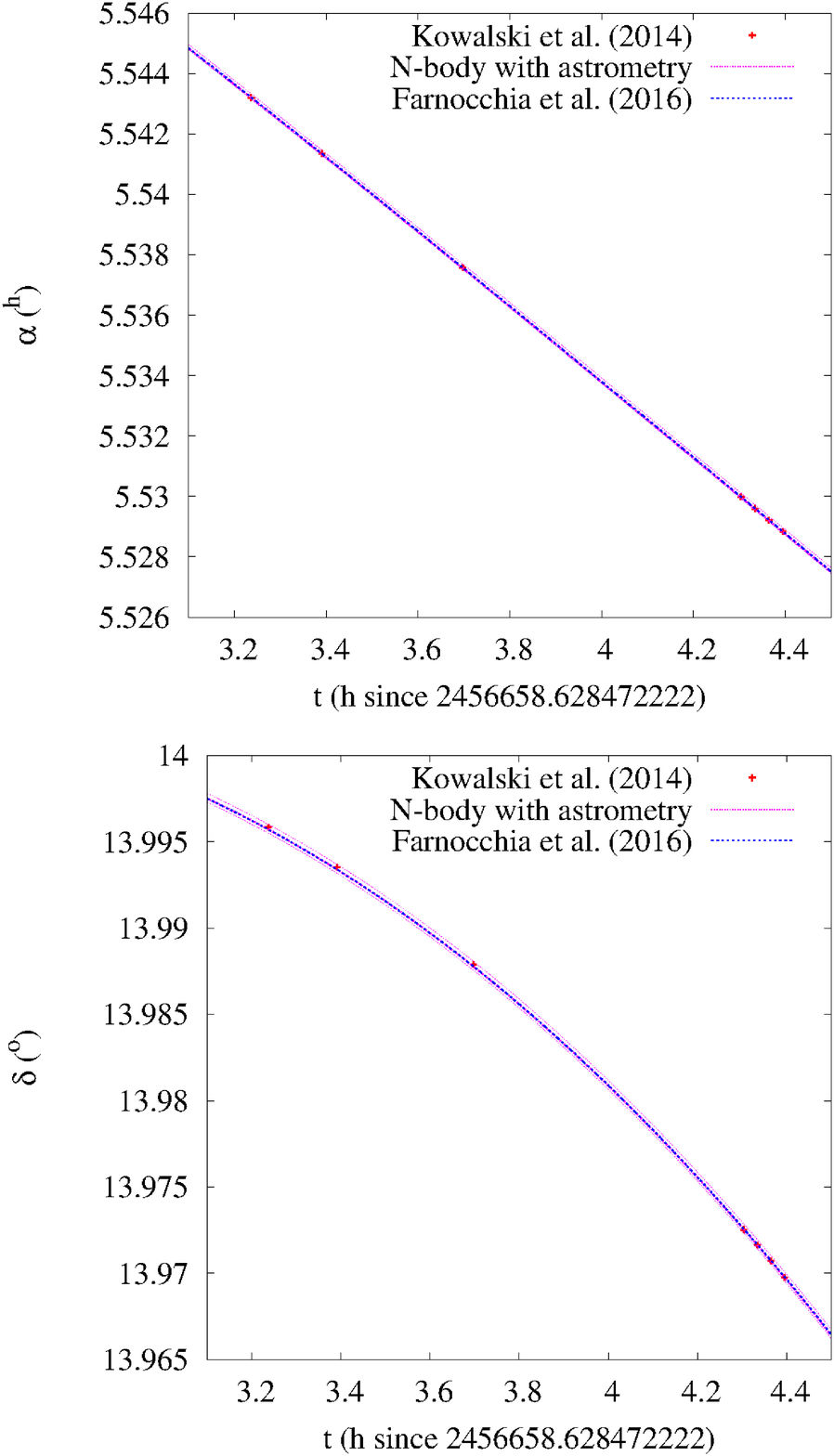}
        \caption{Time evolution of the topocentric equatorial coordinates during the time window defined by the observations in Kowalski et 
                 al. (2014). A relevant integration with a root-mean-square deviation in $\alpha$ of 0\farcs6 and 0\farcs3 in $\delta$ 
                 with respect to the values in Kowalski et al. (2014) is labelled `N-body with astrometry', this solution (Table \ref{ours3}) 
                 is compatible with the improved impact parameters presented in Sect. 3 (within 1$\sigma$). The thinner curves parallel to 
                 it give the maximum and minimum values of the coordinates at the given time for a set of 1,000 control orbits generated 
                 from our favoured solution using the covariance matrix (see Sect. \ref{mccm}).
                }
        \label{radeccorr}
     \end{figure}

        In this section we improve our solutions considering the available astrometry. In order to achieve this goal, we have used a 
        bivariate Gaussian distribution to minimize the deviations between the values of the topocentric coordinates resulting from our 
        candidate orbital solution and the values of the topocentric equatorial coordinates in Kowalski et al. (2014). After a few million 
        trials following the methodology explained above and enforcing consistency with the impact parameters presented in Sect. 3 within 
        1$\sigma$, we obtain the orbital solution displayed in Table \ref{ours3} and plotted in Fig. \ref{radeccorr} under the label 
        `$N$-body with astrometry'. It is only slightly different (the largest difference appears in the value of the orbital inclination) 
        from the previous one (compare values in Table \ref{ours3} against those in the right-hand column of Table \ref{ours2}) but matches 
        the astrometry quite well. The root-mean-square deviation in $\alpha$ is 0\farcs59 and the one in $\delta$ is 0\farcs28. 

        The results from the current solution provided by the JPL Small-Body Database and presented in Farnocchia et al. (2016) are also 
        displayed in Fig. \ref{radeccorr} for comparison; this integration has been performed under the same framework (see Sect. 6.7 for 
        details) as for all the other ones, but using a Cartesian state vector (as initial conditions) computed by the \textsc{Horizons} 
        On-Line Ephemeris System. The two thinner curves, parallel to the one of our best solution, represent the upper and lower boundaries 
        for the values of the topocentric equatorial coordinates of a sample of 1,000 orbits generated using the covariance matrix (see 
        Sect. \ref{mccm} for details) from our own orbital solution. The orbital solution displayed in Table \ref{ours3} is therefore 
        consistent with the available astrometry and consistent within 1$\sigma$ of the values in Table \ref{pierrick}, the mutual delay in 
        impact time is about 15 minutes; in addition, the altitude of the virtual airburst is 47$\pm$4 km that is also consistent with the 
        expectations from infrasound modelling. 

        Applying the same approach but using the impact parameters in Chesley et al. (2015) or Farnocchia et al. (2016) ---the REB solution, 
        see Sects. 2 and 3--- as constraints, we could not find a solution able to comply with the astrometry within 1{\arcsec} rms and the 
        impact solution within 1$\sigma$. It is numerically impossible to satisfy both requirements concurrently. This negative outcome is 
        in agreement with the discussion in Chesley et al. (2015) or Farnocchia et al. (2016). The resulting virtual impact parameters in 
        Chesley et al. (2015) or Farnocchia et al. (2016) are only consistent within 3$\sigma$ of the values in the REB solution due to a 
        significant offset in latitude (see e.g. fig. 1 in Farnocchia et al. 2016).
%
%
%
%

\begin{figure}
        \centering
        \includegraphics[width=\linewidth]{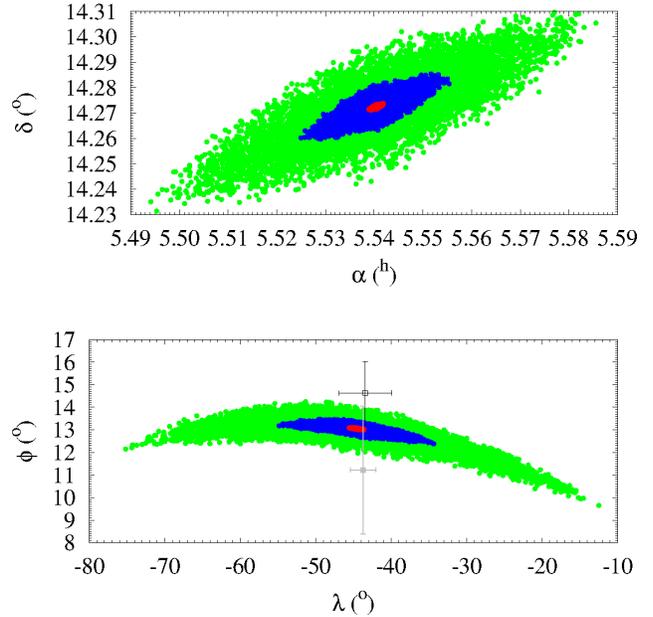}
        \caption{True radiant geocentric equatorial coordinates (top panel) and their associated virtual impact coordinates (bottom panel). 
                 Virtual impacts plotted in green represent those associated with sets of orbital elements within 30$\sigma$ of the orbital 
                 solution in Table \ref{ours3}, those in blue are the result of a 10$\sigma$ spread, and the ones in red are restricted to 
                 1$\sigma$. The impact point derived from the infrasound data in Chesley et al. (2015) appears in black with its 
                 approximate error bars, our determination presented in Sect. 3 is plotted in grey.  
                }
        \label{radiant}
     \end{figure}

\begin{figure}
        \centering
        \includegraphics[width=\linewidth]{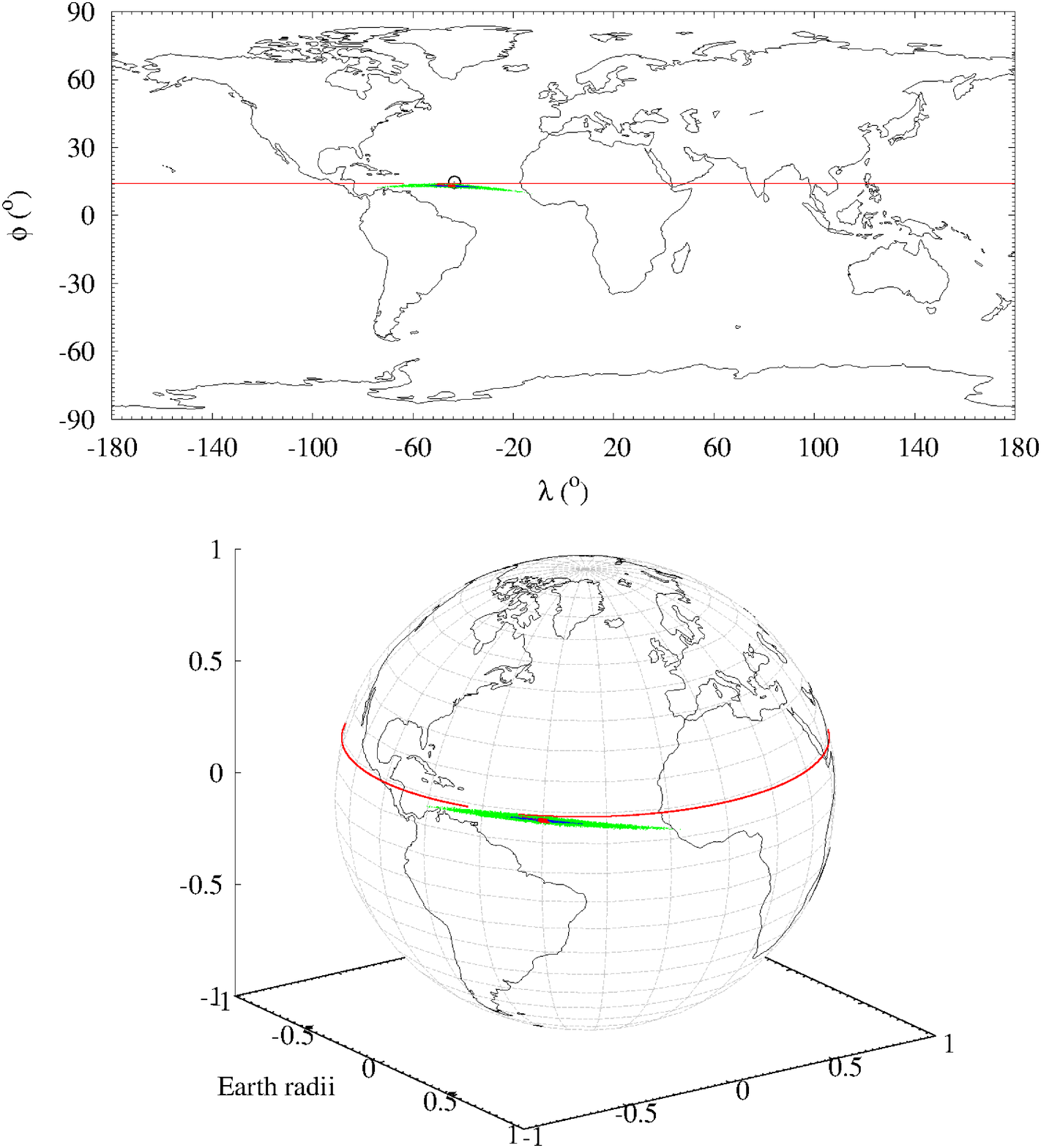}
        \caption{Distribution on the surface of the Earth of the virtual impacts studied in Sect. \ref{li}. Virtual impactor colours are 
                 as in Fig. \ref{radiant}. The impact point derived from the infrasound data described in Chesley et al. (2015) appears as a 
                 circle. This figure is similar to panel b, fig. 5 in Sitarski (1998). The path of risk for one representative orbit is also
                 displayed; it reached an altitude above the surface of the Earth of 47.80 km at coordinates (44\fdg81~W, 13\fdg02~N). The 
                 red curve outlines the flight path (E to W) of the object; it starts above the Caribbean Sea. The location of the impact 
                 point as in Chesley et al. (2015) is also plotted. Only nearly 23 h prior to the strike are displayed.
                }
        \label{needle}
     \end{figure}

        In Fig. \ref{radeccorr}, the curve labelled `N-body with astrometry' is representative of those virtual impactors which are more 
        consistent with the available astrometry, but still compatible with the improved impact parameters presented in Sect. 3 (within 
        1$\sigma$). They determine a volume in orbital parameter space about an orbit that goes evenly between the first and the last 
        observations in Kowalski et al. (2014), this defines an eye-of-a-needle on the sky (see Fig. \ref{radiant}, top panel); any virtual 
        impactor originated from that radiant will comply with the available astrometry to a certain degree (see Fig. \ref{radeccorr}) and 
        it will hit the Earth with some values of the impact parameters consistent with the limits derived in Sect. 3 (see Fig. 
        \ref{radiant}, bottom panel, and Fig. \ref{needle}).
%
%
%
%

        Figure \ref{radiant} shows the results of three experiments consisting of $2\times10^{4}$ test orbits each. The top panel shows the 
        true geocentric equatorial coordinates of the virtual impactors at the beginning of the simulation, i.e. at epoch 2456658.628472222 
        JD TDB. The points in red correspond to test orbits within 1$\sigma$ of the orbital solution in Table \ref{ours3}, those in blue 
        have a 10$\sigma$ spread, and the green ones have 30$\sigma$. However, the points have been generated using uniformly distributed 
        random numbers in order to survey the relevant region of the orbital parameter space evenly. Each virtual impactor generates one 
        point on the bottom panel of the figure following the same colour pattern. The impact point derived from the infrasound data as 
        described in Chesley et al. (2015) and the one presented in Sect. 3 are also plotted with error bars. If the virtual impactors are 
        forced to pass through the astrometry, the simulated impacts are fully statistically consistent ---in terms of coordinates--- with 
        the determination based on infrasound data presented in Sect. 3 but only marginally consistent with the one used in Chesley et al. 
        (2015) due to the values of the latitude which are too far south with respect to that from the REB. This fact clearly shows why our 
        method failed to find a solution able to comply with the astrometry within 1{\arcsec} rms and the REB solution within 1$\sigma$, it 
        is numerically impossible.

        The distribution on the surface of the Earth of the virtual impacts studied here is more clearly displayed in Fig. \ref{needle} 
        where the virtual impacts define an arc extending from the Caribbean Sea to West Africa if deviations as large as 30$\sigma$ from 
        the orbital solution in Table \ref{ours3} are allowed. Figure \ref{needle} also displays the path of risk (the projection of the 
        trajectory of the incoming body on the surface of the Earth as it rotates) associated with a representative most probable solution 
        (see Fig. \ref{radeccorr}). The object was easily observable from Arizona 20 h before impact; the path has been stopped nearly at 
        the time of impact.
%
%
%
%

\begin{figure}
        \centering
        \includegraphics[width=\linewidth]{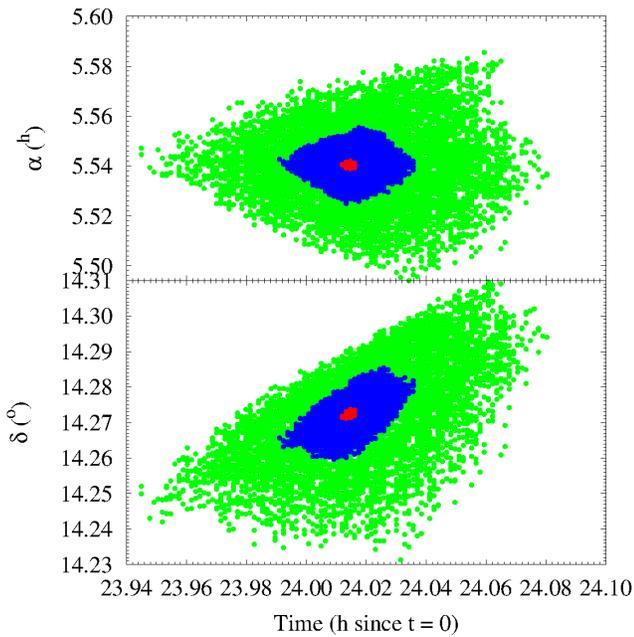}
        \caption{True radiant geocentric equatorial coordinates of the virtual impactors as a function of the impact time. Virtual impactor 
                 colours as in Fig. \ref{radiant}.  
                }
        \label{timing}
     \end{figure}

\begin{figure}
        \centering
        \includegraphics[width=\linewidth]{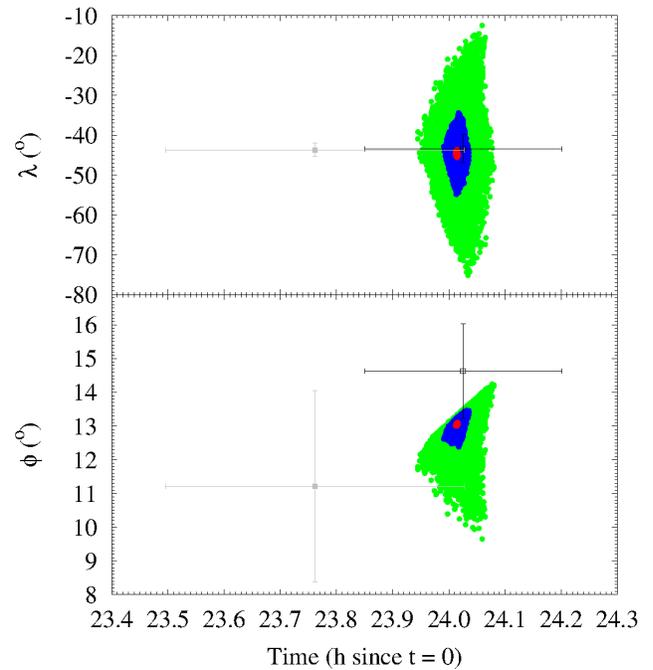}
        \caption{Impact coordinates of the virtual impacts as a function of the impact time. Virtual impactor colours as in Fig. 
                 \ref{radiant}. The impact point derived from the infrasound data in Chesley et al. (2015) appears in black with error bars, 
                 our determination presented in Sect. 3 is plotted in grey. 
                }
        \label{timing2}
     \end{figure}

\begin{figure}
        \centering
        \includegraphics[width=\linewidth]{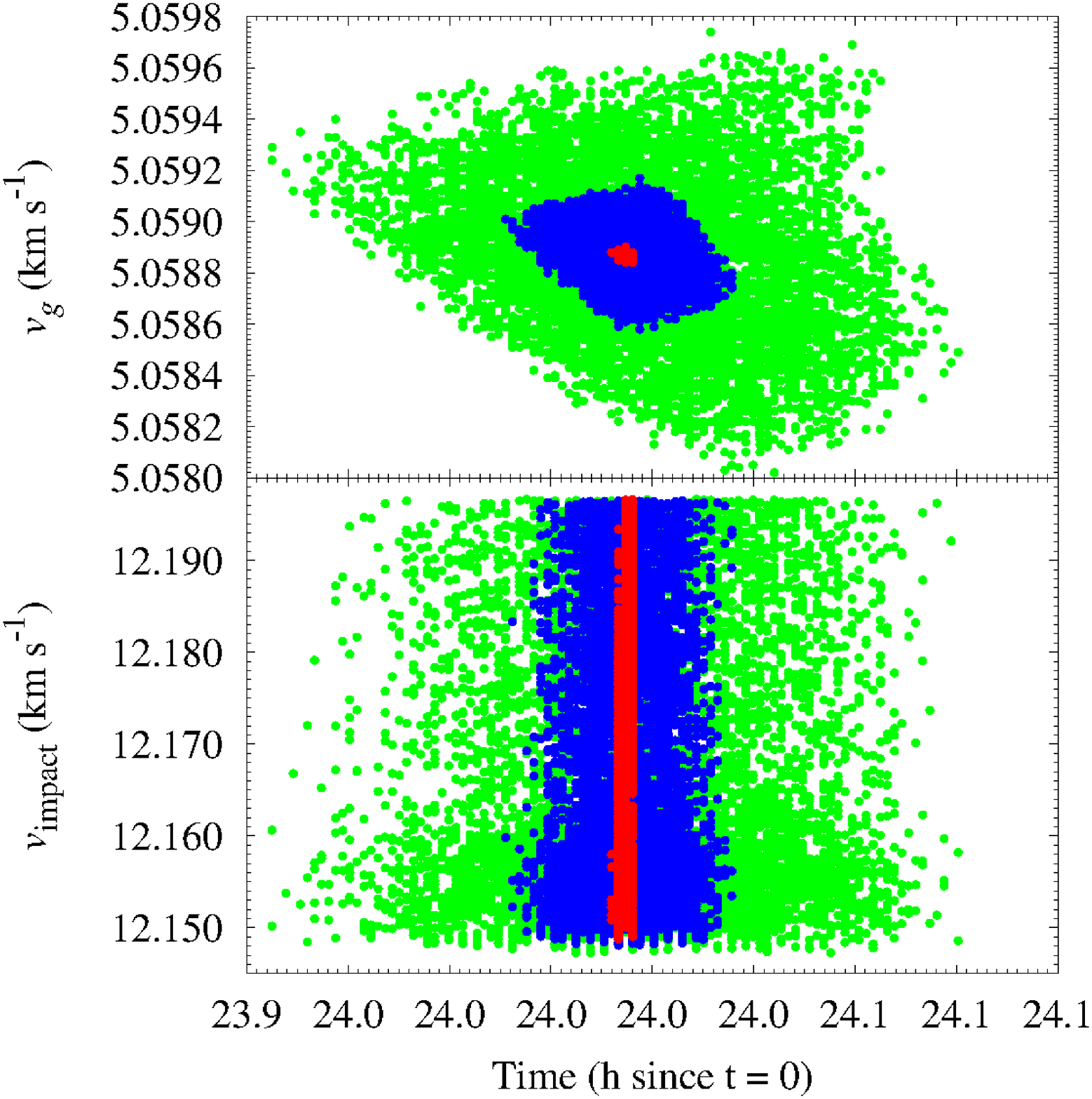}
        \caption{Values of the $v_{\rm g}$ and $v_{\rm impact}$ of the virtual impactors as a function of the impact time. Virtual impactor 
                 colours as in Fig. \ref{radiant}.  
                }
        \label{vs}
     \end{figure}

        Figures \ref{timing} and \ref{timing2} show the results of the same three experiments described above in terms of the impact time.
        Those virtual impactors strictly compatible with the astrometry define a very small range for the associated impact time. The value
        of the impact time derived from infrasound data in Sect. 3 is compatible with virtual impactors from the region consistent
        with the astrometry. Figure \ref{timing2} shows the three pieces of information together and the statistical consistency is quite 
        obvious. The ranges in $v_{\rm g}$ and $v_{\rm impact}$ are plotted in Fig. \ref{vs}. There are no known meteor showers with 
        parameters similar to those in Figs. \ref{timing} and \ref{vs} (see e.g. Jenniskens 2006 or the most up-to-date information in Jopek 
        \& Kanuchov\'a 2014), but the value of $v_{\rm g}$ is rather low to be easily detectable if they do exist. The coordinates of the 
        radiant associated with the orbital solution in Table \ref{ours3} are $\alpha_{0}$ = 5{\fh}540281$\pm$0{\fh}000003 
        (83{\fdg}10421$\pm$0{\fdg}00004) and $\delta_{0}$ = +14{\fdg}27232$\pm$0{\fdg}00005; the values of the velocities are 
        $v_{\rm impact}$ = 12.170$\pm$0.003 km s$^{-1}$ and $v_{\rm g}$ = 5.0589$\pm$0.0001 km s$^{-1}$. 
%
%
%
%
%
%
%
%
%
%
%
%

        These results give a clear picture on how precise an orbital solution should be in order to generate reliable impact predictions in 
        terms of timing and location. We consider that the orbit in Table \ref{ours3} is the best possible and the one that we regard as the 
        most probable pre-impact orbit of 2014~AA because it matches the available astrometry reasonably well, its associated virtual 
        impacts are consistent with the impact solution found using infrasound data in Sect. 3, and it has been computed for an epoch 
        sufficiently distant from the impact time to give an accurate orbital solution, appropriate to study both the past dynamical 
        evolution of 2014~AA and the possible existence of other objects moving in similar orbits among the known NEOs (see Sects. 7 and 8). 

  \section{Our results in context}
     Here and in order to place our results in context, we compare them with those computed by other groups. In the case of the orbital 
     solutions, this comparison allows us to determine if they are consistent with those derived using only astrometric data.
%
%
     \begin{table*}
      \centering
      \fontsize{8}{11pt}\selectfont
      \tabcolsep 0.1truecm
      \caption{Comparison between the values of the topocentric (for observatory code G96, Mt. Lemmon Survey) and geocentric equatorial 
               coordinates (R.A. in hh:mm:sec, Decl. in $\degr$:$\arcmin$:$\arcsec$) of 2014~AA computed from the solution provided by the 
               MPC$^{1}$ and the one available from the JPL (Farnocchia et al. 2016). The original, observational (topocentric) values 
               (Kowalski et al. 2014) are also displayed. (J2000.0 ecliptic and equinox. Sources: MPC and \textsc{Horizons} On-Line 
               Ephemeris System.)
              }
      \resizebox{\linewidth}{0.12\linewidth}{
      \begin{tabular}{lllllllllll}
       \hline
        Source           & \multicolumn{2}{c}{Kowalski et al. (2014)}  & \multicolumn{4}{c}{MPC} & \multicolumn{4}{c}{JPL}  \\
                         & \multicolumn{2}{c}{topocentric}             & \multicolumn{2}{c}{geocentric} & \multicolumn{2}{c}{topocentric} & \multicolumn{2}{c}{geocentric} & \multicolumn{2}{c}{topocentric}  \\
       \hline
        DATE UTC         & \multicolumn{2}{c}{R.A. (J2000) Decl.}      & \multicolumn{2}{c}{R.A. (J2000) Decl.} & \multicolumn{2}{c}{R.A. (J2000) Decl.} & \multicolumn{2}{c}{R.A. (J2000) Decl.} & \multicolumn{2}{c}{R.A. (J2000) Decl.} \\
       \hline
        2014 01 01.26257 & 05:32:35.55 & +13:59:45.0 & 05:32:39.4 & +14:16:32 & 05:32:35.5 & +13:59:45 & 05:32:39.34 & +14:16:14.2 & 05:32:35.56 & +13:59:45.0 \\
        2014 01 01.26896 & 05:32:28.89 & +13:59:36.7 & 05:32:40.4 & +14:16:32 & 05:32:28.8 & +13:59:37 & 05:32:40.20 & +14:16:13.9 & 05:32:28.87 & +13:59:36.6 \\
        2014 01 01.28176 & 05:32:15.27 & +13:59:16.4 & 05:32:42.4 & +14:16:33 & 05:32:15.2 & +13:59:17 & 05:32:41.95 & +14:16:13.2 & 05:32:15.26 & +13:59:16.2 \\
        2014 01 01.30701 & 05:31:47.92 & +13:58:21.1 & 05:32:46.7 & +14:16:34 & 05:31:47.9 & +13:58:21 & 05:32:45.56 & +14:16:12.0 & 05:31:47.92 & +13:58:21.2 \\
        2014 01 01.30828 & 05:31:46.54 & +13:58:17.9 & 05:32:46.9 & +14:16:34 & 05:31:46.5 & +13:58:18 & 05:32:45.75 & +14:16:11.9 & 05:31:46.54 & +13:58:17.9 \\
        2014 01 01.30955 & 05:31:45.15 & +13:58:14.6 & 05:32:47.1 & +14:16:34 & 05:31:45.1 & +13:58:15 & 05:32:45.93 & +14:16:11.8 & 05:31:45.15 & +13:58:14.5 \\
        2014 01 01.31081 & 05:31:43.79 & +13:58:11.1 & 05:32:47.3 & +14:16:34 & 05:31:43.7 & +13:58:11 & 05:32:46.12 & +14:16:11.8 & 05:31:43.79 & +13:58:11.2 \\
       \hline
      \end{tabular}
      }
      \label{proof}
     \end{table*}
%
%

     \subsection{The REB solution}
        The values of the impact parameters of meteoroid 2014~AA as derived from infrasound data in both the REB and our own determination
        are based on data from three detecting stations, one located to the north-west in the Northern Hemisphere and two to the south-west 
        across the Equator (see Fig. \ref{pmialle}). The hypocentre of the airburst is located in the Northern Hemisphere, but the 
        travel-time model used by the IDC system to derive the origin time is global and has been obtained empirically. It does not account 
        for paths crossing the Equator, where stratospheric winds are typically weaker (Brachet et al. 2010; Green \& Bowers 2010; Le Pichon 
        et al. 2012). The REB determination was computed using the global travel-times that are not suited for across the Equator 
        propagation; to compute the origin time, our determination uses the detection time of the closest station to the north-west which is 
        the only one located in the same hemisphere as the hypocentre of the airburst. The origin time in the REB determination is a rather 
        crude estimate of the most probable value.

     \subsection{Kowalski et al. (2014)}
        MPEC 2014-A02 (Kowalski et al. 2014) includes a preliminary orbit for 2014~AA: $a=$1.1660751~AU, $e=$0.2149211, $i=$1\fdg43759, 
        $\Omega=$101\fdg57475, $\omega=$52\fdg35440, and $M=$324\fdg30925, referred to epoch JD 2456658.5. However, no error estimates are 
        given and that prevents a proper quantitative comparison. Assuming that the uncertainties are similar to those quoted above,$^{2}$ 
        this orbital solution seems to be compatible with our findings.

     \subsection{MPC}
        The orbital solution available from the MPC$^{1}$$^{,}$\footnote{\url{http://www.minorplanetcenter.net/db_search/show_object?object_id=2014+AA}} 
        does not include any error estimates, therefore it is difficult to provide a quantitative assessment of its consistency with our 
        results. It is based on the seven astrometric observations available and it is able to reproduce their values (see Table 
        \ref{proof}). The root-mean-square deviation with respect to the values in Kowalski et al. (2014) in right ascension amounts to 
        0\farcs95, the respective root-mean-square deviation in declination is 0\farcs30. In principle, it appears to be fully compatible 
        with our findings both geometric and $N$-body based if the values of the uncertainties are similar to those quoted above.$^{2}$  
 
     \subsection{NEODyS}
        The orbital solution available from the Near Earth Objects Dynamic Site$^{2}$$^{,}$\footnote{\url{http://newton.dm.unipi.it/neodys}} 
        (NEODyS; Chesley \& Milani 1999) is also based on the same seven astrometric observations. Both our geometric and $N$-body solutions 
        are well within the boundaries of the quoted errors.

     \subsection{JPL Small-Body Database}
        Table \ref{elements} shows the orbits of 2014~AA as computed by the JPL's Solar System Dynamics group. Using the orbital solution 
        based on both astrometry and infrasound data (second solution in Table \ref{elements}), the orbital elements of 2014~AA at epoch JD 
        2456658.628472222 (2014 January 1 03:05:00.0000 TDB) are $a$ = 1.164173034310643 AU, $e$ = 0.2134945866378556, $i$ = 
        1\fdg428587405092261, $\Omega$ = 101\fdg6075729620208, $\omega$ = 52\fdg43618813777764, and $M$ = 324\fdg2172353344642 (source: 
        \textsc{Horizons} On-Line Ephemeris System). These orbital elements and the nominal uncertainties ($\Delta a$ = 0.010579 AU, 
        $\Delta e$ = 0.010551, $\Delta i$ = 0\fdg073019, $\Delta \Omega$ = 0\fdg015969, $\Delta \omega$ = 0\fdg49968, and $\Delta M$ = 
        0\fdg44168) indicate that our solution is consistent with this determination. If we compare this solution with the one in Table 
        \ref{ours3}, the relative discrepancies in semi-major axis, eccentricity, inclination, longitude of the ascending node, argument of 
        perihelion and time of perihelion passage are: 0.16\%, 0.88\%, 0.91\%, 0.0010\%, 0.18\%, and 1.1$\times$10$^{-6}$\%. On the other 
        hand, our geometric solution is well within the boundaries of those in Table \ref{elements}. The comparison with the solution 
        currently displayed by the JPL is given in Sect. 6.7. 

     \subsection{Chesley et al. (2015)}
        Chesley et al. (2015) applied systematic ranging (Milani \& Kne\v{z}evi\'c 2005) to estimate the orbit of 2014~AA. This approach 
        uses the recorded observations as input data. In their table 2 they provide the following solution: $a$ = 1.1694951152059 AU, $e$ = 
        0.2185819813893022, $i$ = 1\fdg463031271816491, $\Omega$ = 101\fdg5884374095149, $\omega$ = 52\fdg61778832155812, and 
        $\tau$ = 2456704.205054487 JD TDB. This solution is computed at epoch JD 2456658.8115875920. For this solution the impact happens at 
        2014 January 2, 02:54:20 (UTC), latitude ({\degr}N) = +13\fdg118, and longitude ({\degr}E) = $-$44\fdg207. Therefore, its impact 
        time is over 11 minutes earlier than the one obtained from infrasound data and its reported impact latitude is close to the southern 
        edge of the observational solution (see their fig. 5). Chesley et al. (2015) do not provide an indication of the uncertainty and 
        state that their orbit is their best match between orbital and infrasound constraints (see above). For this solution they compute an 
        impact probability of 99.9\%.

        If we compare this solution with the geometric one in Table \ref{ours1}, left-hand column, we observe the following relative 
        discrepancies in semi-major axis, eccentricity, inclination, longitude of the ascending node, argument of perihelion, and time of 
        perihelion passage: 0.039\%, 0.44\%, 2.1\%, 0.081\%, 0.94\%, and 0.000017\% (using the closest $\tau$). Two completely independent 
        methods produce very similar solutions. This is fully consistent with the results of the quality assessment analysis presented in de 
        la Fuente Marcos \& de la Fuente Marcos (2013) and above for the Almahata Sitta event caused by the meteoroid 2008~TC$_{3}$.

\begin{figure}
        \centering
        \includegraphics[width=\linewidth]{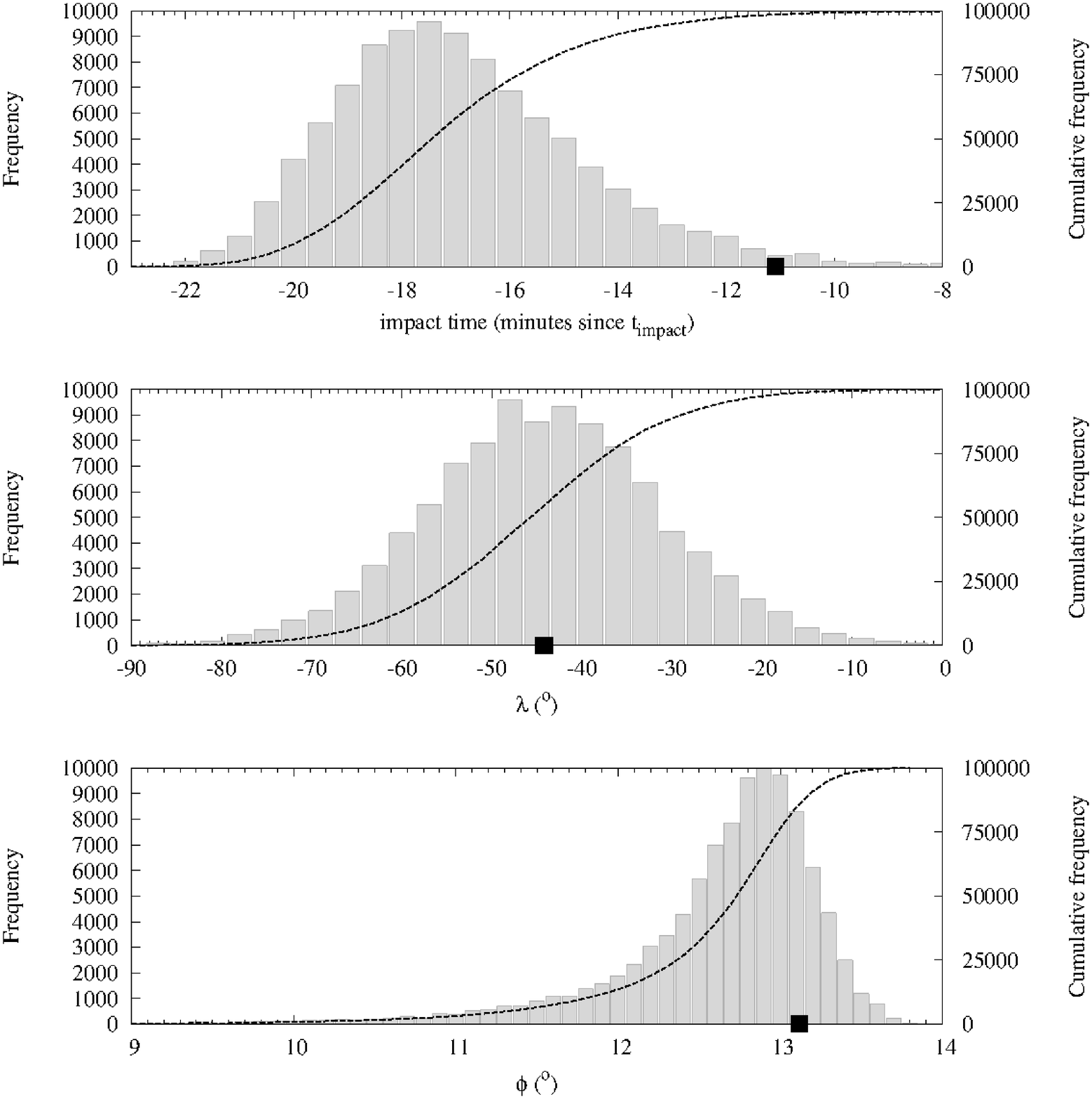}
        \caption{Resulting distributions in impact parameter space for an experiment analogous to the one in Fig. \ref{ctbto} but using the 
                 best match in Chesley et al. (2015) and assumed uncertainties (see the text for details); the values of the best match
                 impact parameters are also indicated (black squares). The impact time is referred to the value based on infrasound data
                 according to Chesley et al. (2015).} 
        \label{chesley}
     \end{figure}

        As for the $N$-body approach, we have performed an experiment similar to the ones described in Sect. 5 but using the solution in 
        table 2 of Chesley et al. (2015). The following uncertainties have been assumed $\sigma_{\rm a}$ = 8.74$\times$10$^{-6}$ AU, 
        $\sigma_{\rm e}$ = 5.73$\times$10$^{-6}$, $\sigma_{\rm i}$ = 4\fdg1364$\times$10$^{-5}$, $\sigma_{\Omega}$ = 
        1\fdg7891$\times$10$^{-6}$, $\sigma_{\omega}$ = 7\fdg6207$\times$10$^{-5}$, and $\sigma_{\tau}$ = 1.4155$\times$10$^{-4}$ JD TDB. 
        These values are those of the currently available orbit of 2008~TC$_{3}$. Using these conditions, 99.999\% of the orbits hit the 
        Earth generating a meteor but nearly 17 minutes (in average) before the impact time in Chesley et al. (2015) and around coordinates 
        latitude ({\degr}N) = +13\degr and longitude ({\degr}E) = $-$44\degr. The outcome of the experiment is summarized in Fig. 
        \ref{chesley}. If we compare this solution with the one from the $N$-body approach corrected with astrometry (Table \ref{ours3}), 
        the relative discrepancies in semi-major axis, eccentricity, inclination, longitude of the ascending node, argument of perihelion 
        and time of perihelion passage are: 0.61\%, 3.19\%, 3.24\%, 0.020\%, 0.53\%, and 3.3$\times$10$^{-7}$\%. Similar differences can be
        found between this orbital solution and the one in Farnocchia et al. (2016). 
%
%
%
%

     \subsection{Farnocchia et al. (2016)}
        Farnocchia et al. (2016) has obtained an improved solution of the pre-impact orbit of 2014~AA. This is the solution currently 
        provided by the JPL Small-Body Database and \textsc{Horizons} On-Line Ephemeris System. Referred to epoch JD 2456658.5, the values 
        of the orbital elements are (see also Table \ref{elements}): $a$ = 1.161570914329746 AU, $e$ = 0.210903385513902, $i$ = 
        1\fdg4109467942359, $\Omega$ = 101\fdg613014674528, $\omega$ = 52\fdg3157504212017, and $M$ = 324\fdg0051879998651. We have 
        downloaded the Cartesian state vector for the epoch JD 2456658.628472222 from the \textsc{Horizons} On-Line Ephemeris System and
        performed a simulation within the same framework used to derive our orbit determination. The values of the geocentric and 
        topocentric coordinates of 2014~AA from this simulation at the times when the original observations of 2014~AA were acquired are 
        shown in Table \ref{compacoor}. The root-mean-square deviation with respect to the values in Kowalski et al. (2014) in right 
        ascension amounts to 0\farcs49, the respective root-mean-square deviation in declination is 0\farcs31. This integration gives 
        $t_{\rm impact}$ = 2456659.629134 JD TDB, $\lambda_{\rm impact}$= $-$44\fdg693, $\phi_{\rm impact}$ = +13\fdg.070, $h_{\rm impact}$ 
        = 54 km, and $v_{\rm impact}$ = 12.1634 km s$^{-1}$. The value of the geocentric impact velocity given in Farnocchia et al. (2016) 
        is 12.17 km s$^{-1}$. On the other hand, the properties of the associated radiant are $\alpha_{0}$ = 5\fh5403, $\delta_{0}$ = 
        14\fdg2723, and $v_{\rm g}$ = 5.05886 km s$^{-1}$. 

\begin{figure}
        \centering
        \includegraphics[width=\linewidth]{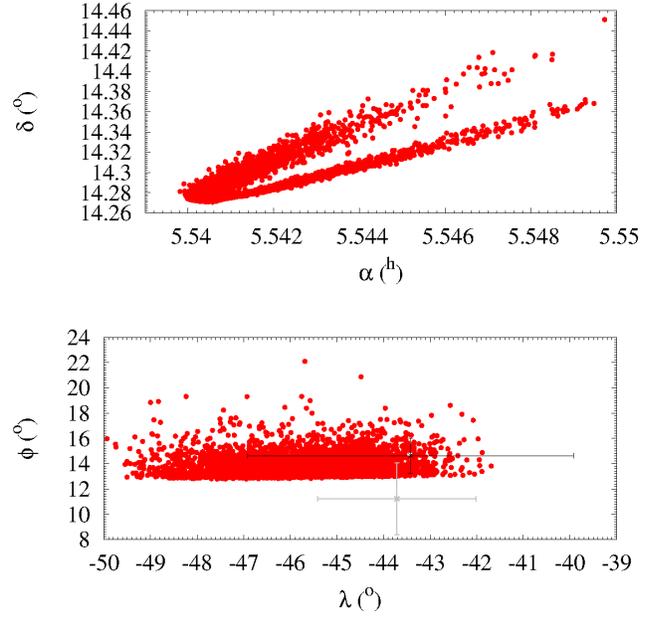}
        \caption{True radiant geocentric equatorial coordinates (top panel) and their associated virtual impact coordinates (bottom panel)
                 for the solution in Farnocchia et al. (2016); see the text for details. The impact point derived from the infrasound data 
                 in Chesley et al. (2015) appears in black with its approximate error bars, our determination presented in Sect. 3 is 
                 plotted in grey.  
                }
        \label{radiantJPL}
     \end{figure}

\begin{figure}
        \centering
        \includegraphics[width=\linewidth]{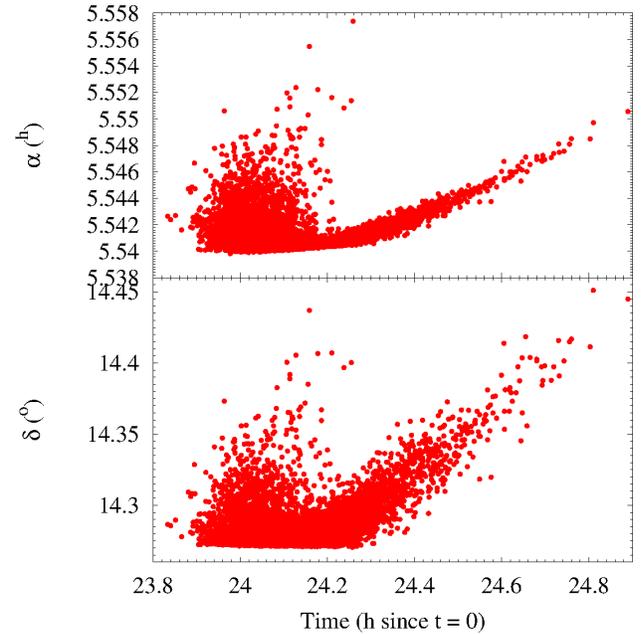}
        \caption{True radiant geocentric equatorial coordinates of the virtual impactors as a function of the impact time for the solution 
                 in Farnocchia et al. (2016).
                }
        \label{timingJPL}
     \end{figure}

\begin{figure}
        \centering
        \includegraphics[width=\linewidth]{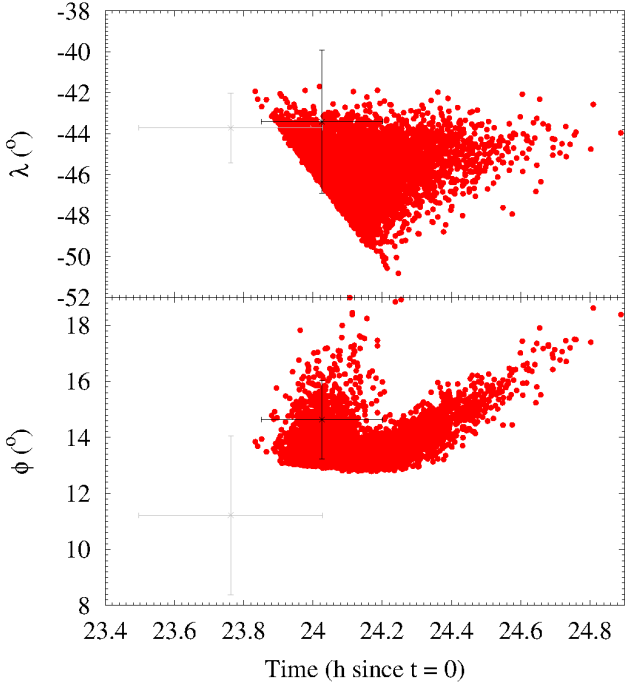}
        \caption{Impact coordinates of the virtual impacts as a function of the impact time for the solution in Farnocchia et al. (2016). 
                 The impact point derived from the infrasound data in Chesley et al. (2015) appears in black with error bars,
                 our determination presented in Sect. 3 is plotted in grey.
                }
        \label{timing2JPL}
     \end{figure}

\begin{figure}
        \centering
        \includegraphics[width=\linewidth]{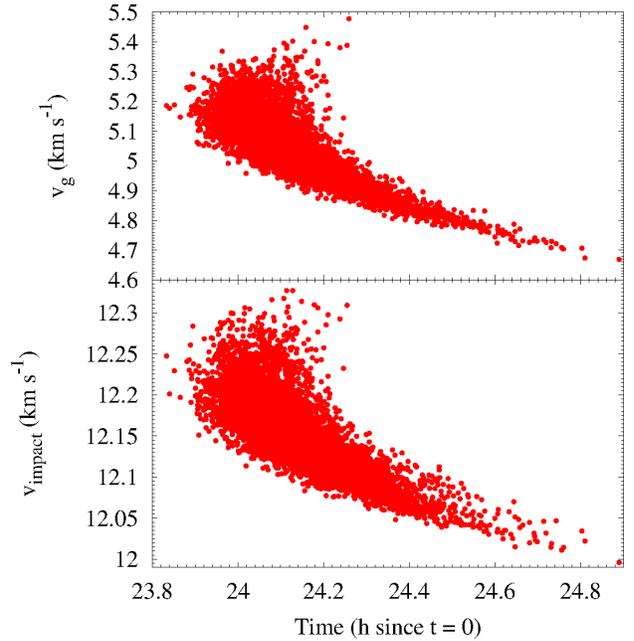}
        \caption{Values of the $v_{\rm g}$ and $v_{\rm impact}$ of the virtual impactors as a function of the impact time for the solution 
                 in Farnocchia et al. (2016). 
                }
        \label{vsJPL}
     \end{figure}

        In addition, we have computed the evolution of 20,000 control orbits generated using the covariance matrix of this solution as 
        provided by the JPL Small-Body Database (see Sect. \ref{mccm}). The results from these simulations are plotted in Figs. 
        \ref{radiantJPL}--\ref{vsJPL} and can be compared directly with those in Figs. \ref{radiant}--\ref{vs}, red points. It is clear that 
        both solutions are reasonably compatible even if they are based on marginally compatible values of the impact parameters (see Sect. 
        3). The values of the relative discrepancies in semi-major axis, eccentricity, inclination, longitude of the ascending node, 
        argument of perihelion and time of perihelion passage are: 0.064\%, 0.34\%, 0.33\%, 0.0043\%, 0.045\%, and 2.8$\times$10$^{-7}$\%. 
        The values of the geocentric and topocentric coordinates of 2014~AA derived from this solution by the JPL at the times when the 
        original observations of 2014~AA were acquired are shown in Table \ref{proof}. The root-mean-square deviation with respect to the 
        values in Kowalski et al. (2014) in right ascension amounts to 0\farcs14, the respective root-mean-square deviation in declination 
        is 0\farcs11. The differences between the root-mean-square deviations as computed by the JPL and the ones from our own calculations
        (see above) using the same input data are small enough to consider our computational approach as robust.
%
%
     \begin{table*}
      \centering
      \fontsize{8}{11pt}\selectfont
      \tabcolsep 0.1truecm
      \caption{Comparison between the values of the geocentric and topocentric (for observatory code G96, Mt. Lemmon Survey) equatorial 
               coordinates of 2014~AA computed from the solution (integrated by us) in Farnocchia et al. (2016) and the best solution from 
               this work at the times when the observations in Table \ref{proof} were acquired (see also Fig. \ref{radeccorr}). The 
               root-mean-square deviation with respect to the values in Kowalski et al. (2014) in right ascension amounts to 0\farcs49 for 
               Farnocchia et al. (2016) and to 0\farcs59 for our solution, the respective root-mean-square deviations in declination are 
               0\farcs31 and 0\farcs28.}
      \begin{tabular}{lllllllll}
       \hline
        Source           & \multicolumn{4}{c}{Farnocchia et al. (2016)} & \multicolumn{4}{c}{This work}  \\
                         & \multicolumn{2}{c}{topocentric} & \multicolumn{2}{c}{geocentric} & \multicolumn{2}{c}{topocentric} & \multicolumn{2}{c}{geocentric} \\
       \hline
        DATE UTC         & \multicolumn{2}{c}{R.A. (J2000) Decl.} & \multicolumn{2}{c}{R.A. (J2000) Decl.} & \multicolumn{2}{c}{R.A. (J2000) Decl.} & \multicolumn{2}{c}{R.A. (J2000) Decl.} \\
       \hline
        2014 01 01.26257 & 05:32:35.55 & +13:59:44.6 & 05:32:39.35 & +14:16:14.1 & 05:32:35.51 & +13:59:44.6 & 05:32:39.31 & +14:16:14.2 \\
        2014 01 01.26896 & 05:32:28.87 & +13:59:36.3 & 05:32:40.21 & +14:16:13.8 & 05:32:28.82 & +13:59:36.3 & 05:32:40.17 & +14:16:13.9 \\
        2014 01 01.28176 & 05:32:15.27 & +13:59:16.0 & 05:32:41.96 & +14:16:13.2 & 05:32:15.22 & +13:59:16.0 & 05:32:41.91 & +14:16:13.2 \\
        2014 01 01.30701 & 05:31:47.96 & +13:58:21.3 & 05:32:45.57 & +14:16:11.9 & 05:31:47.90 & +13:58:21.2 & 05:32:45.51 & +14:16:11.9 \\
        2014 01 01.30828 & 05:31:46.58 & +13:58:18.0 & 05:32:45.76 & +14:16:11.8 & 05:31:46.52 & +13:58:17.9 & 05:32:45.70 & +14:16:11.8 \\
        2014 01 01.30955 & 05:31:45.20 & +13:58:14.7 & 05:32:45.94 & +14:16:11.8 & 05:31:45.13 & +13:58:14.6 & 05:32:45.88 & +14:16:11.7 \\
        2014 01 01.31081 & 05:31:43.83 & +13:58:11.3 & 05:32:46.13 & +14:16:11.7 & 05:31:43.76 & +13:58:11.2 & 05:32:46.07 & +14:16:11.7 \\
       \hline
      \end{tabular}
      \label{compacoor}
     \end{table*}
%
%
%
%
%
%
%
%
%
%
%
%
%
%
%
%
%
%

        Table \ref{proof} shows that geocentric predictions derived from the MPC and JPL solutions (the current JPL solution is the one in 
        Farnocchia et al. 2016) corresponding to the time frame covered by the observations in Kowalski et al. (2014) are incompatible. 
        There is a systematic offset in declination in the range 17\farcs8--22\farcs2 between the geocentric ephemerides derived from these 
        solutions in that time frame. There are also significant differences in right ascension but always $<$17\farcs8. However, the 
        topocentric predictions are both nearly perfect matches of the observational data in Kowalski et al. (2014). Table \ref{compacoor} 
        shows the coordinate predictions from the nominal orbits in Farnocchia et al. (2016) and our own, integrated and processed under the 
        same conditions (see also Fig. \ref{radeccorr}). We observe that the geocentric values from the JPL solution (computed by the JPL) 
        in Table \ref{proof} and those of Farnocchia et al. (2016) in Table \ref{compacoor} are nearly perfect matches (differences below
        0\farcs2); consistently, similar deviations are observed for the topocentric values. The slight discrepancies may be the result of 
        using different formulae for the various conversions. In any case, these very small differences should have no effects on the study 
        of both the past dynamical evolution of 2014~AA and the possible existence of other objects moving in similar orbits among the known 
        NEOs.

  \section{Peeking into the past of 2014 AA}
\begin{figure}
       \centering
        \includegraphics[width=\linewidth]{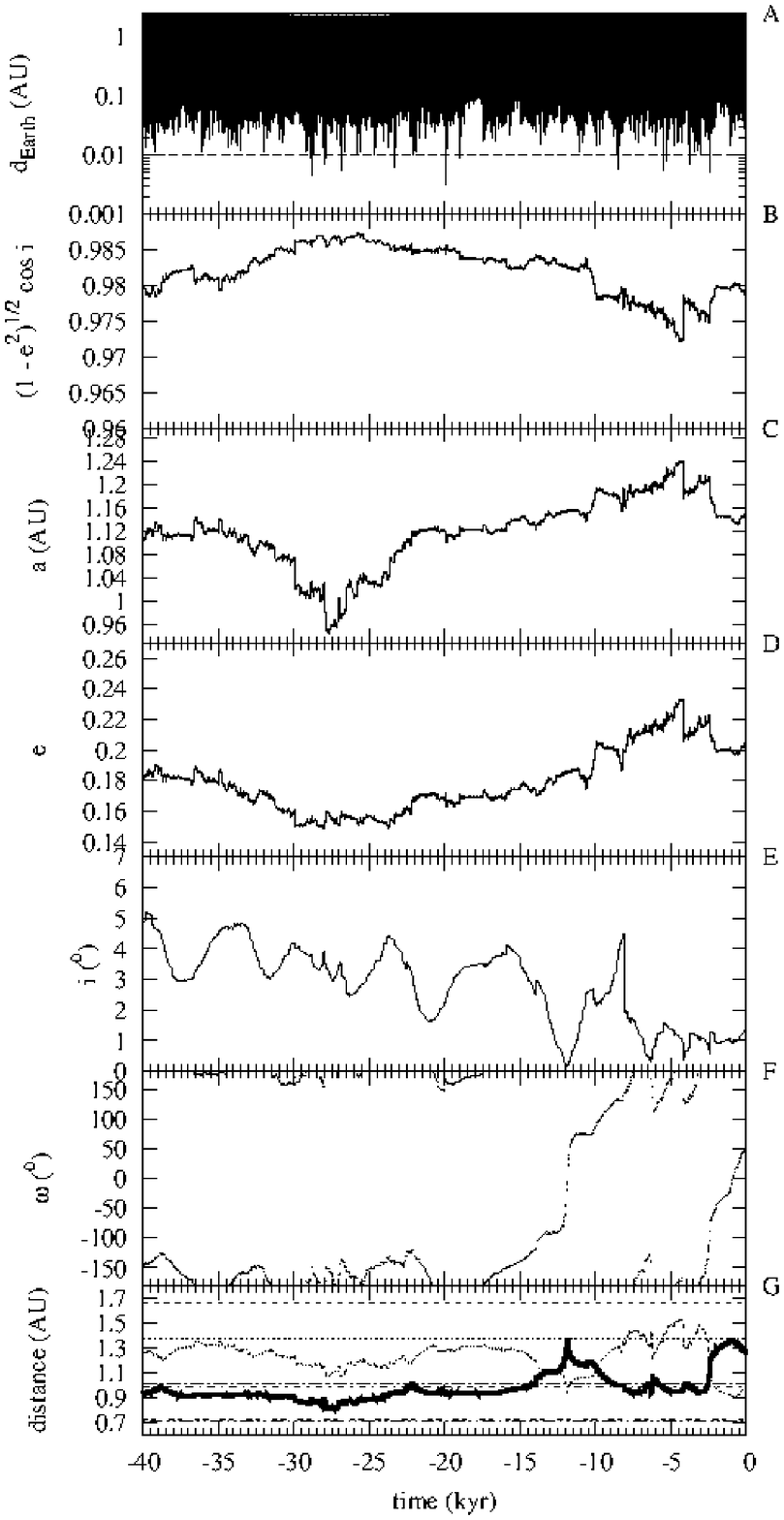}
        \caption{Time evolution of various parameters for the orbital solution of 2014~AA in Table \ref{ours3}. The distance from the Earth 
                 (panel A); the value of the Hill sphere radius of the Earth, 0.0098 AU, is displayed. The parameter $\sqrt{1 - e^2} \cos i$ 
                 (panel B). The orbital elements $a$ (panel C), $e$ (panel D), $i$ (panel E) and $\omega$ (panel F). The distance to the 
                 descending (thick line) and ascending nodes (dotted line) is in panel G. Planetary perihelion and aphelion distances 
                 (Venus, Earth, and Mars) are also shown.
                }
        \label{14AA}
     \end{figure}

     We have used the solution displayed in Table \ref{ours3} and the same $N$-body techniques described above to further investigate the 
     past dynamics of 2014~AA. Figure \ref{14AA} shows that this object has remained in the orbital neighbourhood of our planet for several 
     thousands of years. It was only relatively recently ($\sim$2.5~kyr ago) that it started to undergo close encounters with Mars at 
     aphelion (descending node), although the nodes (e.g. ascending) had been close to Mars in the past (see panel G, Fig. \ref{14AA}). The 
     object experienced very close encounters with the Earth--Moon system in the past and that explains its highly chaotic orbital evolution 
     (see panels A and C, Fig. \ref{14AA}).  
%
%
%
%

     However, the most striking feature in Fig. \ref{14AA} appears in the time evolution of $\omega$ (panel F). This orbital parameter does 
     not circulate smoothly (i.e., take any possible value) as in the case of a passing body and it librated around 180\degr from about 40 
     to 14 kyr ago, then again from 9 kyr to 2 kyr. The object was starting a libration about $\omega\sim0\degr$ when the impact took place.
     These are the signposts of one of the variants of the Kozai resonance (Kozai 1962). An argument of perihelion librating around 180\degr 
     implies that the associated object reaches perihelion at approximately the same time it crosses the Ecliptic from North to South (the 
     descending node). Perihelion at the ascending node is linked to $\omega\sim0\degr$; meteoroid 2014~AA found our planet at the ascending 
     node. When the Kozai resonance occurs at low inclinations, the argument of perihelion librates around 0\degr or 180\degr (see e.g. 
     Milani et al. 1989). Michel \& Thomas (1996) confirmed that, at low inclinations, the argument of perihelion of some NEOs can librate 
     around either 0\degr or 180\degr. This topic received additional attention from Namouni (1999). 
 
     The Kozai-controlled past evolution of this object is firmly established, we have integrated 50 control orbits derived from the orbital 
     solution in Table \ref{ours3}, assuming a normal distribution in orbital parameter space, and all of them exhibit this behaviour during 
     the last 15 kyr or so. The only difference is in the timing of the episodes, when it switches from libration around 180\degr to other 
     states. Some control orbits, in the dynamical neighbourhood of the one plotted, exhibit libration of the argument of perihelion around 
     180\degr for most of the time interval displayed in Fig. \ref{14AA}. The evolution of the value of the distance from the nodes to the 
     Sun (panel G in Fig. \ref{14AA}) is clearly coupled with the behaviour of $\omega$. The values of eccentricity and inclination are also 
     coupled when $\omega$ librates ($\sqrt{1 - e^2} \cos i \approx$ constant, see panel B in Fig. \ref{14AA}) although the oscillation in 
     $e$ is difficult to observe due to superposition of secular resonances (see below).  
%
%
      \begin{figure}
        \centering
         \includegraphics[width=\linewidth]{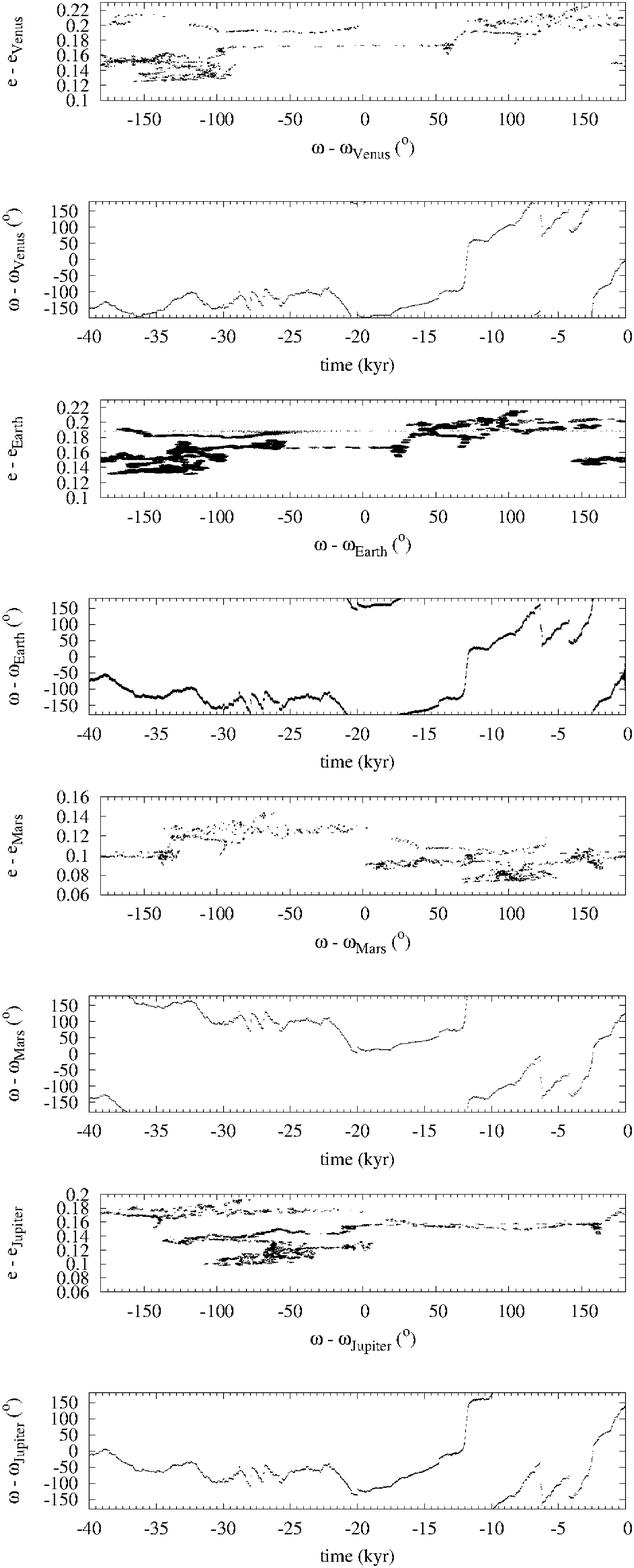}
         \caption{The $e_{\rm r} \omega_{\rm r}$-portrait for the orbital solution of 2014~AA relative to Venus, the Earth, Mars, and 
                  Jupiter. Data as in Fig. \ref{14AA}.
                 }
         \label{fEWaa}
      \end{figure}
%
%

     The overall evolution of 2014~AA in Fig. \ref{14AA} appears to be rather chaotic, perhaps secularly chaotic. This secular evolution is 
     best studied in the $e_{\rm r} \omega_{\rm r}$-plane, where $e_{\rm r} = e - e_{\rm p}$ and $\omega_{\rm r} = \omega - \omega_{\rm p}$, 
     $e_{\rm p}$ and $\omega_{\rm p}$ are, respectively, the eccentricity and argument of perihelion of a given planet (Namouni 1999). Our 
     $e_{\rm r} \omega_{\rm r}$ maps look very irregular (see Fig. \ref{fEWaa}) and 2014~AA suffers secular interactions that induce 
     librations of its relative argument of perihelion with respect to our planet but also to Venus, Mars, and Jupiter. The topic of 
     overlapping secular resonances and its effects on the dynamics of asteroids was first studied by Michel (1997) in the particular case 
     of objects moving in Venus horseshoe orbits. He concluded that overlapping of secular resonances is possible, complicating the dynamics 
     of horseshoe orbits significantly. Asteoroid 2014~AA does not appear to have experienced co-orbital (horseshoe or any other type) 
     episodes with our planet in the immediate past. Figure \ref{fEWaa} shows that the secular behaviour of this object with respect to 
     Venus and the Earth was rather similar, with oscillation of their respective $\omega_{\rm r}$ around 180\degr until about 13 kyr before 
     impact. For Mars, the coupling is also obvious during the same period of time, although its $\omega_{\rm r}$ librates about 90\degr for 
     most of the displayed time and around $-$90{\degr} as the impact time approaches. As for Jupiter, it librated around $-$90{\degr} for 
     most of the displayed time. The actual source of the observed secular behaviour is Jupiter. If Jupiter is not included in the 
     calculations, the secular chaos vanishes. 

     The calculations in Namouni (1999) were made within the context of the restricted elliptic three-body problem. The dynamical situation 
     here is significantly more complicated. An external perturber, that is not interacting directly with 2014~AA, induces the secular 
     behaviour observed and the entire system of overlapping secular resonances drives the evolution of the meteoroid. If Jupiter is 
     removed, the librations of the relative argument of perihelion of 2014~AA with respect to the other planets cease immediately. This 
     behaviour is expected as the terrestrial planets share the effect of the secular perturbation from Jupiter (see e.g. Ito \& Tanikawa 
     1999, 2002; Tanikawa \& Ito 2007). The reported secular behaviour is quite consistent across control orbits. This web of overlapping 
     secular resonances appears to keep the semi-major axis of this object confined within the neighbourhood of the Earth for extended 
     periods of time. This may also have played a role in facilitating the eventual impact. Given its marginal stability (the value of the 
     semi-major axis remains fairly stable during some of the displayed evolution), this subdomain of the NEO orbital parameter realm may 
     host additional objects and this interesting possibility is explored in the following section.

  \section{Possible related objects and their stability}
     It has been argued that some of the recent, most powerful Earth impacts may be associated with resonant groups of NEOs and/or very 
     young meteoroid streams (de la Fuente Marcos \& de la Fuente Marcos 2015a). Both 2008~TC$_{3}$ and 2014~AA caused atmospheric impact 
     events or airbursts that released an amount of energy equivalent to about one kiloton of Trinitrotoluene (TNT) explosives (Jenniskens 
     et al. 2009; Chesley et al. 2015); therefore, they can be included among the low-yield members of the group of recent, most powerful 
     Earth impacts with the Chelyabinsk Event occupying the top of the scale (Brown et al. 2013). Since 2000, the CTBTO infrasound sensors 
     of the IMS network (Le Pichon et al. 2012) have detected at least 26 events related to asteroid impacts with an individual energy in 
     excess of 1 (and up to $\sim$500) kt of TNT 
     equivalent.\footnote{\url{http://newsroom.ctbto.org/2014/04/24/ctbto-detected-26-major-asteroid-impacts-in-earths-atmosphere-since-2000/}} 
     The most extraordinary event recorded so far by the IMS network is the Chelyabinsk superbolide, on 2013 February 15 (Brown et al. 2013; 
     Le Pichon et al. 2013; Pilger et al. 2015). 

     \subsection{Dynamically-related objects?}
        Assuming that 2014~AA may have been a fragment of a larger body and/or that other objects could be moving in similar orbits if they 
        are trapped in some web of secular resonances like the one described in the previous section, here we try to single out candidates
        to be following somewhat similar orbits. The dynamical evolution of these potentially similar minor bodies is further studied to 
        confirm or reject a putative dynamical link between 2014~AA and the selected candidates. 

        In order to identify suitable candidates we use the $D$-criteria of Southworth \& Hawkins (1963), $D_{\rm SH}$, Lindblad \& 
        Southworth (1971), $D_{\rm LS}$ (in the simplified form of eqn. 1 in Lindblad 1994 or eqn. 1 in Foglia \& Masi 2004), Drummond 
        (1981), $D_{\rm D}$, and the $D_{\rm R}$ from Valsecchi et al. (1999). A search among NEOs currently catalogued (as of 2016 July 31) 
        by the JPL Small-Body Database\footnote{\url{http://ssd.jpl.nasa.gov/sbdb.cgi}} using these criteria gives the list in Table 
        \ref{candidates}. The objects are sorted by ascending $D_{\rm LS}$. Only objects with $D_{\rm LS}$ and $D_{\rm R} < 0.05$ are shown, 
        which is a somewhat arbitrary but conservative choice within the NEO context (see e.g. de la Fuente Marcos \& de la Fuente Marcos 
        2016). The $D$-criteria have been computed with respect to the solution displayed in Table \ref{ours3}. We must emphasize that the 
        use of the various $D$-criteria is a helpful device to single out candidates suitable for further study; we are not assuming that a 
        low value of one or more of the $D$-criteria computed using osculating orbital elements must necessarily imply any physical or 
        dynamical link between two given objects.
%
%
     \begin{landscape}
     \begin{table}
      \centering
      \fontsize{8}{11pt}\selectfont
      \tabcolsep 0.09truecm
      \caption{Orbital elements, orbital periods ($P_{\rm orb}$), perihelia ($q = a \ (1 - e)$), aphelia ($Q = a \ (1 + e)$), number of
               observations ($n$), data-arc, absolute magnitudes ($H$) and MOID of minor bodies with orbits similar to that of the meteoroid 
               2014~AA (as in Table \ref{ours3}). The various $D$-criteria ($D_{\rm SH}$, $D_{\rm LS}$, $D_{\rm D}$ and $D_{\rm R}$) are 
               also shown. The objects are sorted by ascending $D_{\rm LS}$ (eqn. 1 in Lindblad 1994). Only objects with $D_{\rm LS}$ and 
               $D_{\rm R} < 0.05$ are shown. The orbit of 2014~AA is the solution displayed in Table \ref{ours3} that is 
               referred to the epoch JD 2456658.628472222. The orbits of the other objects are referred to the Epoch 2457600.5 
               (2016-July-31.0) TDB (Barycentric Dynamical Time) with the exceptions of 2009 SH$_{1}$ that is referred to 2455092.5 
               (2009-September-18.0) and 2015 MZ$_{53}$ that is referred to 2457194.5 (2015-June-21.0). Data as of 2016 July 31.}
      \begin{tabular}{lllllllllllllllll}
       \hline
          Asteroid       & $a$ (AU)  & $e$        & $i$ (\degr) & $\Omega$ (\degr) & $\omega$ (\degr) & $P_{\rm orb}$ (yr) & $q$ (AU) & $Q$ (AU)
                         & $n$ & arc (d) & $H$ (mag) & MOID (AU)
                         & $D_{\rm SH}$ & $D_{\rm LS}$ & $D_{\rm D}$ & $D_{\rm R}$ \\
       \hline
    {\bf 2014 AA}        & {\bf 1.16231}   & {\bf 0.21161}    & {\bf 1.41559}     & {\bf 101.60863}        &  {\bf 52.33925}        &  {\bf 1.26}              & {\bf 0.91635}  & {\bf 1.40827} 
                         & {\bf --}  & {\bf --}      & {\bf 30.90}     & {\bf 0.0000005}
                         & {\bf --}           & {\bf --}           & {\bf --}          & {\bf --         }\\
       \hline
         2011 JV$_{10}$  & 1.13988   & 0.20225    & 1.40510     & 221.37001        & 297.53422        &  1.22              & 0.90934  & 1.37043 
                         & 18  & 2       & 29.70     & 0.00130
                         & 0.04762      & 0.01170      & 0.02739     & 0.04182     \\
         2011 GJ$_{3}$   & 1.14129   & 0.20439    & 0.84382     & 331.06066        & 308.53533        &  1.22              & 0.90801  & 1.37456 
                         & 38  & 20      & 26.20     & 0.00357
                         & 0.37201      & 0.01487      & 0.14677     & 0.04106     \\
         2012 DJ$_{54}$  & 1.17610   & 0.22986    & 1.99084     & 336.88562        & 120.04517        &  1.28              & 0.90577  & 1.44644 
                         & 25  & 17      & 28.60     & 0.00163
                         & 0.21820      & 0.02336      & 0.08311     & 0.02792     \\
         2015 MZ$_{53}$  & 1.14419   & 0.18837    & 2.20730     & 259.66059        & 298.31300        &  1.22              & 0.92865  & 1.35972 
                         &  7  &  2      & 27.50     & 0.00054
                         & 0.16437      & 0.02971      & 0.07886     & 0.02157     \\
         2007 HC         & 1.15510   & 0.20717    & 3.15638     & 216.84639        &  57.97253        &  1.24              & 0.91580  & 1.39440 
                         & 31  &  6      & 25.20     & 0.00331
                         & 0.37076      & 0.03071      & 0.14257     & 0.02658     \\
         2012 UY$_{68}$  & 1.17489   & 0.22823    & 2.90143     &  70.31433        &  35.75187        &  1.27              & 0.90675  & 1.44303 
                         & 46  & 24      & 25.00     & 0.01743
                         & 0.18231      & 0.03226      & 0.07059     & 0.03450     \\
         2013 RV$_{9}$   & 1.16511   & 0.19902    & 3.50067     & 332.82949        & 108.32269        &  1.26              & 0.93323  & 1.39699 
                         & 88  & 183     & 23.60     & 0.02268
                         & 0.25691      & 0.04204      & 0.09239     & 0.02910     \\
         2013 UM$_{9}$   & 1.21328   & 0.24618    & 2.93258     &  38.92410        & 283.29746        &  1.34              & 0.91459  & 1.51197 
                         & 32  & 13      & 24.80     & 0.01826
                         & 0.45899      & 0.04358      & 0.22727     & 0.03576     \\
         2013 HO$_{11}$  & 1.19448   & 0.23414    & 3.71878     &  59.05671        & 245.90995        &  1.31              & 0.91481  & 1.47416 
                         & 81  & 251     & 23.00     & 0.00564
                         & 0.43503      & 0.04611      & 0.19423     & 0.02404     \\
         2009 SH$_{1}$   & 1.19832   & 0.24599    & 3.32567     & 354.92542        & 294.87464        &  1.31              & 0.90355  & 1.49310 
                         & 11  &  1      & 29.40     & 0.00399
                         & 0.43131      & 0.04957      & 0.18978     & 0.04105     \\
         2004 XK$_{3}$   & 1.22784   & 0.26053    & 1.48182     &  57.91377        & 304.67526        &  1.36              & 0.90795  & 1.54772 
                         & 240 & 1464    & 24.40     & 0.00095
                         & 0.46053      & 0.04964      & 0.22403     & 0.04411     \\
       \hline
      \end{tabular}
      \label{candidates}
     \end{table}
     \end{landscape}
%
%
%
%
         \begin{table*}
          \centering
          \fontsize{8}{11pt}\selectfont
          \tabcolsep 0.15truecm
          \caption{Heliocentric Keplerian orbital elements of 2011~JV$_{10}$, 2011~GJ$_{3}$, 2012 DJ$_{54}$, and 2013~NJ$_{4}$ used in this 
                   research. Values include the 1$\sigma$ uncertainty. The orbits are computed at epoch JD 2457600.5 that corresponds to 
                   0:00 TDB on 2016 July 31 (J2000.0 ecliptic and equinox). Source: JPL Small-Body Database.
                  }
          \begin{tabular}{llllll}
           \hline
                                                              &   &     2011~JV$_{10}$    &     2011~GJ$_{3}$     &    2012 DJ$_{54}$       
                                                                  &     2013~NJ$_{4}$     \\ 
           \hline
            Semi-major axis, $a$ (AU)                         & = &   1.1399$\pm$0.0002   &   1.1413$\pm$0.0003   &   1.1761$\pm$0.0002  
                                                                  &   1.13324$\pm$0.00007 \\
            Eccentricity, $e$                                 & = &   0.2023$\pm$0.0002   &   0.2044$\pm$0.0003   &   0.2299$\pm$0.0002             
                                                                  &   0.20253$\pm$0.00008 \\
            Inclination, $i$ (\degr)                          & = &   1.4051$\pm$0.0012   &   0.8438$\pm$0.0015   &   1.9908$\pm$0.0014             
                                                                  &   1.3195$\pm$0.0004   \\
            Longitude of the ascending node, $\Omega$ (\degr) & = & 221.370$\pm$0.002     & 331.06$\pm$0.03       & 336.8856$\pm$0.0002               
                                                                  & 115.6066$\pm$0.0007   \\
            Argument of perihelion, $\omega$ (\degr)          & = & 297.534$\pm$0.004     & 308.54$\pm$0.02       & 120.0452$\pm$0.0003                
                                                                  & 238.5093$\pm$0.0007   \\
            Mean anomaly, $M$ (\degr)                         & = & 155.4$\pm$0.3         &  69.1$\pm$0.6         & 208.6$\pm$0.3               
                                                                  & 145.81$\pm$0.08       \\
            Perihelion, $q$ (AU)                              & = &   0.90934$\pm$0.00007 &   0.90801$\pm$0.00015 &   0.90577$\pm$0.00005               
                                                                  &   0.90373$\pm$0.00004 \\
            Aphelion, $Q$ (AU)                                & = &   1.3704$\pm$0.0002   &   1.3746$\pm$0.0004   &   1.4464$\pm$0.0002             
                                                                  &   1.36275$\pm$0.00008 \\
            Absolute magnitude, $H$ (mag)                     & = &  29.7                 &  26.2                 &  28.6          
                                                                  &  27.4                 \\
           \hline
          \end{tabular}
          \label{family}
         \end{table*}
%
%
%
%
     \begin{figure*}
       \centering
        \includegraphics[width=0.49\linewidth]{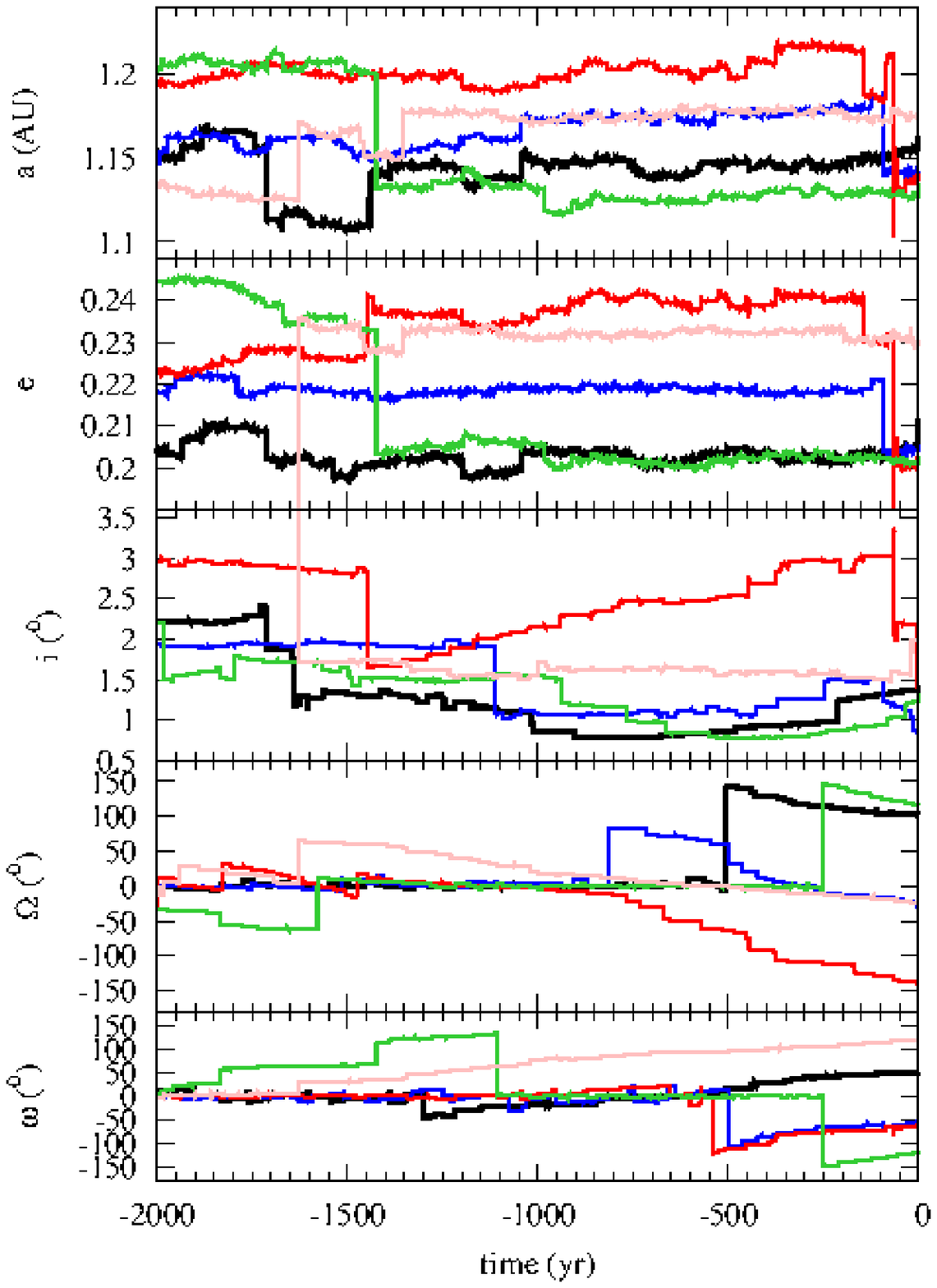}
        \includegraphics[width=0.49\linewidth]{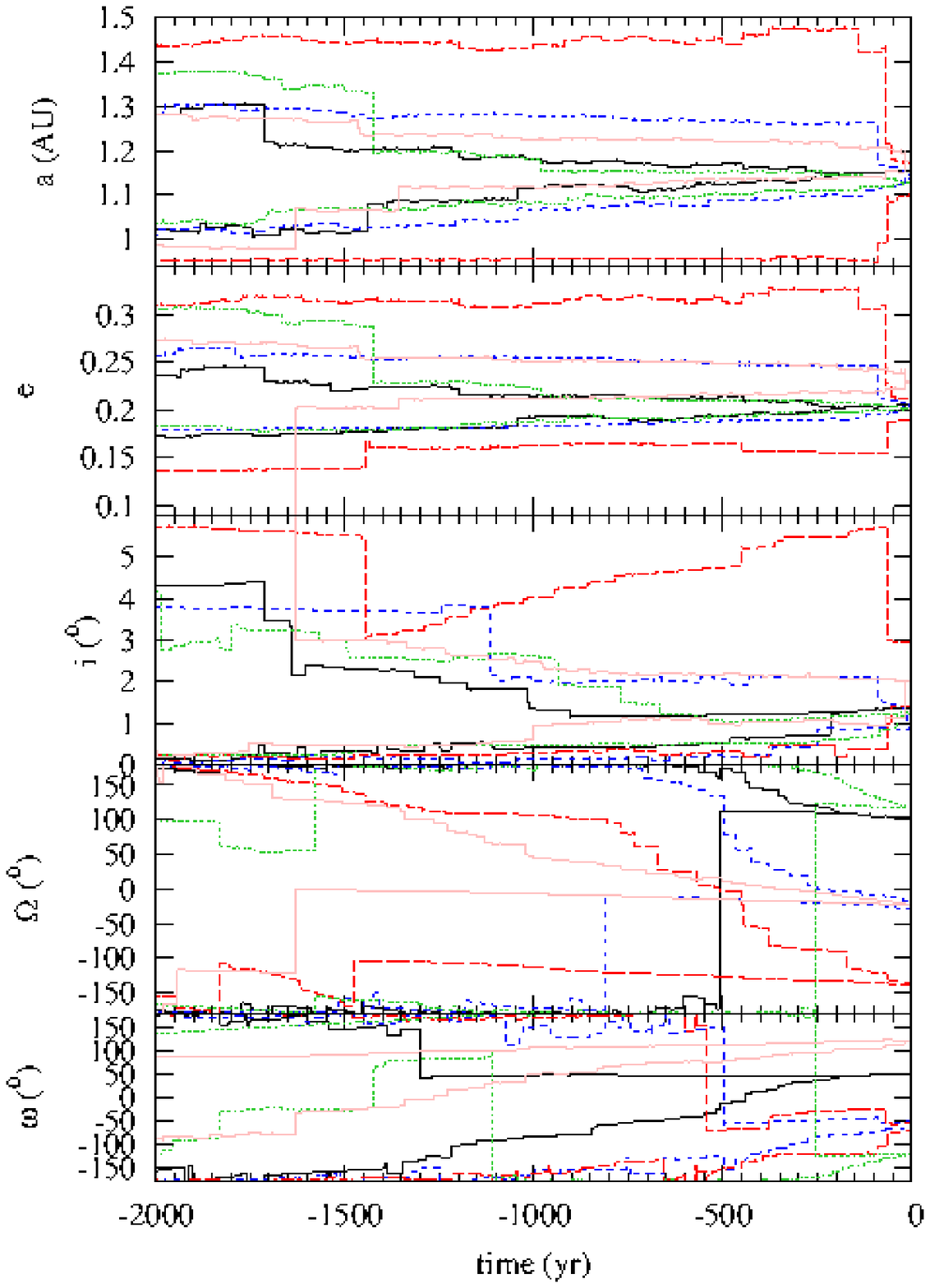}\\
        \caption{Time evolution of the orbital elements $a$, $e$, $i$, $\Omega$, and $\omega$ of 2011~JV$_{10}$ (red line), 2011~GJ$_{3}$
                 (blue line), 2012 DJ$_{54}$ (pink line), 2013~NJ$_{4}$ (green line), and 2014~AA (black line) as described by the solution 
                 displayed in Table \ref{ours3}. The left-hand panels show the average evolution of 100 control orbits, the right-hand 
                 panels show the ranges in the values of the parameters at the given time.
                }
        \label{14AAc}
     \end{figure*}
%
%

        Apollo asteroids 2011~GJ$_{3}$ (McMillan et al. 2011), 2011~JV$_{10}$ (Kowalski et al. 2011), and 2012 DJ$_{54}$ (Micheli et al. 
        2012) follow very similar orbits at present. The relative $D_{\rm SH}$, $D_{\rm LS}$, $D_{\rm D}$, and $D_{\rm R}$ of 2011~GJ$_{3}$ 
        and 2011~JV$_{10}$ are 0.35488, 0.01011, 0.13679, and 0.00119, respectively. In the case of 2011~GJ$_{3}$ and 2012 DJ$_{54}$ the 
        values are 0.435352, 0.0324687, 0.221674, and 0.0286270. Their orbits are rather uncertain but they might be related, with 
        2011~JV$_{10}$ and 2012 DJ$_{54}$ perhaps being relatively recent fragments of 2011~GJ$_{3}$. But having similar orbits at present 
        time is not enough to claim a relationship, dynamical or otherwise; a representative set of orbits must be integrated to show that 
        the dynamical evolution over a reasonable amount of time is also similar (see e.g. Porubcan et al. 2006; Jopek \& Williams 2013). 

        Figure \ref{14AAc} shows the short-term evolution of the orbital elements $a$, $e$, $i$, $\Omega$, and $\omega$ of 2011~GJ$_{3}$, 
        2011~JV$_{10}$, 2012 DJ$_{54}$, 2013~NJ$_{4}$, and 2014~AA. The left-hand panels show the average evolution of 100 control orbits 
        (see Sect. \ref{mccm} for additional details), the right-hand panels show the ranges in the values of the parameters at the given 
        time. The orbits of 2014~AA and 2011~GJ$_{3}$ are alike, their past orbital evolution also being quite similar (see Fig. 
        \ref{14AAc}). Not included in Table \ref{candidates} ---because it does not comply with the restriction $D_{\rm LS}$ and $D_{\rm R} 
        < 0.05$ with respect to 2014~AA--- is meteoroid 2013~NJ$_{4}$ (Wainscoat et al. 2013; $a$ = 1.13324 AU, $e$ = 0.20253, $i$ = 
        1{\fdg}31952, $\Omega$ = 115{\fdg}60660, $\omega$ = 238{\fdg}50930, $H$ = 27.40 mag, MOID = 0.0046 AU) that also has low relative 
        $D_{\rm SH}$, $D_{\rm LS}$, $D_{\rm D}$ and $D_{\rm R}$ with respect to 2011~GJ$_{3}$: 0.24901, 0.00953, 0.08515, and 0.02152. Given 
        the uncertainty of their current orbital solutions it cannot be discarded that 2011~GJ$_{3}$, 2011~JV$_{10}$, 2012 DJ$_{54}$, 
        2013~NJ$_{4}$, and 2014~AA (see Table \ref{family}) are the result of an asteroid break-up that took place perhaps 1,800 to 1,400 yr 
        ago (see Fig. \ref{14AAc}). 

\begin{figure}
        \centering
        \includegraphics[width=\linewidth]{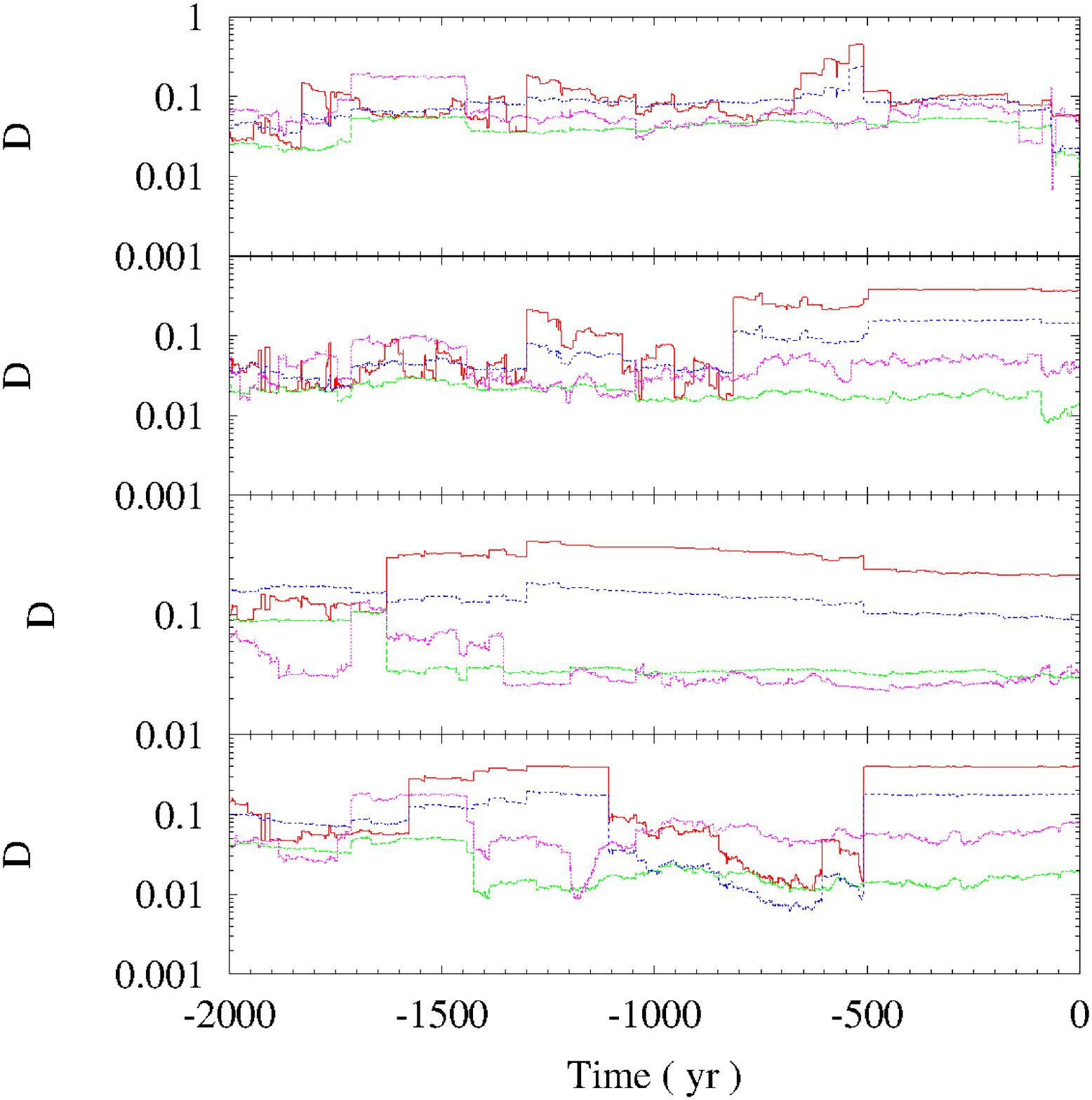}
         \caption{Average time evolution of the various $D$-criteria ---$D_{\rm SH}$ (red), $D_{\rm LS}$ (green), $D_{\rm D}$ (blue), and
                  $D_{\rm R}$ (pink)--- for 2011~JV$_{10}$ (top panel), 2011~GJ$_{3}$ (second to top panel), 2012 DJ$_{54}$ (third to top
                  panel), and 2013~NJ$_{4}$ (bottom panel) with respect to 2014~AA as described by the solution displayed in Table 
                  \ref{ours3}. The values have been computed using the data in Figure \ref{14AAc}, left-hand panels.
                 }
         \label{Ds}
      \end{figure}

        Figure \ref{Ds} shows the average time evolution of the various $D$-criteria for 2011~GJ$_{3}$, 2011~JV$_{10}$, 2012 DJ$_{54}$, and 
        2013~NJ$_{4}$ with respect to 2014~AA as described by the data in Fig. \ref{14AAc}, left-hand panels. This type of analysis is 
        customarily used to link meteors and NEOs (see e.g. Trigo-Rodr\'{\i}guez et al. 2007; Olech et al. 2015). From the figures, a 
        catastrophic disruption event around 1,600 yr ago cannot be ruled out. In terms of statistics, these objects are probable dynamical 
        relatives: the ranges of their orbital parameters, $a$, $e$, and $i$, fully overlap after less than 100 yr of backwards integration. 
        Schunov\'a et al. (2012) have shown that a robust statistical estimate of a dynamical relationship between objects part of the 
        near-Earth asteroid (NEA) population is only possible for groups of four and more objects although they could not find any 
        statistically significant group of dynamically related objects among those known at that time. All the objects in the candidate 
        group of dynamically related asteroids proposed here were discovered after the completion of the analysis in Schunov\'a et al. 
        (2012) that considered 7,563 NEOs (through 2011~DW). According to that analysis, this group of objects may be an asteroid cluster 
        and perhaps have a common origin as the values of their $D_{\rm SH}$ are $<0.060$ for most or all of them nearly 1,600 yr ago. 
        Although the available orbital solutions for these objects are rather poor (see Table \ref{family}), the evidence provided by Figs. 
        \ref{14AAc} and \ref{Ds} in favour of a common origin is certainly encouraging (but still far from conclusive). In any case, it must 
        be emphasized that false asteroid clusters may be identified as a result of significant orbital element uncertainties (Schunov\'a et 
        al. 2012). Here, we have tried to reduce this effect by using the average time evolution of the various $D$-criteria (see Fig. 
        \ref{Ds}).
%
%
%
%

\begin{figure*}
       \centering
        \includegraphics[width=\linewidth]{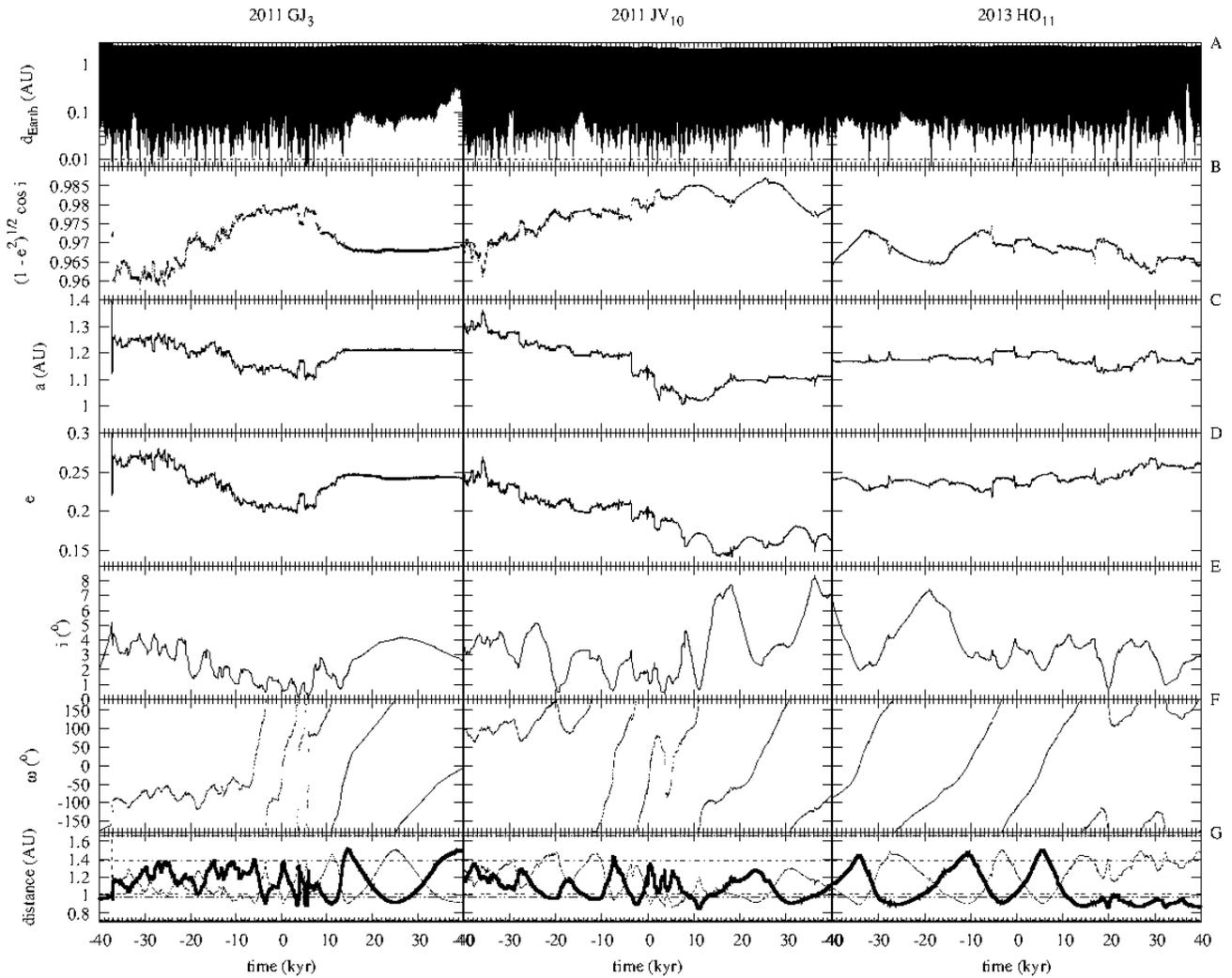}
        \caption{Time evolution of various parameters for the nominal orbital solutions of 2011~GJ$_{3}$, 2011~JV$_{10}$, and 2013 HO$_{11}$. 
                 The distance from the Earth (panel A); the value of the Hill sphere radius of the Earth, 0.0098 AU, is displayed. The 
                 parameter $\sqrt{1 - e^2} \cos i$ (panel B). The orbital elements $a$ (panel C), $e$ (panel D), $i$ (panel E) and $\omega$ 
                 (panel F). The distance to the descending (thick line) and ascending nodes (dotted line) is in panel G. Planetary 
                 perihelion and aphelion distances (Earth and Mars) are also shown. These integrations are referred to the JD 2456600.5 
                 epoch.
                }
        \label{like}
     \end{figure*}

        Asteroid 2011~JV$_{10}$ reached perigee on 2011 May 5 at a geocentric distance of 0.0023 AU. In spite of its small size ($H$ = 29.7 
        mag or $\sim$7 m) it attracted considerable attention because it became an obvious example of the Red Baron dynamical scenario 
        (Adamo 2011) in which a small body approaches the Earth from out of the Sun's glare, as the parent body of the Chelyabinsk 
        superbolide did (Popova et al. 2013). Red Baron scenario events are rather frequent and objects moving in 2011~JV$_{10}$-like orbits 
        appear to be prone to them. In addition to their Kozai-like dynamics (compare Figs. \ref{14AA} and \ref{like}), this group of 
        objects share a number of peculiar dynamical features. They reach perigee within one or two months of reaching perihelion. Earth 
        approaches can occur before or after perihelion. When they occur after perihelion, these objects approach the Earth from its day 
        side. In this configuration, the incoming object cannot be discovered until after perigee as it was the case of 2011~JV$_{10}$. Even 
        if they approach from the night side, when the encounter takes place at the ascending node (as in the case of 2014~AA) the object 
        will move south from the Ecliptic. That area of the sky receives less attention than the northern one because there are less 
        telescopes (and less land masses) south from the equator. After the encounter, they become part of the day-time sky and, therefore, 
        no longer observable from the ground. This suggests that these objects are relatively difficult to detect if they are all as small 
        as 2011~GJ$_{3}$ (17--38 m) or smaller; their windows of optimal visibility would be too short ---perhaps no more than a few days. 
        This is consistent with the fact that most of them have very short arcs, eight of them have arcs shorter than a month. Asteroid 
        2011~GJ$_{3}$ may have a size similar to that of the Chelyabinsk impactor. 
%
%
%
%

        Among the objects included in Table \ref{candidates}, the largest is 2013~HO$_{11}$ (Ries et al. 2013) with an absolute magnitude of 
        23.0 (diameter 75--169 m) and a MOID of 0.006 AU. Its orbit is also one of the most statistically robust. Figure \ref{like} shows 
        the orbital evolution of 2011~GJ$_{3}$, 2011~JV$_{10}$, and 2013 HO$_{11}$. As 2014~AA did, the three objects may remain within the 
        immediate neighbourhood of our planet for dozens of thousands of years. Asteroids 2011~GJ$_{3}$ and 2011~JV$_{10}$ exhibit 
        Kozai-like dynamics; 2013 HO$_{11}$ will experience this behaviour in the future but now its argument of perihelion circulates. This 
        object is currently outside the web of overlapping secular resonances pointed out above. From a strictly dynamical standpoint, 
        2013~HO$_{11}$ and most of the objects in Table \ref{candidates} are not like the five objects 2011~GJ$_{3}$, 2011~JV$_{10}$, 2012 
        DJ$_{54}$, 2013~NJ$_{4}$, or 2014~AA; they are not subjected to Kozai resonances now but they may have been Kozai resonators in the 
        past or become ones in the future. Taking into account that the orbits of the five objects are poorly constrained (perhaps with the 
        exception of 2014~AA), it is not possible to confirm a putative common origin for these objects, but it cannot be ruled out either. 

        At this point one may argue that, within a large sample of minor bodies, it is always possible to identify groups of a few objects 
        with values of the various $D$-criteria as low as the threshold used here and this has no real dynamical implications. Therefore, 
        what is the statistical significance of our findings, if any? In other words, given an orbital solution like the one displayed in 
        Table \ref{ours3}, how high is the theoretical likelihood of finding one or more objects with $D_{\rm LS}$ and $D_{\rm R} < 0.05$? 
        The NEOSSat-1.0 orbital model (Greenstreet et al. 2012) is widely regarded as one of the best models available to describe the 
        orbital distribution of the NEO population. Synthetic data from this model do not contain any physically related objects, but they 
        may include dynamically related virtual objects as the model is the result of extensive numerical integrations. Within the synthetic 
        data, groups of objects following similar dynamical pathways could still be found because the integrations can reproduce the web of 
        overlapping resonances that permeates the region. In order to check the statistical significance of our findings, we have used the 
        codes described in Greenstreet et al. (2012)\footnote{http://www.phas.ubc.ca/$\sim$sarahg/n1model/} with the same standard input 
        parameters to generate sets of orbital elements of about 15,000 virtual objects (the NEOs currently known amount to 14,759 objects). 
        These datasets have been processed using the same algorithm applied above to real data in order to single out the objects in Table 
        \ref{candidates}. 

        Let us assume that the size and orbital elements of asteroids are uncorrelated. Ignoring the values of the absolute magnitude 
        (NEOSSat-1.0 was originally developed for NEOs with $H<18$~mag) and only performing the processing in terms of orbits, the search 
        produced no results. We had to double the value of the threshold to obtain an average of one virtual object. Although not fully 
        conclusive, this experiment suggests that our findings may be somewhat robust and that objects in this group could be truly 
        (dynamically and/or genetically) related. However, genetically related asteroids can only be confirmed spectroscopically and none of 
        these objects have been observed spectroscopically yet. In this context, it is rather surprising that if we apply the same approach 
        to 2008~TC$_{3}$ a few compatible virtual objects can be readily found. Table \ref{candidates} includes 11 objects with $D_{\rm LS}$ 
        and $D_{\rm R} < 0.05$ with respect to the orbital solution in Table \ref{ours3}; a similar analysis for 2008~TC$_{3}$ using real 
        data produces 24 candidates. However, using larger synthetic datasets, the results suggest that objects moving in 2008~TC$_{3}$-like 
        orbits are nearly three times more likely to exist than those following 2014~AA-like orbits. Not accounting for observational 
        biases, this may be tentatively interpreted as pointing to an intrinsically different origin for 2008~TC$_{3}$ and 2014~AA.

        Schunov\'a et al. (2014) have studied the expected dynamical signature of a catastrophic asteroidal disruption during a very close 
        Earth approach. These authors have found that the minimum size of a progenitor capable of producing an observable NEO family (a 
        group of objects genetically, not just dynamically, related) is about 350 m in diameter. Such asteroid family would be observable 
        for about 2000 yr and include a few million fragments with sizes in the 1--10 m range. Schunov\'a et al. (2014) have also found that 
        formation of tidally-disrupted NEO families is enhanced for objects following orbits with 0.5 AU $<a<$ 1.25 AU, $e<$0.5, and 
        $i<$5\degr which is the case of the group of minor bodies discussed here. If the objects moving in 2014~AA-like orbits are 
        physically related, i.e. they are the result of a catastrophic disruption event, many more should be detected over the next few 
        decades. They could be intrinsically difficult to detect though (see above). However, if this group of objects is just a resonant 
        family (or even a random grouping) without a common physical origin, the number of objects moving in such orbits could be relatively 
        small. Genetic asteroid families have been known for about a century (see e.g. Hirayama 1918; Zappal\`a et al 1990, 1994); dynamical 
        asteroid families or dynamical groups are asteroids temporarily trapped in a mean motion or secular resonance (or a web of 
        overlapping ones). Genetic families may exist within dynamical groups, like in the case of the Hildas (Bro\v{z} \& Vokrouhlick\'y 
        2008). 

        It may be argued that any conclusions drawn from a set of orbital solutions based on few observations and short arcs are nothing but 
        mere speculation. However, we are dealing here with objects that are very small and because of this can only be discovered and 
        recovered when they pass very close to our planet. Their observational windows are therefore exceptionally short and in some cases
        well spaced due to the influence of secular resonances (Kozai-like behaviour). If the existence of objects moving in similar orbits
        is systematically neglected based on their relatively poor orbital solutions, their orbits are not going to be improved in the 
        future simply because they are not going to attract any attention. In the present case, we have a meteoroid that actually hit the 
        Earth and a few comparably small bodies that move in rather similar orbits with small MOIDs. Asteroid 2012 DJ$_{54}$ is included in 
        the list of potential future Earth impact events compiled by the JPL Sentry System with an impact probability of 
        2.0$\times10^{-4}$;\footnote{\url{http://neo.jpl.nasa.gov/risk/2012dj54.html}} asteroid 2009 SH$_{1}$ is also included with an 
        impact probability of 7.0$\times10^{-8}$.\footnote{\url{http://neo.jpl.nasa.gov/risk/2009sh1.html}} These facts clearly deserve 
        further attention as the data may hint at a relatively recent asteroid break-up. In addition, the past orbital evolution of these 
        objects strongly suggest that they have remained in the neighbourhood of the Earth--Moon system and away from the main asteroid belt 
        for many thousands of years. If the objects have a common origin, the fragmentation episode that created them may have happened 
        relatively recently in astronomical terms (see Fig. \ref{14AAc}). This means that catastrophic disruption events, perhaps due to 
        rotational disruptions (Denneau et al. 2015), may not only be taking place in the main belt but also among the closest NEOs. The 
        possible production of meteoroids within the immediate neighbourhood of our planet has an obvious and direct effect on the 
        evaluation of the overall asteroid impact hazard (e.g. Schunov\'a et al. 2014). Fortunately, although this is of considerable 
        theoretical interest, most of these fragments are small enough to be of less concern in practice.

        It may also be argued that ignoring the Yarkovsky and YORP effects for these objects may seriously hamper our qualitative 
        understanding of their dynamics; however, the integrations completed here hint at semi-major axis drifts $<$ 0.3 AU over time-scales 
        of dozens of kyr. The largest predicted Yarkovsky drift rates are $\sim$10$^{-7}$ AU yr$^{-1}$ (see e.g. Farnocchia et al. 2013); a 
        simple estimate shows that, in order to produce a semi-major axis drift comparable to those observed in Figs. \ref{14AA} or 
        \ref{like} at the largest Yarkovsky drift rate, several Myr are required. The uncertainties in the orbital parameters of these 
        objects grow only moderately after 1 kyr or so (see Fig. \ref{14AAc}, right-hand panels), therefore the previous discussion on their 
        long-term orbital evolution as well as the comparisons made above are likely valid.

     \subsection{Average short-term orbital evolution: MCCM} \label{mccm}
        Figure \ref{14AAc} shows the short-term evolution of the orbital elements $a$, $e$, $i$, $\Omega$, and $\omega$ of the objects 
        studied here, including the probable orbit of 2014~AA in Table \ref{ours3}. In the figures, we show the average results of the 
        evolution of 100 control orbits and their ranges (minimum and maximum) in the values of the parameters at a given time. The initial 
        orbital elements of each control orbit have been computed varying them randomly, within the ranges defined by their mean values and 
        standard deviations. As pointed out above, this is equivalent to considering a number of different virtual minor planets moving in 
        similar orbits. This approach is reasonable if the orbital solution is the result of stochastic simulations, but it is arguable if 
        that solution is associated with a set of observations obtained for a single minor planet. In this case, the fact that the elements 
        affect each other cannot be neglected and the covariance matrix should be applied. 

\begin{figure}
        \centering
        \includegraphics[width=\linewidth]{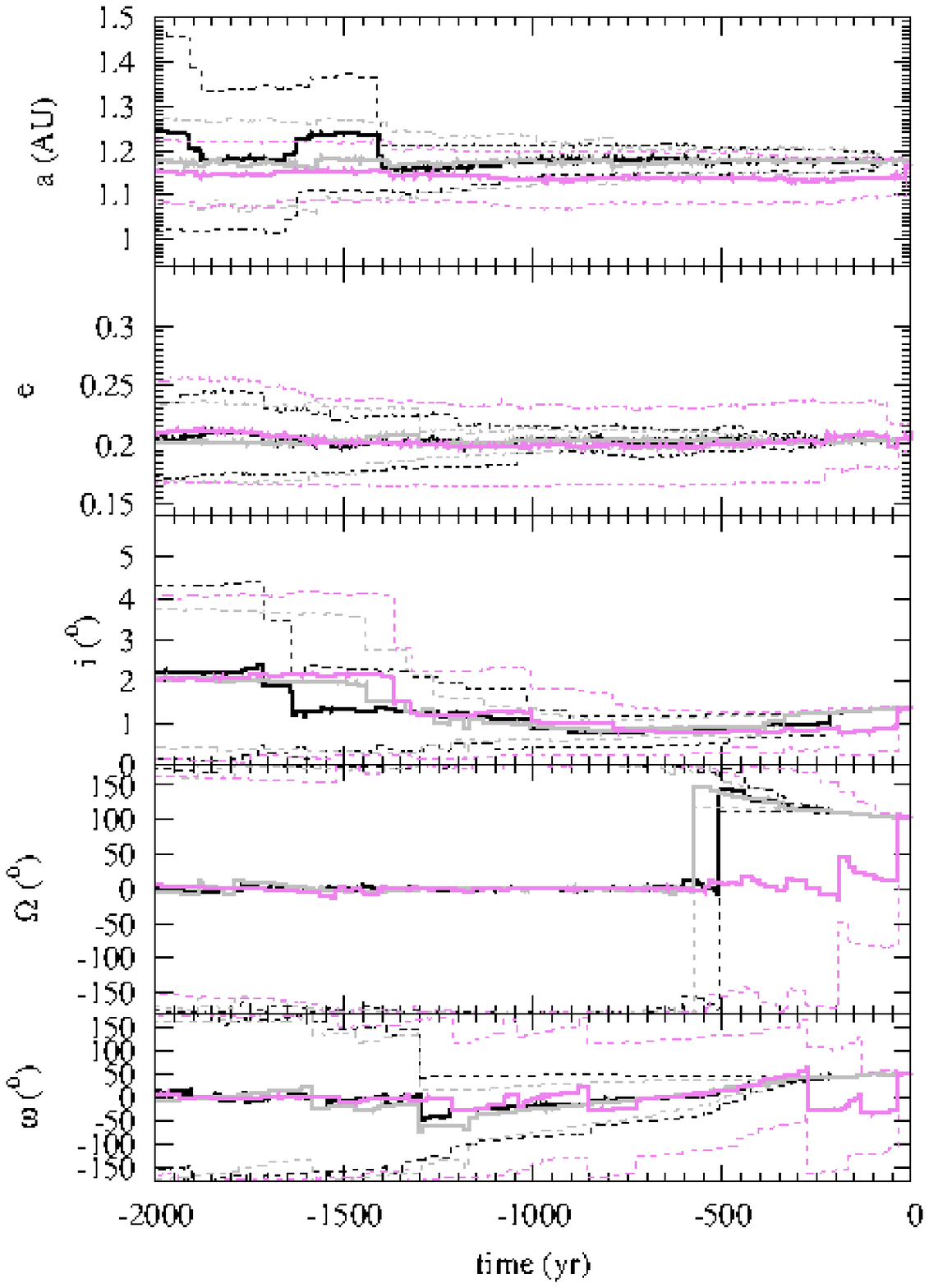}
         \caption{Time evolution of the orbital elements $a$, $e$, $i$, $\Omega$, and $\omega$ of 2014~AA. In black, we plot data derived 
                  from the orbit in Table \ref{ours3} (same data as in Fig. \ref{14AAc}), in grey we show the results based on 
                  MCCM for the same orbital solution (see the text for details), and in violet we display the results based on MCCM for 
                  the orbit available from the JPL Small-Body Database (see the text for details, Farnocchia et al. 2016). In this figure 
                  both average values and their ranges are plotted.
                 }
         \label{disper14AA}
      \end{figure}

\begin{figure}
        \centering
        \includegraphics[width=\linewidth]{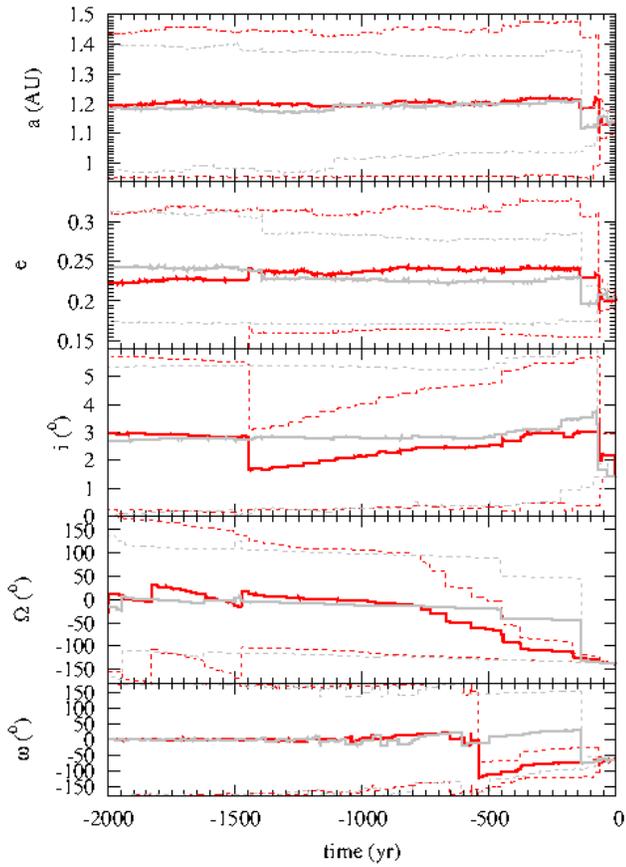}
         \caption{Time evolution of the orbital elements $a$, $e$, $i$, $\Omega$, and $\omega$ of 2011~JV$_{10}$. In red, we replot the
                  data in Fig. \ref{14AAc}, in grey we show the results based on MCCM (see the text for details). In this figure both
                  average values and their ranges are plotted.
                 }
         \label{disper11JV10}
      \end{figure}

\begin{figure}
        \centering
        \includegraphics[width=\linewidth]{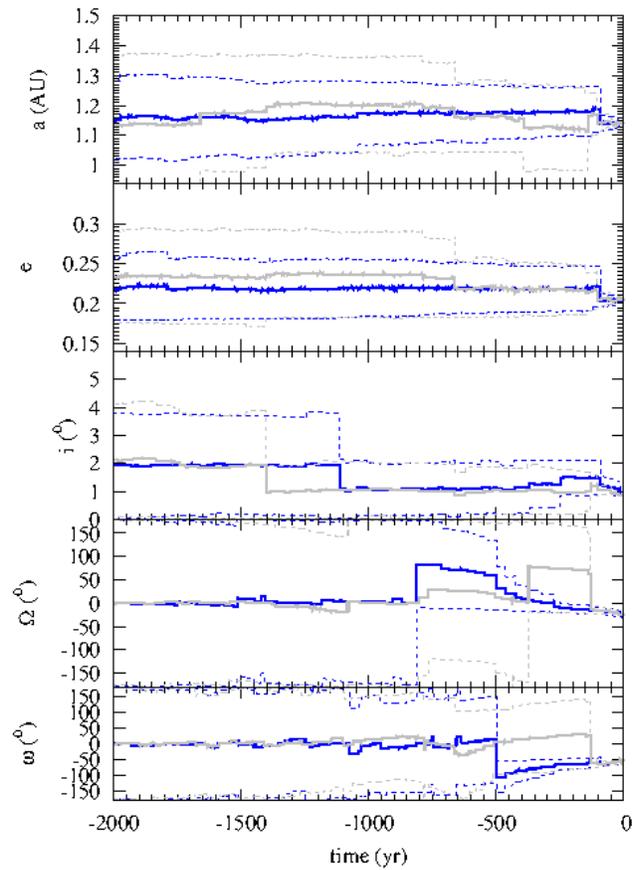}
         \caption{Same as Fig. \ref{disper11JV10} but for 2011~GJ$_{3}$.
                 }
         \label{disper11GJ3}
      \end{figure}

\begin{figure}
        \centering
        \includegraphics[width=\linewidth]{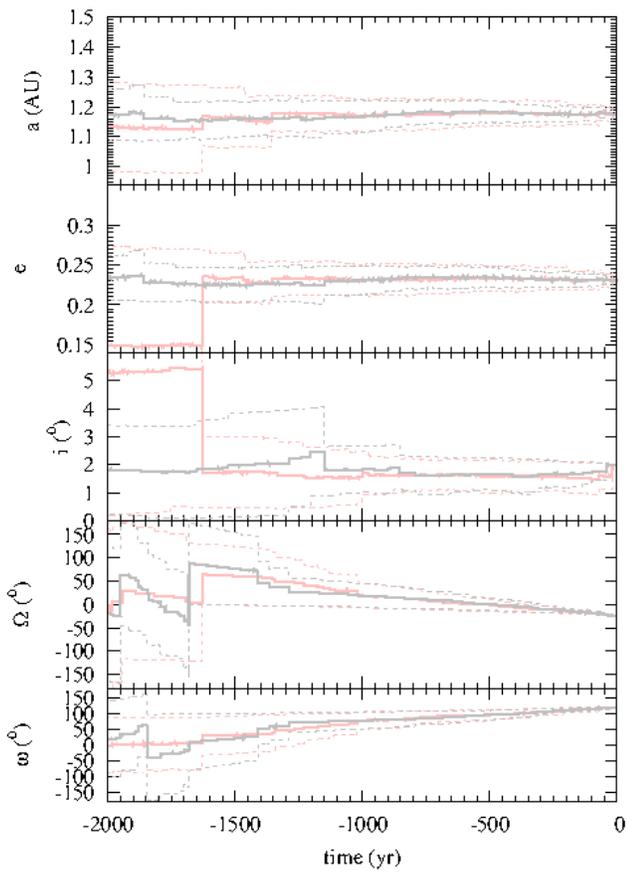}
         \caption{Same as Fig. \ref{disper11JV10} but for 2012 DJ$_{54}$.
                 }
         \label{disper12DJ54}
      \end{figure}

\begin{figure}
        \centering
        \includegraphics[width=\linewidth]{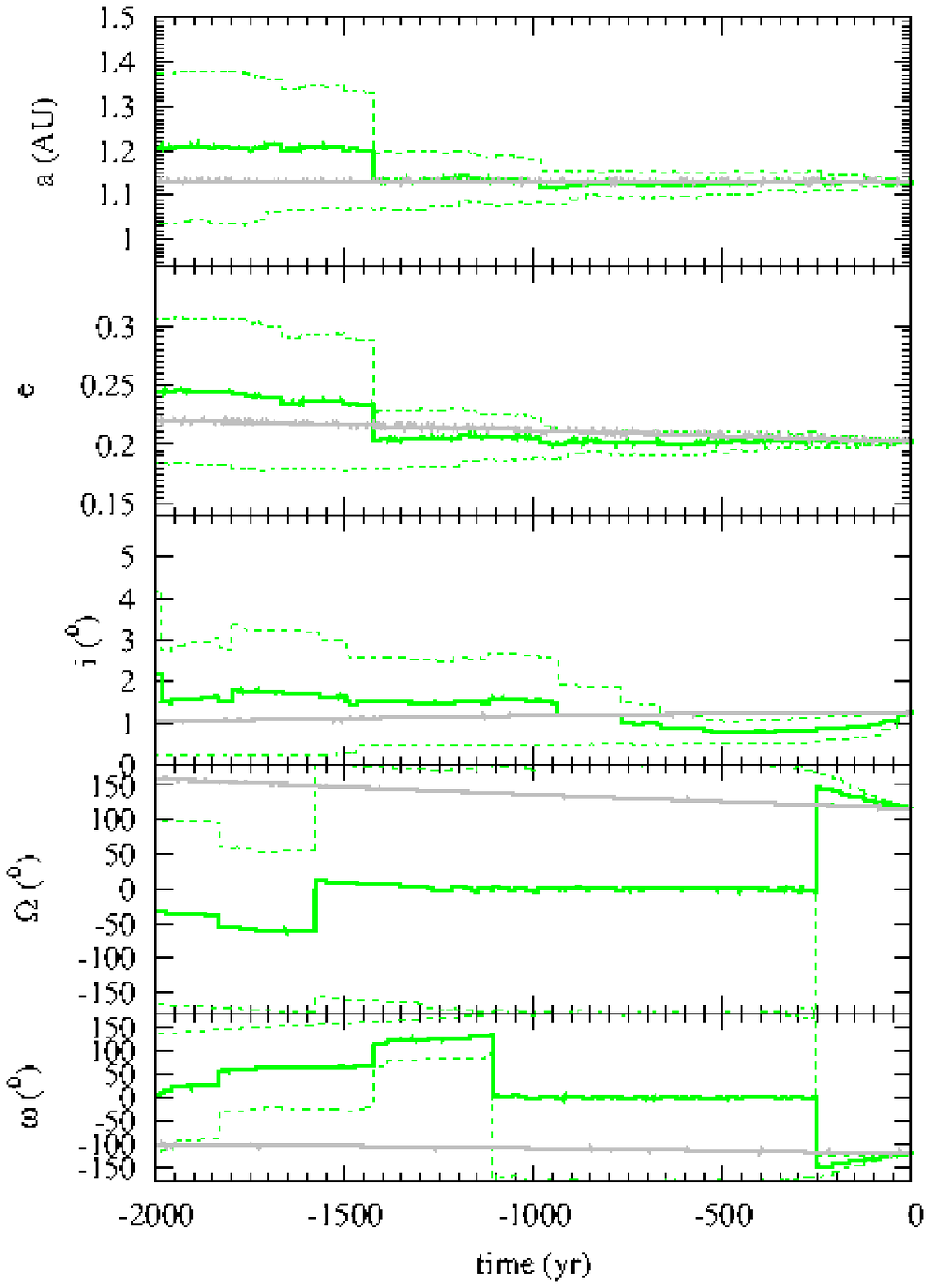}
         \caption{Same as Fig. \ref{disper11JV10} but for 2013~NJ$_{4}$.
                 }
         \label{disper13NJ4}
      \end{figure}

        As a consistency test, we have used an implementation of the classical Monte Carlo using the Covariance Matrix (MCCM; Bordovitsyna 
        et al. 2001; Avdyushev \& Banschikova 2007) approach to recompute the past orbital evolution of these objects generating control 
        orbits with initial parameters from the nominal orbit adding random noise on each initial orbital element making use of the 
        covariance matrix (for details, see de la Fuente Marcos \& de la Fuente Marcos 2015b). The covariance matrix of the orbit in Table 
        \ref{ours3} has been computed as described in e.g. Press et al. (2007) using a sample of 19 best solutions in terms of astrometry as 
        described above. The other covariance matrices have been retrieved from the JPL Small-Body Database. Our results are given in Figs. 
        \ref{disper14AA}--\ref{disper13NJ4} and they show that, in general, the difference is not very significant. Figure \ref{disper14AA} 
        compares the evolution of the solution in Table \ref{ours3} (both in terms of standard deviations, black curves, and covariance 
        matrix, grey curves) with the one available from the JPL Small-Body Database (Farnocchia et al. 2016) derived from the covariance 
        matrix available from the JPL Small-Body Database. These solutions are based on different values of the impact parameters, but the 
        short-term evolution of their orbits is fairly similar. Therefore, the detailed discussion made above on the past short-term 
        evolution of 2014~AA as described by our favoured orbital solution applies to the one derived in Farnocchia et al. (2016) as well. 
        For most of the orbits studied here, the average orbital evolution of samples obtained from the standard deviations matches well the 
        one derived for samples generated using the covariance matrix. The most dramatic difference is found for 2013~NJ$_{4}$. Our 
        calculations appear to have uncovered an unexpected stable island in the surveyed volume of the orbital parameter space. Objects 
        moving inside that region are largely unperturbed, with resonances cancelling each other out (see Fig. \ref{stable}). 
%
%
%
%
%
%
%
%
%
%
%
%
%
%
%
%
%
%
%
%
     \subsection{A stable island in a sea of PHAs}
\begin{figure}
       \centering
        \includegraphics[width=\linewidth]{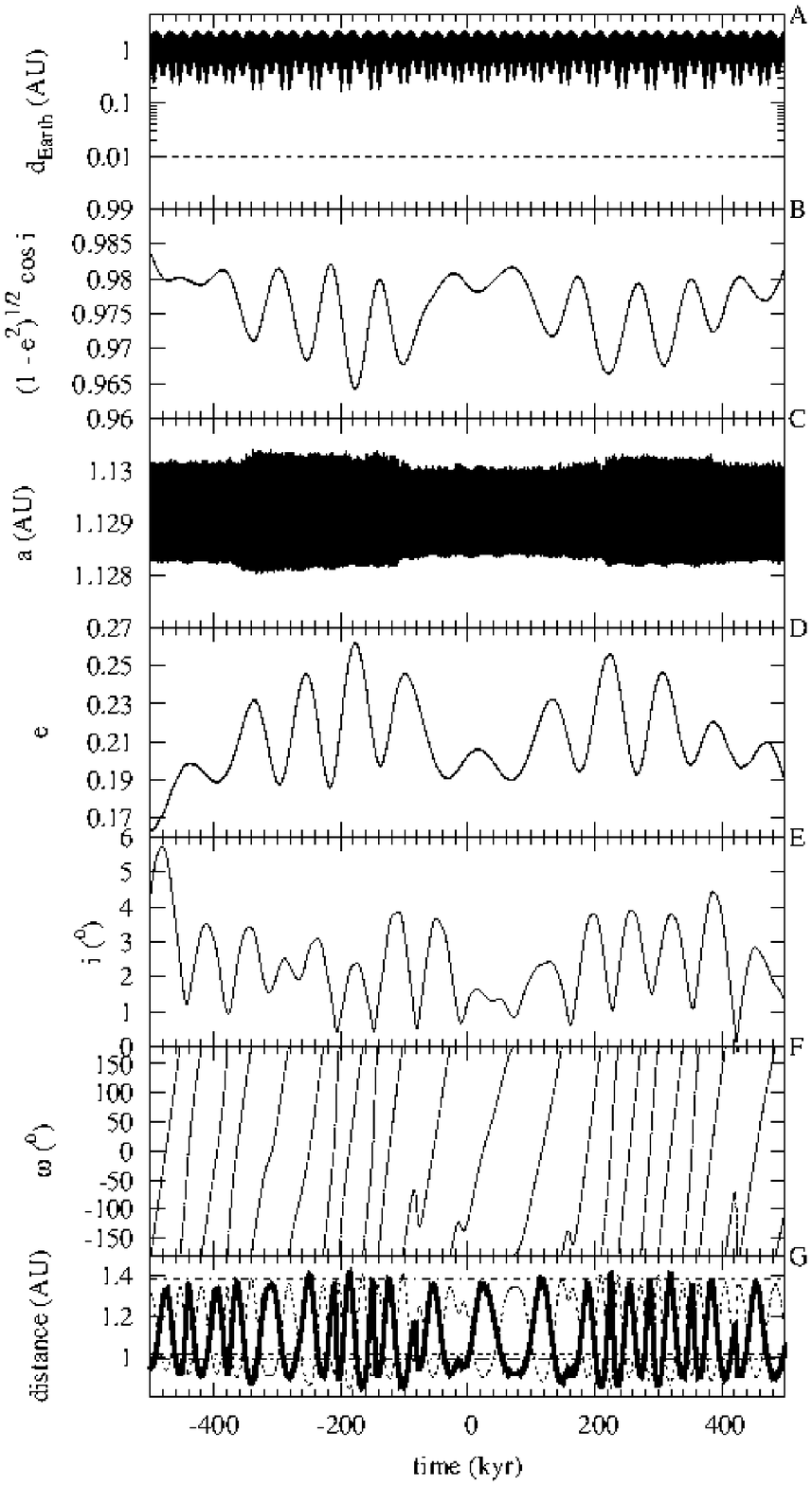}
        \caption{Same as Fig. \ref{14AA} but for a representative orbit belonging to the stable island described in the text. 
                }
        \label{stable}
     \end{figure}

        The previous section compares the short-term orbital evolution of the objects studied above using initial conditions derived with 
        and without the application of the covariance matrix. Surprisingly, the evolution of meteoroid 2013~NJ$_{4}$ as computed using 
        control orbits derived from the covariance matrix exhibits Trojan-like stability. The standard deviations associated with the 
        osculating orbital elements for this object are unusually small for an orbit with an Earth MOID of just 0.0046 AU that is comparable 
        to those of Potentially Hazardous Asteroids (PHAs). The relatively long-term evolution ($\pm$500 kyr) of a representative instance 
        of such an orbit is displayed in Fig. \ref{stable}. The presence of this stable island within their orbital subdomain is another 
        argument in favour of singling out this group of objects among the general NEO population. A dynamically stable island may act as a 
        long-term source of drifting small bodies. 
%
%
%
%

  \section{Conclusions}
     The aim of this work was to perform an independent determination of the pre-impact orbit of meteoroid 2014~AA using an improved set of
     impact parameters for the airburst event. This has been accomplished by applying two different techniques: geometric Monte Carlo and 
     $N$-body calculations. Our results are consistent with those obtained by other authors using other techniques. The results of a search 
     for minor bodies moving in similar orbits among already known objects and subsequent $N$-body simulations suggest that 2014~AA might 
     have formed during a relatively recent asteroid break-up. If this somewhat speculative interpretation is correct, 2014~AA would have 
     been a fragment of a parent body and a (probably large) group of meteoroids of similar composition moving in trajectories analogous to 
     that of 2014~AA might exist. Asteroid 2014~AA was comparable in size to 2008~TC$_{3}$, the single other example of an impacting object 
     observed prior to atmospheric entry. However, their pre-impact orbital evolutions were rather different. The dynamical evolution of 
     2014~AA and related objects is also not similar to that of the parent body of the Crete bolide (2002 June 6, Brown et al. 2002) 
     observed over the Mediterranean Sea or the recent Chelyabinsk Event (see e.g. de la Fuente Marcos \& de la Fuente Marcos 2013, 2014; de 
     la Fuente Marcos et al. 2015). Our conclusions can be summarized as follows.
     \begin{itemize}
        \item The values of the impact parameters of meteoroid 2014~AA as derived from infrasound data are: ($\lambda_{\rm impact}$, 
              $\phi_{\rm impact}$, $t_{\rm impact}$) = (-43{\fdg}7$\pm$1{\fdg}7, +11{\fdg}2$\pm$2{\fdg}8, 2456659.618$\pm$0.011 JD UTC). 
              These values are consistent with the available astrometry.
        \item In the decades preceding its impact, 2014~AA followed an orbit ($a$~=~1.1623 AU, $e$~=~0.2116, $i$~=~1$\fdg$4156, 
              $\Omega$~=~101$\fdg$6086, and $\omega$~=~52$\fdg$3393) with perihelion ($q$ = 0.9164 AU) inside the orbit of the Earth and 
              aphelion ($Q$ = 1.4083 AU) beyond Mars' perihelion. These values are equally consistent with the available astrometry.
        \item Meteoroid 2014~AA was subjected to a Kozai resonance prior to its collision with the Earth and it may have remained in the
              dynamical neighbourhood of our planet for many thousands of years; however, it may also be a relatively recent fragment 
              spawned from another NEA.
        \item A search for objects moving in orbits similar to that of 2014~AA gives several tentative candidates. All these objects have 
              remained in the neighbourhood of our planet for thousands of years and some of them, 2014~AA included, could be fragments from 
              a recent break-up.
        \item Our analysis of the past orbital evolution of 2014~AA and related objects suggests that asteroidal disruption events might not 
              only be taking place in the main belt but also among the closest NEOs. If confirmed, this finding would imply that the 
              asteroid impact hazard associated with bodies small enough to be of less concern could be higher than commonly thought.
     \end{itemize}

  \acknowledgments
     The authors thank the referee, T. J. Jopek, for his constructive, detailed and very helpful reports, S. J. Aarseth for providing one of 
     the codes used in this research and for comments on early versions of this work, S. R. Chesley for providing his results on the 
     pre-impact orbit of 2014~AA prior to publication, D. Farnocchia for his input on early versions of this work, J. D. Giorgini for 
     providing the details of the orbit computed by the JPL, Bill Gray for sharing the early results of his impact calculations, and S. R. 
     Proud for sharing the results of his analysis of some weather satellite imagery. This work was partially supported by the Spanish 
     `Comunidad de Madrid' under grant CAM S2009/ESP-1496. Some of the calculations discussed in this paper were completed on the `Servidor 
     Central de C\'alculo' of the Universidad Complutense de Madrid. This research has made use of NASA's Astrophysics Data System, the 
     ASTRO-PH e-print server, the MPC data server, and the NEODyS information service.

  \newpage
  \appendix
  \section{Orbital elements of the Earth around the time of impact}
     Chesley et al. (2015) have released the impact time and the hypocentre location for the 2014~AA impact (see their table 1) as included 
     in the REB of the IDC of the CTBTO for 2014 January 2. The impact time was 2014 January 2 at 3:05:25 UTC with an uncertainty of 632 s. 
     The impact location coordinates were latitude ({\degr}N) equal to +14\fdg6326 and longitude ({\degr}E) of $-$43\fdg4194. Therefore, the 
     actual impact with the atmosphere took place at epoch 2456659.629537 Julian Date, Barycentric Dynamical Time. The uncertainty is about 
     10 minutes. The osculating orbital elements of the Earth within $\pm$150 s of the detection are given in Table \ref{Earth}. These 
     values have been computed by the SSDG, \textsc{Horizons} On-Line Ephemeris System.  
%
%
     \begin{landscape}
     \begin{table}
      \centering
      \fontsize{8}{11pt}\selectfont
      \tabcolsep 0.06truecm
      \caption{Orbital elements of the Earth around JD 2456659.629537 = A.D. 2014-Jan-02 03:06:32.00 TDB (Source: JPL \textsc{Horizons} 
               system). Data as of 2016 April 12.}
      \begin{tabular}{cccccccc}
       \hline
          Epoch JD TDB      & TDB        &
          $a$ (AU)          & $e$                 & $i$ (\degr)          & $\Omega$ (\degr)  & $\omega$ (\degr)  & $f$ (\degr)       \\
       \hline
          2456659.627777778 & 03:04:00.0 &
          1.000972380029170 & 0.01761703261345213 & 0.001880262202725939 & 196.1937930138271 & 266.8968539832863 & 358.3453037378698 \\
          2456659.628472222 & 03:05:00.0 &
          1.000972354161487 & 0.01761700936572361 & 0.001880231304943617 & 196.1824664060930 & 266.9084313181994 & 358.3457610767323 \\
          2456659.629166667 & 03:06:00.0 &
          1.000972328266890 & 0.01761698609092967 & 0.001880200483856643 & 196.1711389039578 & 266.9200095419956 & 358.3462184211128 \\
          2456659.629537037 & 03:06:32.0 &
          1.000972314445433 & 0.01761697366663995 & 0.001880184077304693 & 196.1650972042493 & 266.9261849576632 & 358.3464623403758 \\
          2456659.629861111 & 03:07:00.0 &
          1.000972302345379 & 0.01761696278907163 & 0.001880169739474883 & 196.1598105087923 & 266.9315886532928 & 358.3466757710223 \\
          2456659.630555556 & 03:08:00.0 &
          1.000972276396955 & 0.01761693946014976 & 0.001880139071808320 & 196.1484812220189 & 266.9431686506603 & 358.3471331264694 \\
          2456659.631250000 & 03:09:00.0 &
          1.000972250421621 & 0.01761691610416575 & 0.001880108480869461 & 196.1371510449252 & 266.9547495327997 & 358.3475904874647 \\
       \hline
      \end{tabular}
      \label{Earth}
     \end{table}
     \end{landscape}
%
%

  \section{Cartesian state vectors at epoch JD TDB 2456658.628472222 = A.D. 2014-Jan-1 03:05:00.0000 TDB}
     In order to facilitate verification of our results by other astrodynamicists, we show in Table \ref{Cartesian} the Cartesian state 
     vectors of the physical model used in all the calculations presented here. These values have been computed by the SSDG, 
     \textsc{Horizons} On-Line Ephemeris System at epoch JD TDB 2456658.628472222 = A.D. 2014-Jan-01 03:05:00.0000 TDB, this instant is 
     considered as $t = 0$ across this work unless explicitly stated. Positions and velocities are referred to the barycentre of the Solar 
     System. 
%
%
     \begin{landscape}
     \begin{table}
      \centering
      \fontsize{8}{10pt}\selectfont
      \tabcolsep 0.05truecm
      \caption{Cartesian state vectors at epoch JD TDB 2456658.628472222 that corresponds to 03:05:00.0000 TDB on 2014 January 1 (Source: 
               JPL \textsc{Horizons} system, data as of 2016 April 12). The sample Cartesian vector for 2014~AA corresponds to the nominal 
               orbit in Table \ref{ours3}.}
      \begin{tabular}{llllllll}
       \hline
          Body              & Mass (kg)    &
          $X$ (AU)          & $Y$ (AU)            & $Z$ (AU)             & $V_{\rm X}$ (AU/day)  & $V_{\rm Y}$ (AU/day)  & $V_{\rm Z}$ (AU/day)       \\
       \hline
          Sun                     & 1.988544E+30 &
           9.876557315045510E-04 & -2.277483052076684E-03 & -9.309130493714816E-05 &  6.061452466600682E-06 &  2.330421391102359E-06 & -1.404175947545831E-07 \\
          Mercury                 & 3.302E+23    &
           1.234712512947228E-01 & -4.353820916108924E-01 & -4.671846192551433E-02 &  2.143843797991616E-02 &  9.091989724464501E-03 & -1.223863011823374E-03 \\
          Venus                   & 48.685E+23   &
          -5.149511742679028E-02 &  7.151279710261775E-01 &  1.276613490638690E-02 & -2.023520017529031E-02 & -1.582388301347185E-03 &  1.146311987350358E-03 \\
          Earth                   & 5.97219E+24  &
          -1.768146919670241E-01 &  9.648703577630898E-01 & -1.251184107370999E-04 & -1.720245307304753E-02 & -3.173746202131230E-03 & -3.541894002146146E-07 \\
          Moon                    & 734.9E+20    & 
          -1.765843263680927E-01 &  9.624988926026269E-01 &  5.765116799479567E-05 & -1.657120668505812E-02 & -3.100270805332192E-03 &  2.659631061553224E-05 \\
          Mars                    & 6.4185E+23   & 
          -1.512135183594610E+00 &  6.930582722833578E-01 &  5.161676323389820E-02 & -5.312436625323014E-03 & -1.151792918278832E-02 & -1.109714131381030E-04 \\
          (1) Ceres               & 9.393E+20    &
          -2.532553587303031E+00 &  2.006361724318938E-02 &  4.677363530307643E-01 & -4.977519833731861E-04 & -1.106269591780204E-02 & -2.549781096676565E-04 \\
          (2) Pallas              & 2.108E+20    &
          -1.446164856316815E+00 &  1.343011887296409E+00 & -8.086050115131924E-01 & -9.795008211501307E-03 & -6.689790295651018E-03 &  5.443421539610216E-03 \\
          (4) Vesta               & 2.59076E+20  &
          -2.280182334607567E+00 &  2.789898573544310E-01 &  2.689336672805309E-01 & -3.557060360882084E-04 & -1.140592407389319E-02 &  3.860042747257529E-04 \\
          Jupiter                 & 1898.13E+24  & 
          -1.330758770698398E+00 &  5.016245845201770E+00 &  8.864460103589159E-03 & -7.385422526176123E-03 & -1.576169860785649E-03 &  1.718627296846660E-04 \\
          Saturn                  & 5.68319E+26  &
          -6.883730900143171E+00 & -7.078257667923089E+00 &  3.970324150311733E-01 &  3.695935133299771E-03 & -3.904438907150792E-03 & -7.891989877674008E-05 \\
          Uranus                  & 86.8103E+24  &
           1.964589537999795E+01 &  3.920793902887398E+00 & -2.399582673853207E-01 & -7.984941344501358E-04 &  3.673711191904133E-03 &  2.400439209208154E-05 \\
          Neptune                 & 102.41E+24   &
           2.706460640088594E+01 & -1.289376969420147E+01 & -3.582088655530658E-01 &  1.328799214346798E-03 &  2.852447990866922E-03 & -8.914679327416514E-05 \\
          Pluto-                  & 1.45712E+22  &
           6.258166791449586E+00 & -3.192714938519402E+01 &  1.606170877827874E+00 &  3.141405466621021E-03 & -2.968297792659612E-05 & -9.055043888369047E-04 \\
                       Charon     &              &
                                 &                        &                        &                        &                        &                        \\
                       barycentre &              &
                                 &                        &                        &                        &                        &                        \\
       \hline
          Nominal                 & 4.0E+04      &
          -1.7644464556E-01      &  9.6798951463E-01      & -6.2289614E-04         & -1.756498228E-02       & -6.03688848E-03        &  4.5521881E-04         \\
                  impactor        &              &
                                 &                        &                        &                        &                        &                        \\
       \hline
      \end{tabular}
      \label{Cartesian}
     \end{table}
     \end{landscape}
%
%

\end{document}